\documentclass[prd,twocolumn,nofootinbib,superscriptaddress,amsmath,amssymb]{revtex4-1}

\usepackage{mathtools}
\usepackage{graphics}
\usepackage{graphicx}
\usepackage{dcolumn}
\usepackage{bm}
\usepackage{dsfont} 
\usepackage{amsmath,amssymb}
\usepackage{hyperref}
\usepackage{tabularx}
\usepackage{epstopdf}
\usepackage[normalem]{ulem}
\usepackage[usenames]{color}
\usepackage{multirow}
\usepackage{makecell}
\usepackage{diagbox}
\usepackage{float}
\allowdisplaybreaks

\hypersetup{
    colorlinks=true,
    linkcolor=blue,
    filecolor=magenta,      
    urlcolor=blue,
    citecolor=blue
}
 
\newcommand{\R}{{\mbox{\tiny R}}}
\newcommand{\A}{{\mbox{\tiny A}}}

\newcommand{\Krange}[2]{#1 MeV $\leq K_0 \leq$ #2 MeV}
\newcommand{\Lrange}[2]{#1 MeV $\leq L_0 \leq$ #2 MeV}
\newcommand{\Ksymrange}[2]{#1 MeV $\leq K_\mathrm{sym,0} \leq$ #2 MeV}
\newcommand{\Mrange}[2]{#1 MeV $\leq M_0 \leq$ #2 MeV}

\newcolumntype{C}[1]{>{\centering\arraybackslash}m{#1}}

\begin{document}

\title{Constraining nuclear matter parameters with GW170817}

\author{Zack Carson}
\affiliation{%
 Department of Physics, University of Virginia, Charlottesville, Virginia 22904, USA
}%

\author{Andrew W. Steiner}
\affiliation{%
 Department of Physics and Astronomy, University of Tennessee, Knoxville, TN 37996, USA
}%
\affiliation{%
 Physics Division, Oak Ridge National Laboratory, Oak Ridge, TN 37831, USA
}%

\author{Kent Yagi}
\affiliation{%
 Department of Physics, University of Virginia, Charlottesville, Virginia 22904, USA
}%

\date{\today}

\begin{abstract}

The tidal measurement of gravitational waves from the binary neutron star merger event GW170817 allows us to probe nuclear physics that suffers less from astrophysical systematics compared to neutron star radius measurements with electromagnetic wave observations.
A recent work found strong correlation among neutron-star tidal deformabilities and certain combinations of nuclear parameters associated with the equation of state. 
These relations were then used to derive bounds on such parameters from GW170817 assuming that the relations and neutron star masses are known exactly.
Here, we expand on this important work by taking into account a few new considerations: (1) a broader class of equations of state; (2) correlations with the mass-weighted tidal deformability that was directly measured with GW170817;
(3) how the relations depend on the binary mass ratio;
(4) the uncertainty from equation of state variation in the correlation relations; (5) adopting the updated posterior distribution of the tidal deformability measurement from GW170817.    
Upon these new considerations, we find GW170817 90\% confidence intervals on nuclear parameters (the incompressibility $K_0$, its slope $M_0$ and the curvature of symmetry energy $K_{\mathrm{sym},0}$ at nuclear saturation density) to be \Krange{69}{352}, \Mrange{1371}{4808}, and \Ksymrange{-285}{7}, which are more conservative than previously found with systematic errors more properly taken into account.

\end{abstract}

\maketitle


\section{Introduction}\label{sec:intro}

One of the largest mysteries in nuclear physics comes from the determination of the equation of state (EoS) of ultra-dense nuclear matter, found exclusively in neutron stars (NSs). 
Many useful relations, such as the one between mass and radius, depend strongly on the EoS, and are vital to the study of nuclear physics to constrain EoSs for supranuclear matter and model-independent parameters that characterize such EoSs.
Indeed, the mass-radius measurement of NSs via X-ray observations have been used to obtain constraints on nuclear matter EoSs~\cite{guver,ozel-baym-guver,steiner-lattimer-brown,Lattimer2014,Ozel:2016oaf}.

Recently, gravitational waves (GWs) from a binary NS merger have been detected (GW170817)~\cite{TheLIGOScientific:2017qsa}, which can also be used to probe nuclear physics~\cite{Abbott2018,Abbott:2018exr,Paschalidis2018,Burgio2018,Malik2018}.
This is mainly because as two NSs in a binary system inspiral due to GW emission, each of them become tidally deformed in response to the tidal gravitational field created by the companion. Such a tidal effect is characterized by  the tidal deformability~\cite{Flanagan2008} which depends strongly on the underlying EoSs. In fact, the leading tidal parameter entering in the gravitational waveform is given by a mass-weighted combination of the two tidal deformabilities $\tilde{\Lambda}$ associated with each NS.
The LIGO Scientific Collaboration and the Virgo Collaboration (LVC) recently placed a 90\% credible bound on $\tilde \Lambda$ as $70 \leq \tilde{\Lambda} \leq 720$~\cite{Abbott2018} (see~\cite{De:2018uhw} for similar bounds).
Coughlin \textit{et al.}~\cite{Coughlin:2018fis} further combined numerical relativity simulations with electromagnetic counterpart signals for GW170817 and derived  $279 \leq \tilde{\Lambda} \leq 822$ (a similar bound was also derived in Radice \emph{et al}.~\cite{Radice2018}).
Such bounds on $\tilde \Lambda$ have also been mapped to those on the NS radius~\cite{Annala:2017llu,Lim:2018bkq,Bauswein:2017vtn,De:2018uhw,Most:2018hfd}.

Given that all of the EoSs proposed so far use certain approximations, one informative approach is to directly measure nuclear physics parameters which parameterize EoSs in a model-independent way. One way to obtain such a parameterization is to Taylor expand the energy per nucleon of asymmetric nuclear matter about the saturation density\footnote{Other ways of parameterizing EoSs include piecewise polytropes~\cite{Read2009,Lackey:2014fwa,Carney:2018sdv} and spectral EoSs~\cite{Lindblom:2010bb,Lindblom:2012zi,Lindblom:2013kra,Lindblom:2018rfr,Abbott:2018exr}. See also~\cite{Landry:2018prl} for a non-parametric inference of EoSs with GW170817.}. Taylor-expanded coefficients include the symmetry energy's slope $L_0$, the incompressibility $K_0$, its slope $M_0$ and the curvature of symmetry energy $K_{\mathrm{sym},0}$.

Interestingly, approximate universal relations exist among nuclear physics parameters mentioned above and NS radius at a given mass~\cite{Alam2016} (see e.g.~\cite{Sotani:2013dga,Silva:2016myw} for other universal relations involving nuclear parameters). 
The authors found that while individual nuclear parameters are only weakly correlated with the stellar radius, linear combinations of the form $K_0+\alpha L_0$ and $M_0+\beta L_0$ become highly correlated, where $\alpha$ and $\beta$ are chosen such that the correlation becomes maximum.

Such work was recently extended by Malik \textit{et al}.~\cite{Malik2018} by considering correlations with individual NS tidal deformabilities.
By taking these relations to be exact and assuming individual NS masses from GW170817 to be $m_1=1.40 M_{\odot}$ and $m_2=1.33 M_{\odot}$, Ref.~\cite{Malik2018} utilized existing measurements on tidal deformability from GW170817~\cite{Abbott2017,Radice2018} and $L_0$~\cite{Abbott2018,Oertel2017,Lattimer2014} to derive constraints on the nuclear incompressibility and the symmetry energies' curvature at saturation density to be $2254 \text{ MeV} \leq M_0 \leq 3631 \text{ MeV}$ and $-112 \text{ MeV} \leq K_{\text{sym},0} \leq -52 \text{ MeV}$, respectively.

This important first-step work of Ref.~\cite{Malik2018} needs to be improved in various ways. In this paper, we propose an extension upon this work by taking into account at least the following five points of interest.
First, we consider a broader class of EoSs by phenomenologically varying nuclear parameters.
Second, we consider correlations among the mass-weighted tidal deformability (instead of the individual tidal deformabilities) and nuclear parameters for various mass ratios. This allows us to eliminate the need to choose specific NS masses $m_1$ and $m_2$, as was done in Ref.~\cite{Malik2018}.
Third, instead of assuming perfect linear regression between nuclear parameters and $\tilde{\Lambda}$, the uncertainty from scatter (corresponding to the EoS variation in the approximate universal relations) is taken into account, including the covariances among parameters.
Fourth, we use the recent updated posterior distribution of the dominant tidal deformability $\tilde \Lambda$ by LVC~\cite{Abbott2018}.
Finally, we investigate constraints on the incompressibility $K_0$ in addition to its slope $M_0$ and the curvature of symmetry energy $K_{\text{sym},0}$.

\subsection{Executive Summary}
\begin{figure}
\begin{center} 
\includegraphics[width=8.5cm]{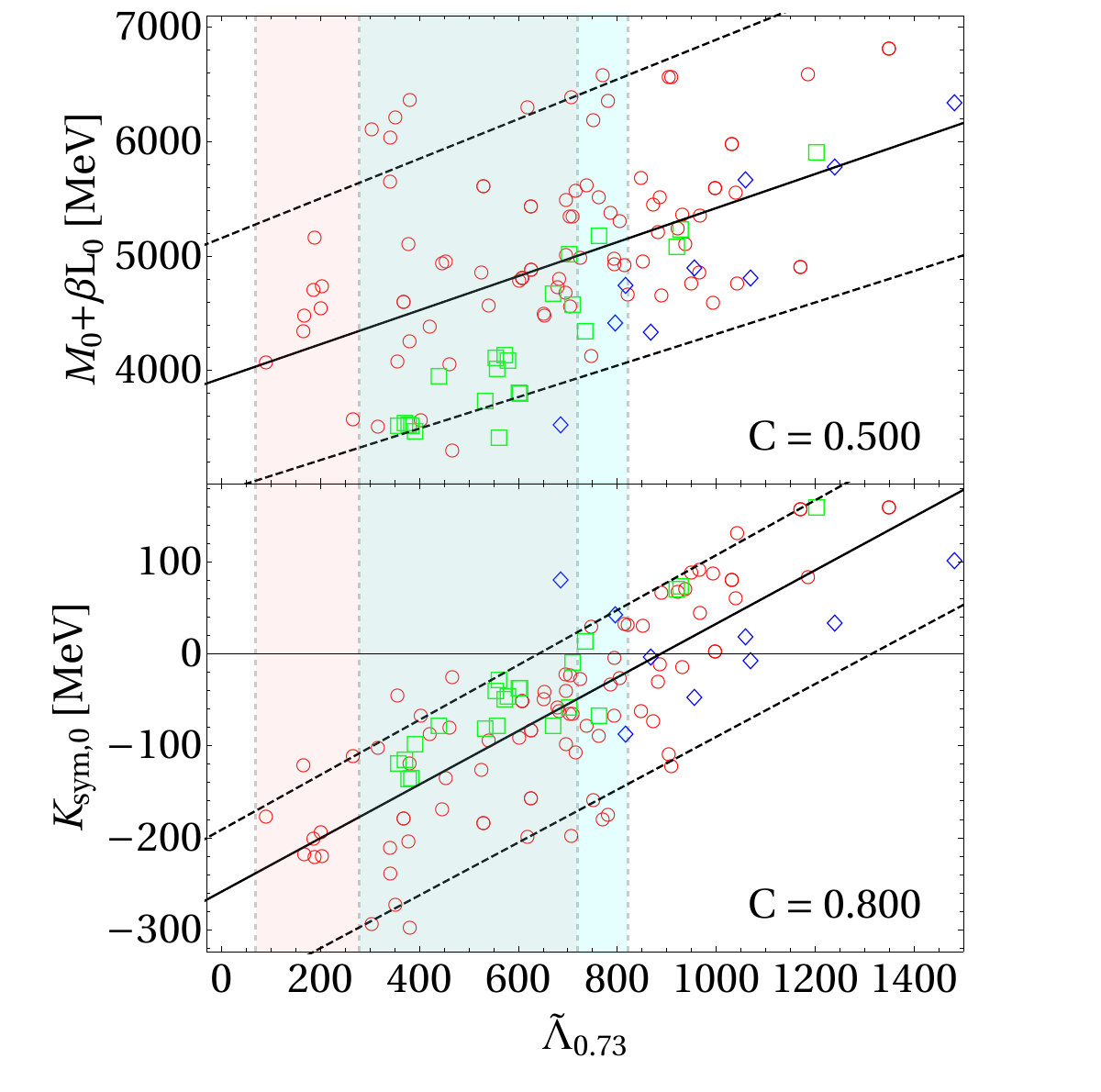}
\end{center}
\caption{
(Top) Correlations between mass-weighted average tidal deformability $\tilde{\Lambda}$ and linear combination of nuclear parameters  (the slope of the incompressibility $M_0$ and the symmetry energy's slope $L_0$) for a chirp mass of $\mathcal{M} = 1.188M_\odot$ corresponding to GW170817, using Skyrme EoSs (green square), relativistic mean field (RMF) EoSs (blue diamond), and phenomenologically varied EoSs (red circle). 
The first two classes were also considered in Ref.~\cite{Malik2018} while the last class is considered here for the first time.
Mass ratio is chosen to be $q=0.87$, consistent with GW170817, though such correlations are insensitive to $q$.
The shaded cyan and magenta regions represent the measurement constraints on $\tilde{\Lambda}$ from GW170817~\cite{Abbott2018,Coughlin:2018fis}.
The solid black line represents the best fit line through the data, while the dashed lines correspond to the lines drawn with 90\% error bars on y-intercept and slope. The Pearson correlation coefficient $C$ measures the amount of correlation ($C=1$ being the absolute correlation and $C=0$ being no correlation) 
The constant $\beta$ for the linear combination $M_0 + \beta L_0$ is chosen to be $\beta=24.28$ such that the correlation between observables becomes $50\%$.
(Bottom) Similar to the top panel but for the curvature of symmetry energy $K_{\text{sym},0}$.
}
\label{fig:combinedKsymCorrelations}
\end{figure}

Let us summarize important results for busy readers.
First, we find new universal relations between $\tilde{\Lambda}$ and $K_0$, $M_0$, or $K_{\text{sym},0}$ (bottom panel of Fig.~\ref{fig:combinedKsymCorrelations}) for a number of mass ratios allowed by GW170817.
Contrary to previous work, we find low-order nuclear parameters $K_0$ and $M_0$ to have very poor correlations, due to the inclusion of a broad new class of EoSs.

Additionally, we studied similar universal relations between $\tilde{\Lambda}$ and linear combinations of nuclear parameters (top panel of Fig.~\ref{fig:combinedKsymCorrelations}), such as $K_0 + \alpha L_0$, $M_0 + \beta L_0$, and $K_{\text{sym},0} + \gamma L_0$.
We found that such relations typically have a stronger correlation than that in the case of individual nuclear parameters.
This is consistent with the findings of Ref.~\cite{Malik2018} on correlations between nuclear parameters and \emph{individual} tidal deformabilities, though the correlations presented here are much lower than that reported in the previous work.
Contrary to Ref.~\cite{Malik2018} where coefficients are chosen such that correlation is maximal, we choose coefficients $\alpha = 2.27$, $\beta = 24.28$, and $\gamma = 0$.
To avoid the propagation of uncertainties from $L_0$, we manually choose $\alpha$ and $\beta$ to be as small as possible, while keeping in mind that the correlation with $\tilde\Lambda$ must be large enough to determine bounds on nuclear parameters.
We arbitrarily choose $\alpha$ and $\beta$ such that correlations are $0.50$ to give one example of the derived bounds.
The parameter $\gamma$ was chosen to be $0$ in order to neglect the additional uncertainty accrued by the addition of $L_0$, possible in this case only due to the high correlations between $K_{\text{sym},0}$ and $\tilde\Lambda$.

\begin{figure}
\begin{center} 
\includegraphics[width=\columnwidth]{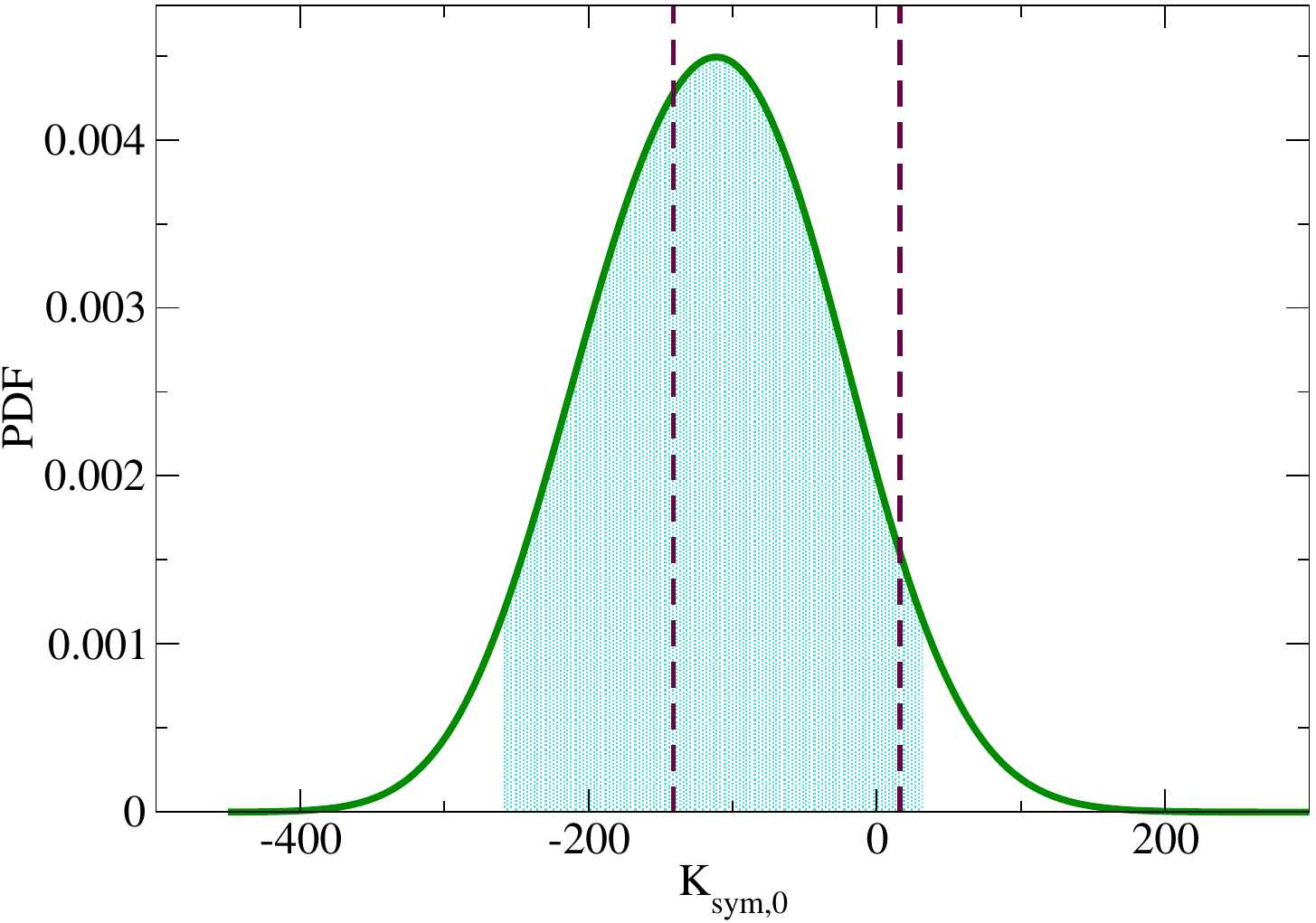}
\end{center}
\caption{
Posterior probability distribution on the curvature of symmetry energy $K_{\text{sym},0}$ as derived in Sec.~\ref{sec:posteriors}. These results utilized a distribution on $\tilde\Lambda$ given by Ref.~\cite{Abbott:LTposterior} as prior information. The shaded region represents the 90\% confidence interval on the data.
Also shown by dashed vertical lines are the constraints found in Ref.~\cite{Malik2018} for priors on the mass-weighted tidal deformability $70 \leq \tilde{\Lambda} \leq 720$ and the symmetry energy's slope \Lrange{30}{86}.
Notice how the inclusion of EoS variation uncertainty from a larger class of EoSs weakens the bounds found in Ref.~\cite{Malik2018}.}
\label{fig:constraints}
\end{figure}

Figure~\ref{fig:constraints} presents 90\% confidence interval on $K_{\text{sym},0}$ after GW170817, based on the universal relation in Fig.~\ref{fig:combinedKsymCorrelations}.
In the computation of these above bounds, the posterior probability distribution on $\tilde\Lambda$ as derived by the LIGO Collaboration~\cite{Abbott:LTposterior} was used.
In particular, we find such bounds to be \Ksymrange{-285}{7}~\footnote{The constraint on $K_{\text{sym,0}}$ bears a close resemblance to that in Refs.~\cite{Margueron:Ksym,Mondal:Ksym}.} at the $90\%$ confidence interval.
Additionally, we find bounds on $K_0$ and $M_0$ to be \Krange{69}{352} and \Mrange{1371}{4808}.
Such results are much weaker than the results found in Ref.~\cite{Malik2018}, born from the inclusion of systematic errors from a broader class of EoSs and the scatter uncertainty from EoS variation on universal relations.
These results lead us to conclude that it is important to account for the large systematic errors accrued from a wider range of valid EoSs and EoS variation in the approximate universal relations.

The organization of this paper is as follows.
We begin with complementary background material on NS tidal deformability in Sec.~\ref{sec:tidal}. 
We continue on discussing the standard asymmetric nuclear matter parameters, and their resulting EoSs and mass-radius relations in Sec.~\ref{sec:nuclear}. 
We next examine the correlations between nuclear matter parameters and mass-weighted tidal deformability in Sec.~\ref{sec:correlations} and further use these results to derive constraints on such nuclear parameters in Sec.~\ref{sec:constraints}. 
We conclude in Sec.~\ref{sec:conclusion} by discussing our results and give possible avenues for future work.
Throughout this paper, we have adopted geometric units of $G=c=1$, unless otherwise stated.

\section{Neutron star tidal deformability}\label{sec:tidal}
We begin by reviewing how one can extract internal structure information of NSs via GW measurement.
In the presence of a neighboring tidal field $\mathcal{E}_{ij}$, such as the binary NS system found in GW170817, NSs tidally deform away from sphericity and acquire a non-vanishing quadrupole moment $Q_{ij}$ that is characterized by the \textit{tidal deformability} $\lambda$~\cite{Flanagan2008,hinderer-love,Yagi2013}:
\begin{equation}
Q_{ij} = - \lambda \mathcal{E}_{ij}.
\end{equation}
Such tidal deformability can be made dimensionless as:
\begin{equation}
\Lambda \equiv \frac{\lambda}{M^5},
\end{equation}
with $M$ representing the stellar mass.
$\Lambda$ can be calculated via the following expression~\cite{hinderer-love,damour-nagar,Yagi2013}:
\begin{align}
\begin{split}
\Lambda &= \frac{16}{15} (1-2\bar{C})^2[2+2\bar{C}(y_\R-1)-y_\R]\\
& \times \{2\bar{C}[6-3y_\R+3\bar{C}(5y_\R-8)]\\ 
&+4\bar{C}^3[13-11y_\R+\bar{C}(3y_\R-2)+2\bar{C}^2(1+y_\R)]\\ 
&+3(1-2\bar{C})^2[2-y_\R+2\bar{C}(y_\R-1)]\ln{(1-2\bar{C})}\}^{-1}.
\end{split}
\end{align}
Here $\bar{C} \equiv M / R$ is the stellar compactness with $R$ representing the NS radius, and $y_\R \equiv y(R)$ with $y(r) \equiv r h'(r)/h(r)$, where a prime stands for taking a derivative with respect to the radial coordinate $r$. $h$ represents the quadrupolar part of the $(t,t)$ component of the metric perturbation satisfying the following differential equation:
\begin{align}\label{eq:deq} 
\begin{split}
& h''+ \Big\{ \frac{2}{r} + \Big\lbrack \frac{2m}{r^2}+4 \pi r (p - \epsilon ) \Big\rbrack e^{\lambda} \Big\} h'\\
&+ \Big\{4 \pi \Big\lbrack 5 \epsilon + 9p + (p+ \epsilon) \frac{d \epsilon}{dp} \Big\rbrack e^{\lambda}- \frac{6}{r^2}e^{\lambda} - \Big(\frac{d \nu}{dr} \Big)^2 \Big\}h =0,
\end{split}
\end{align}
with background metric coefficients $e^{\nu} = g_{tt}$ and $e^{\lambda} = (1-2m/r)^{-1} = g_{rr}$, while $p$ and $\epsilon$ represent pressure and energy density respectively.

The above differential equation can be solved as follows.
First, one needs to prepare unperturbed background solutions by choosing a specific EoS, or $p(\epsilon)$, and solve a set of Tolman-Oppenheimer-Volkoff (TOV) equations with a chosen central density (or pressure) and appropriate boundary conditions (the exterior metric being the Schwarzschild one). The stellar radius is determined from $p(R)=0$ while the mass is given by $M=m(R) = 4\pi \int_0^R \epsilon(r)\, r^2 dr$.
Having such solutions at hand, one then plugs them into Eq.~\eqref{eq:deq} and solves it with the boundary condition $y(0)=2$~\cite{hinderer-love}. 

Because there are two NSs in a binary, two tidal deformabilities $\Lambda_1$ and $\Lambda_2$ associated with each star enter in the gravitational waveform.
However, extracting such parameters independently is challenging due to the strong correlation between them\footnote{One way to cure this problem is to use universal relations between them~\cite{Yagi:2015pkc,Yagi:2016qmr,De:2018uhw,Zhao:2018nyf}.}. 
Thus, one can instead measure the dominant tidal parameter in the waveform, corresponding to the mass-weighted average tidal deformability given by~\cite{Flanagan2008}:
\begin{equation}
\tilde{\Lambda} = \frac{16}{13} \frac{(1+12q) \Lambda_1+(12+q)q^4\Lambda_2}{(1+q)^5},
\end{equation}
where $q \equiv m_2/m_1 (<1)$ is the mass ratio between two stars.

\section{Nuclear matter parameters and equations of state}\label{sec:nuclear}

\subsection{Asymmetric Nuclear Matter Parameters}\label{sec:parameters}

Here we review a generic method of parameterizing EoSs. 
Our starting point is expanding the energy per nucleon $e$ of asymmetric nuclear matter with isospin symmetry parameter $\delta \equiv (n_n-n_p)/n$ (with $n_p$ and $n_n$ representing the proton and neutron number densities respectively and $n \equiv n_p + n_n$) about $\delta=0$ (symmetric nuclear matter case) as~\cite{Vidana2009}:
\begin{equation}
e(n,\delta)=e(n,0)+S_2(n)\delta^2+ \mathcal{O}(\delta^4),
\end{equation}
where $e(n,0)$ corresponds to the energy of symmetric nuclear matter.
$e(n,0)$ and $S_2(n)$ can then be characterized by once again expanding about the \textit{saturation density} $n_0$ as:
\begin{align}
\begin{split}
e(n,0)&=e_0+\frac{K_0}{2} y^2 + \frac{Q_0}{6}y^3 + \mathcal{O}(y^4),\\
S_2(n)&=J_0+L_0 y + \frac{K_\mathrm{sym,0}}{2} y^2 + \mathcal{O}(y^3),
\end{split}
\end{align}
where $y \equiv (n - n_0)/3 n_0$. 
Here, the coefficients are known as the energy per particle $e_0$, incompressibility coefficient $K_0$, third derivative of symmetric matter $Q_0$, symmetry energy $J_0$, its slope $L_0$, and its curvature $K_\mathrm{sym,0}$ at saturation density, respectively. Following Refs.~\cite{Alam2014,Malik2018}, we further introduce the slope of the incompressibility:
\begin{align}
M_0 &= Q_0 + 12 K_0.
\end{align}

In this paper, we investigate correlations between the various nuclear parameters $L_0$, $K_0$ $M_0$, $K_{\text{sym},0}$ and the mass-weighted average tidal deformability $\tilde{\Lambda}$ in order to derive bounds on nuclear parameters from GW170817.  Bounds on $M_0$ and $K_\mathrm{sym,0}$ have previously been derived in Ref.~\cite{Malik2018} using GW170817, which we revisit in this paper. Current experiments and astrophysical observations place bounds on $L_0$ as $40\text{ MeV} < L_0 < 62\text{ MeV}$~\cite{Lattimer2013,Lattimer2014,Tews2017}, and $30 \text{ MeV} < L_0 < 86\text{ MeV}$~\cite{Oertel2017}.

\subsection{Equations of State}\label{sec:eos}
The structure of a NS and its tidal interactions in a binary system rely heavily on the underlying EoS of nuclear matter. 
Because of this, we employ a wide range of 120 different nuclear models in our analysis. 
These EoSs can be classified into three broad categories: 24 non-relativistic EoSs with Skyrme-type interaction, 9 RMF EoSs, and 88 EoSs derived through phenomenological variation.
Following Ref.~\cite{Read2009}, the high-density core EoSs listed above are all matched to the low-density EoS of Douchin and Haensel~\cite{Douchin:2001sv} at the transition density $\epsilon_{\text{tr}}$ such that the pressures are equivalent.

The EoSs in the first two classes are used also in~\cite{Alam2016,Malik2018}.
The Skyrme models used here are: SKa, SKb~\cite{Koehler1976}, SkI2, Sk13, SkI4, SkI5~\cite{Reinhard1995}, SkI6~\cite{Nazarewicz1996}, Sly230a~\cite{Chabanat1997}, Sly2, Sly9~\cite{Chabanat1995}, Sly4~\cite{Chabanat1998}, SkMP~\cite{Bennour1989}, SkOp~\cite{Reinhard1999}, KDE0V1~\cite{Agrawal2005}, SK255, SK272~\cite{Agrawal2003}, Rs~\cite{Friedrich1986}, BSK20, BSK21~\cite{Goriely2010}, BSK22, BSK23, BSK24, BSK25, BSK26~\cite{Goriely2013}.
On the other hand, the RMF models selected are BSR2, BSR6~\cite{Dhiman2007,Agrawal2010}, GM1~\cite{Glendenning1991}, NL3~\cite{Lalazissis1997}, NL3$\omega \rho$~\cite{Carriere2003}, TM1~\cite{Sugahara1994}, DD2~\cite{Typel2010}, DDH$\delta$~\cite{Gaitanos2004},  DDME2~\cite{Typel1999}.

One of the new EoS classes that we consider is the phenomenological
EoSs (PEs). To construct these EoSs, we followed the
formalism of Ref.~\cite{Margueron2018} by randomly sampling nuclear parameters $J_0$, $K_0$, $L_0$, $Q_0$ and $K_\mathrm{sym,0}$ as found in Table I of the above reference.  Following this, nonphysical EoSs with acausal structure ($v_s > c$), or having decreasing pressure as a function of density were removed.

\begin{figure}
\begin{center} 
\includegraphics[width=\columnwidth]{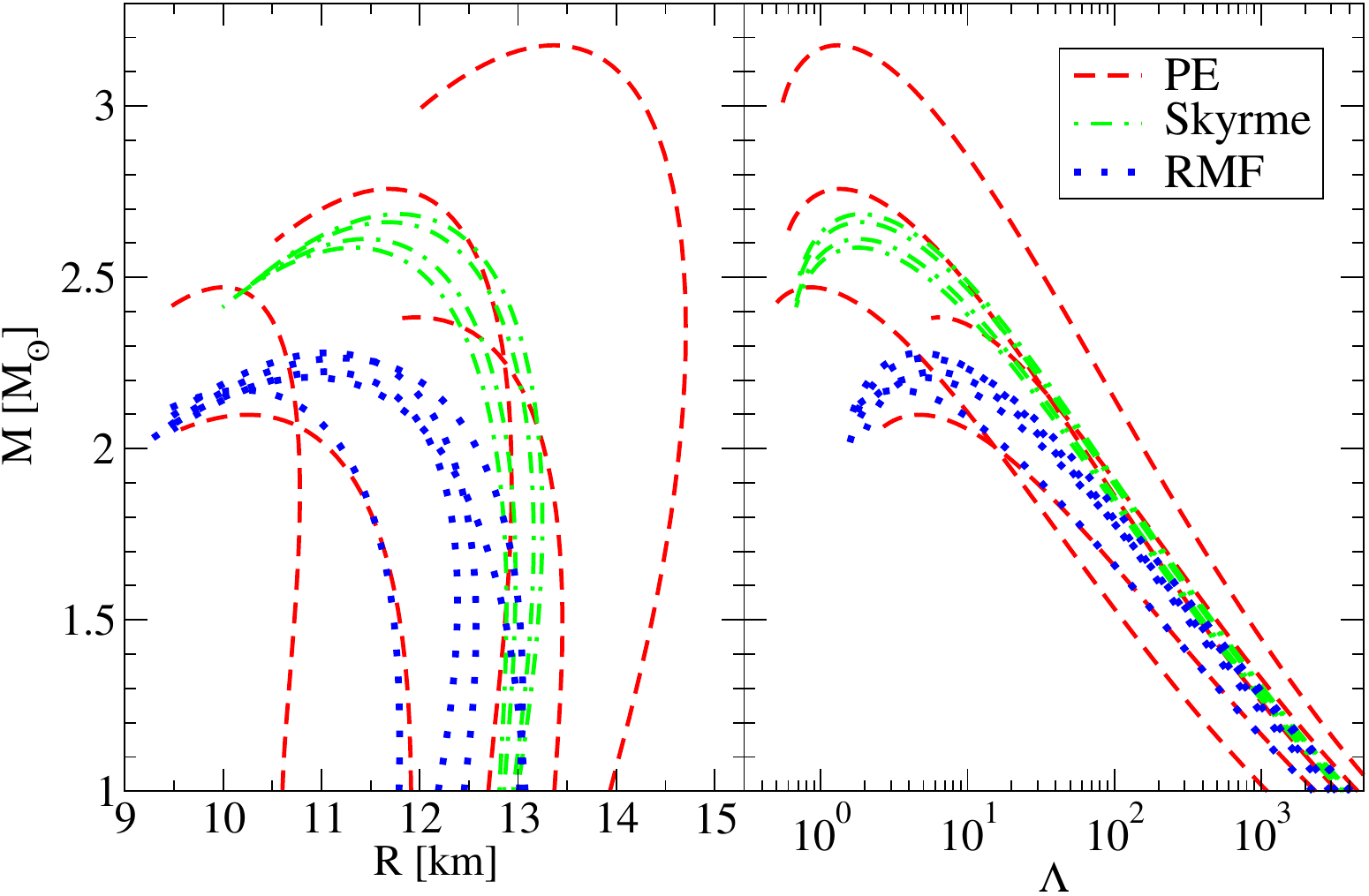}
\end{center}
\caption{Neutron star mass as a function of radius (left) and tidal deformability $\Lambda$ (right) for a representative set of the EoSs used in our analysis, separated into groups of phenomenological (red dashed), RMF  (green dotted-dashed) and Skyrme-type (blue dotted). 
Observe how Skyrme and RMF EoSs follow self-consistent behavior, while PEs see a wide variance in properties such as maximum mass and radius, due to the nature of the random sampling in nuclear parameters.
}
\label{fig:M_RL}
\end{figure} 

Figure~\ref{fig:M_RL} presents the relations among the NS mass, radius and tidal deformability for selected EoSs in different classes mentioned above. Observe that RMF EoSs tend to produce NSs with larger radii and maximum mass than those for Skyrme-types, while the PE ones generate NSs with a wide range of properties.

\section{Correlations between tidal deformability and Nuclear Parameters}\label{sec:correlations}

In this section, we study correlations among nuclear parameters and tidal deformability, where the latter can be measured from GW observations. 
The amount of correlation between two variables $x$ and $y$ with $N$ data points can be quantified by the Pearson correlation coefficient $C$ defined by:
\begin{equation}\label{eq:correlation}
C(x,y) = \frac{\sigma_{xy}}{\sqrt{\sigma_{xx} \sigma_{yy}}},
\end{equation}
where the covariances $\sigma_{xy}$ are given by:
\begin{equation}\label{eq:covariance}
\sigma_{xy} = \frac{1}{N} \sum\limits^N_{i=0}x_i y_i - \frac{1}{N^2}\Big( \sum\limits^N_{i=0}x_i \Big) \Big( \sum\limits^N_{i=0}y_i \Big).
\end{equation}
$C= 1$ represents absolute correlation, while $C= 0$ corresponds to having no correlation. 

\subsection{$\tilde \Lambda$ versus Nuclear Parameters}

\begin{figure*}
\begin{center} 
\includegraphics[width=.96\textwidth]{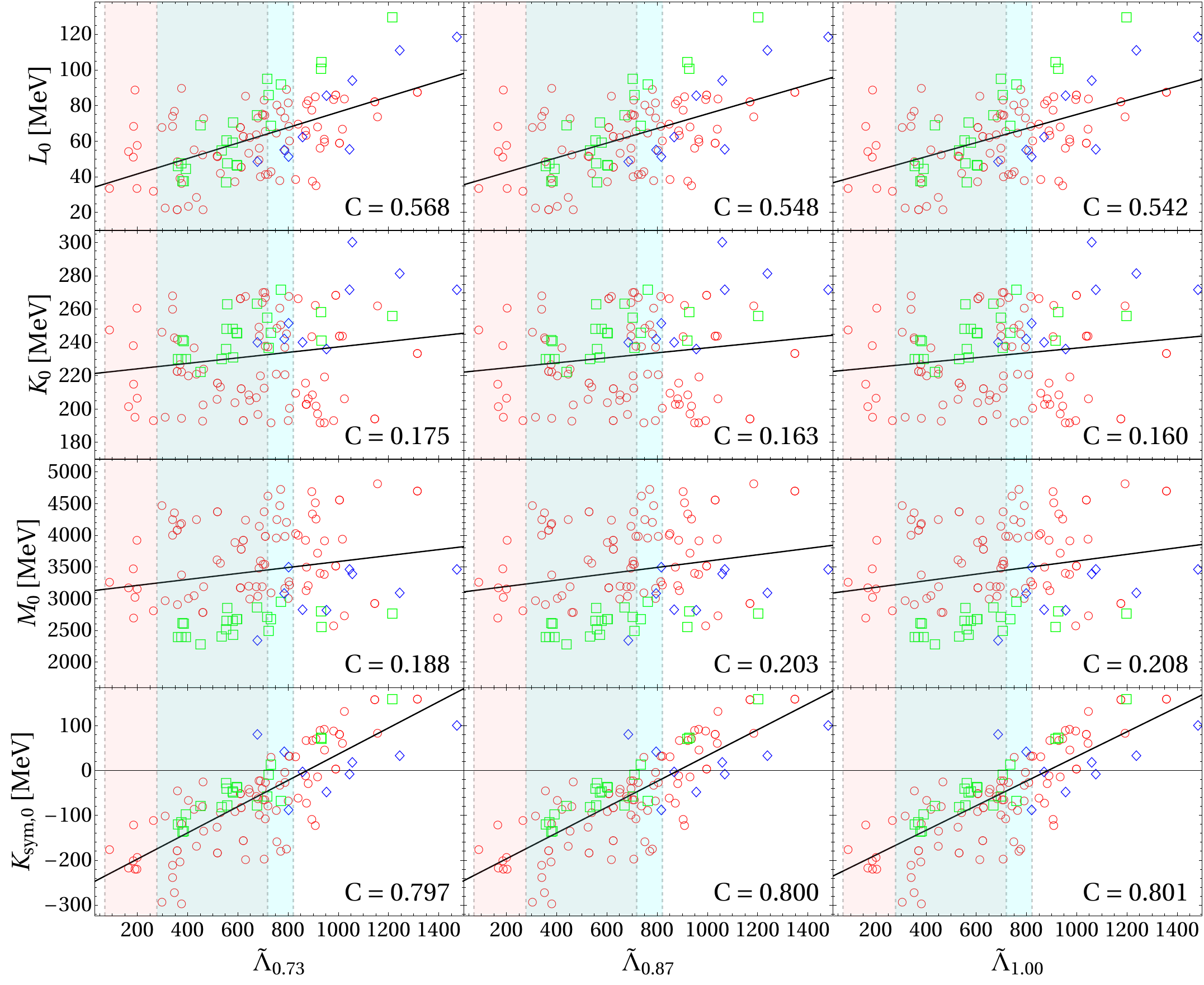}
\end{center}
\caption{
Correlations between nuclear parameters $L_0$, $K_0$, $M_0$, $K_{\text{sym},0}$ and the mass-weighted average tidal deformability $\tilde{\Lambda}_q$ for a chirp mass of $\mathcal{M} = 1.188M_\odot$ corresponding to GW170817, using Skyrme EoSs (green square), RMF EoSs (blue diamond), and PEs (red circle). Mass ratios are chosen as $q=0.73$ (left), $0.875$ (middle), and $1.00$ (right) consistent with GW170817.
The shaded cyan and magenta regions represent the measurement constraints on $\tilde{\Lambda}$ from GW170817~\cite{Abbott2018,Coughlin:2018fis}.
The solid black line in each panel represents the best fit line through the data, and the Pearson correlation coefficient $C$ measures the amount of correlation ($C=1$ being the absolute correlation and $C=0$ being no correlation).
}
\label{fig:nuclear}
\end{figure*} 

Reference~\cite{Malik2018} first studied the universal relations between nuclear parameters and the tidal deformability for isolated neutron stars. 
The authors then map this to the GW measurement on $\tilde \Lambda$ by using yet another universal relation between $\tilde \Lambda$ and $\Lambda_{1.4}$ for a specific choice of masses in a binary neutron star that is consistent with GW170817. 
However, the mass ratio $q \in \lbrack 0.73, 1.00\rbrack$~\cite{TheLIGOScientific:2017qsa} for this event has not been measured very precisely (the lower bound of this constraint has recently been improved to $0.8$ in Ref.~\cite{Coughlin:2018fis}), and the question arises as to whether such relation holds for various $q$. 
As we show in Appendix~\ref{appendix:LTL14}, indeed the universal relation is highly insensitive to the choice of $q$. 
This suggests that there are universal relations between nuclear parameters and $\tilde \Lambda$ for a given chirp mass $\mathcal{M}$ which has been measured with high accuracy for GW170817. 
Finding these universal relations is the focus of this section. 
Universal relations involving $\tilde \Lambda$ are, in some sense, practically more useful than those with $\Lambda_{1.4}$, because the former is a quantity which can be directly measured from GW observations.

Figure~\ref{fig:nuclear} shows the correlations between nuclear parameters ($L_0$, $K_0$, $M_0$, $K_{\text{sym},0}$), and the mass-weighted average tidal deformability $\tilde{\Lambda}_q$ evaluated at mass ratios of $q=0.73$, $0.87$ and $1.00$. 
The linear regression shown in each panel represents the best fit line describing the relation between nuclear parameters and $\tilde{\Lambda}$. 
Observe that $K_0$ and $M_0$ show very poor correlations, resulting from a disconnect between PEs and EoSs found in Ref.~\cite{Malik2018}.
On the other hand, higher order parameter $K_{\text{sym},0}$ sees a fairly strong correlation of $\sim 0.80$.
It is noted that PEs typically have values of $K_0$ that are much lower than those for Skyrme or RMF EoSs, while $M_0$ is much higher, and $L_0$ and $K_{\text{sym},0}$ are very similar. Let us emphasize that we have restricted to physically valid PEs which have increasing pressure, and this is why we do not have PEs with e.g. $M_0 < 2500$ MeV~\footnote{This does \emph{not} mean that Skyrme and RMF EoSs with $M_0 < 2500$ are nonphysical.}.
The above finding indicates a necessity in using a large number of EoSs as nuclear parameters can take on a much wider range of values than considered in~\cite{Malik2018}.
Observe also that the behavior of the scattering and the amount of correlation found in Fig.~\ref{fig:nuclear} is not very sensitive to $q$.  This can also be seen from Fig.~\ref{fig:correlation}, where correlations between various nuclear parameters and $\tilde{\Lambda}$ are plotted as a function of mass ratio q.

\subsection{$\tilde \Lambda$ versus linear combinations of nuclear parameters}\label{sec:lin-comb}

References~\cite{Alam2016,Malik2018} report that correlations among nuclear parameters and NS observables become stronger if one considers certain combinations of the former, which we study here. In Refs.~\cite{Oertel2017,Lattimer2014,Tews2017}, tight constraints on the slope of the symmetry energy $L_0$ were derived. Thus we focus on constraining the incompressibility $K_0$, its slope $M_0$, and the symmetry energies' curvature $K_{\text{sym},0}$, utilizing prior bounds on $L_0$ and $\tilde{\Lambda}$ by considering linear combinations of the form $K_0+\alpha L_0$, $M_0+\beta L_0$, and $K_{\text{sym},0}+\gamma L_0$ with some coefficients $\alpha$, $\beta$ and $\gamma$. In previous literature~\cite{Alam2016,Malik2018}, these coefficients are chosen such that correlations become maximum.

\begin{figure}
\begin{center} 
\includegraphics[width=\columnwidth]{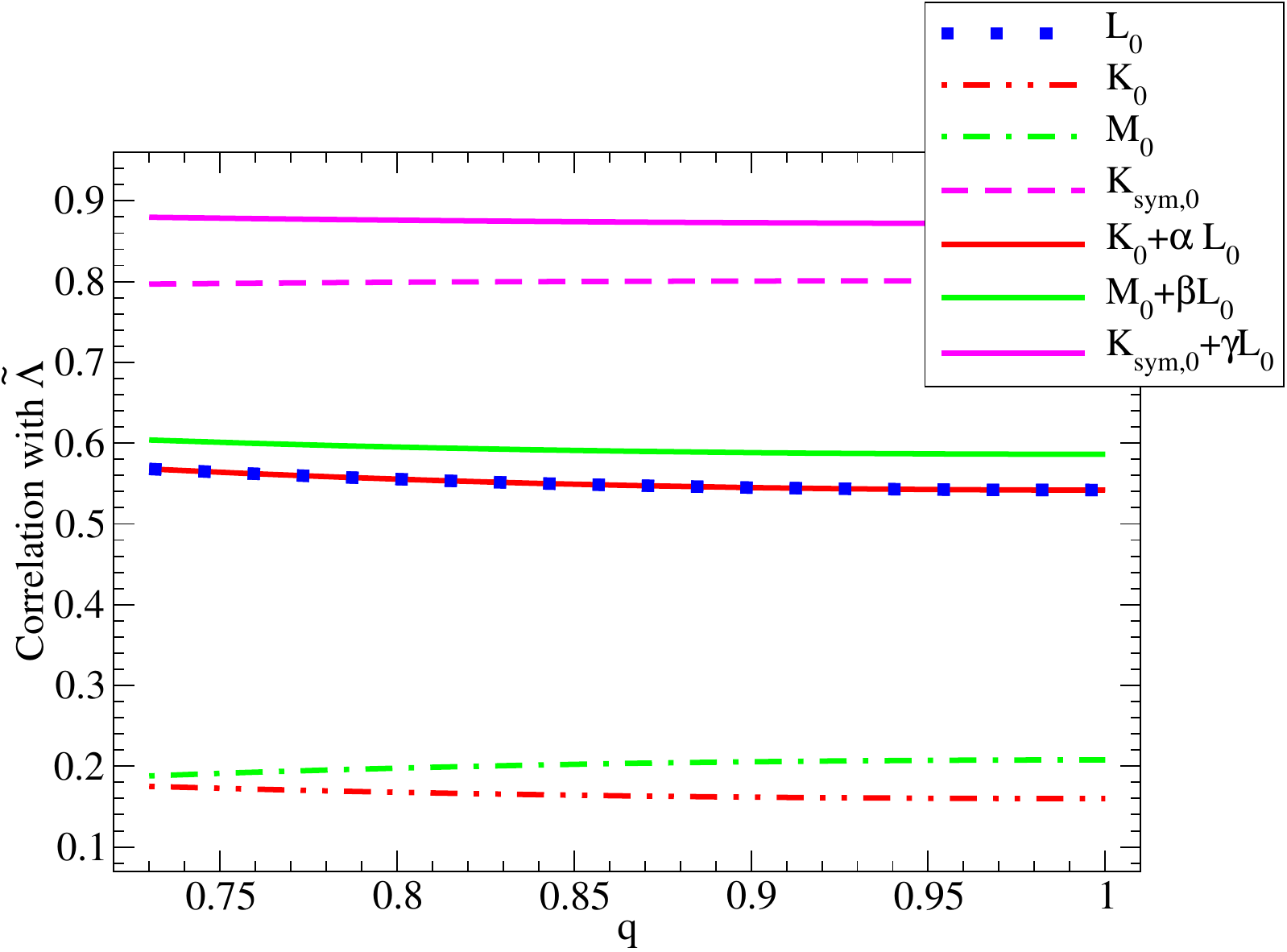}
\end{center}
\caption{Correlations with $\tilde{\Lambda}$ as a function of mass ratio $q$ for $K_0+\alpha L_0$, $M_0+\beta L_0$, and $K_{\text{sym},0}+\gamma L_0$ for $\mathcal{M} = 1.188M_\odot$.
These are much stronger than those involving single nuclear parameters, which is also shown for reference. 
Here we choose $\alpha=2.27$ and $\beta=24.28$ giving 50\% correlations in the universal relations, while we choose $\gamma=2.63$ such that the correlation is maximized (see Sec.~\ref{sec:constraints} for more details).
Observe that correlations do not change significantly with $q$ across a wide range of mass ratios.
}
\label{fig:correlation}
\end{figure} 

\begin{figure}
\begin{center} 
\includegraphics[width=\columnwidth]{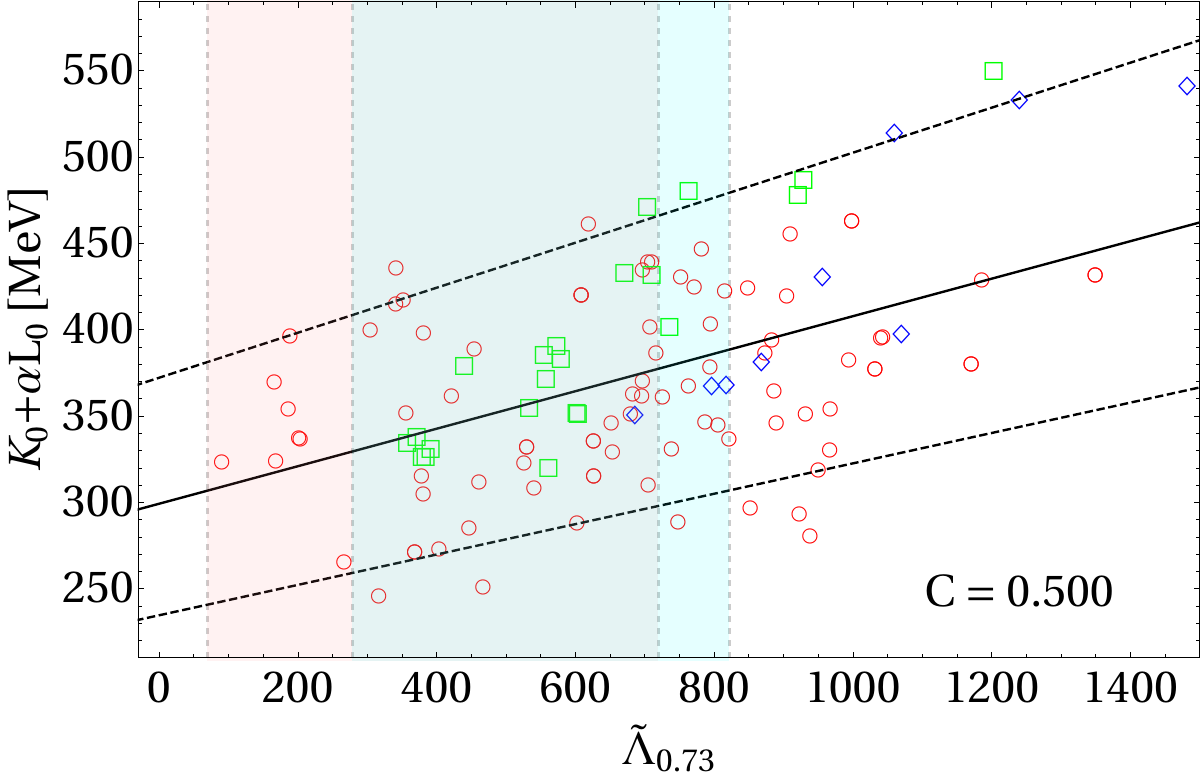}
\end{center}
\caption{
Similar to Fig.~\ref{fig:combinedKsymCorrelations} but for a linear combination of nuclear parameters $K_0+\alpha L_0$. 
As discussed in Section~\ref{sec:constraints}, the linear coefficient $\alpha$ is chosen to be $\alpha=2.27$ such that 50\% correlation is achieved in the universal relations.
Observe that the correlations here are much stronger than those involving single nuclear parameters, as in Fig.~\ref{fig:nuclear}. 
}
\label{fig:combinedLinearCorrelations}
\end{figure}

Figure~\ref{fig:correlation} presents the correlations between $\tilde{\Lambda}$ and linear combinations of nuclear parameters as a function of mass ratio $q$. 
We found that the values of $\alpha$ and $\beta$ which give maximal correlation are unnecessarily large. For practical purposes, we choose here $\alpha =2.27$ and $\beta = 24.28$, such that a correlation of 50\% in the universal relations is achieved. For $\gamma$, we use $\gamma = 2.63$ which maximizes the correlation, as was done previously (see Sec.~\ref{sec:constraints} for more details).
For reference, we also show correlations involving single nuclear parameters. 
Observe that the former correlations are much stronger than the latter (except for $K_0 + \alpha L_0$ whose correlation is comparable to that of $K_0$) and remain to be strong over the acceptable region of mass ratio.
This implies that our choice of $q$ when calculating bounds on nuclear parameters does not matter significantly.
Therefore, we consider universal relations evaluated at the central mass ratio of $q=0.87$, shown in Figs.~\ref{fig:combinedKsymCorrelations} and~\ref{fig:combinedLinearCorrelations}.
Also notice how linear combinations involving high-order nuclear parameter $K_{\text{sym},0}$ continue to significantly outperform lower-order parameters.

\section{Constraints on nuclear matter parameters}\label{sec:constraints}

\begin{figure*}
\begin{center} 
\includegraphics[width=.32\textwidth]{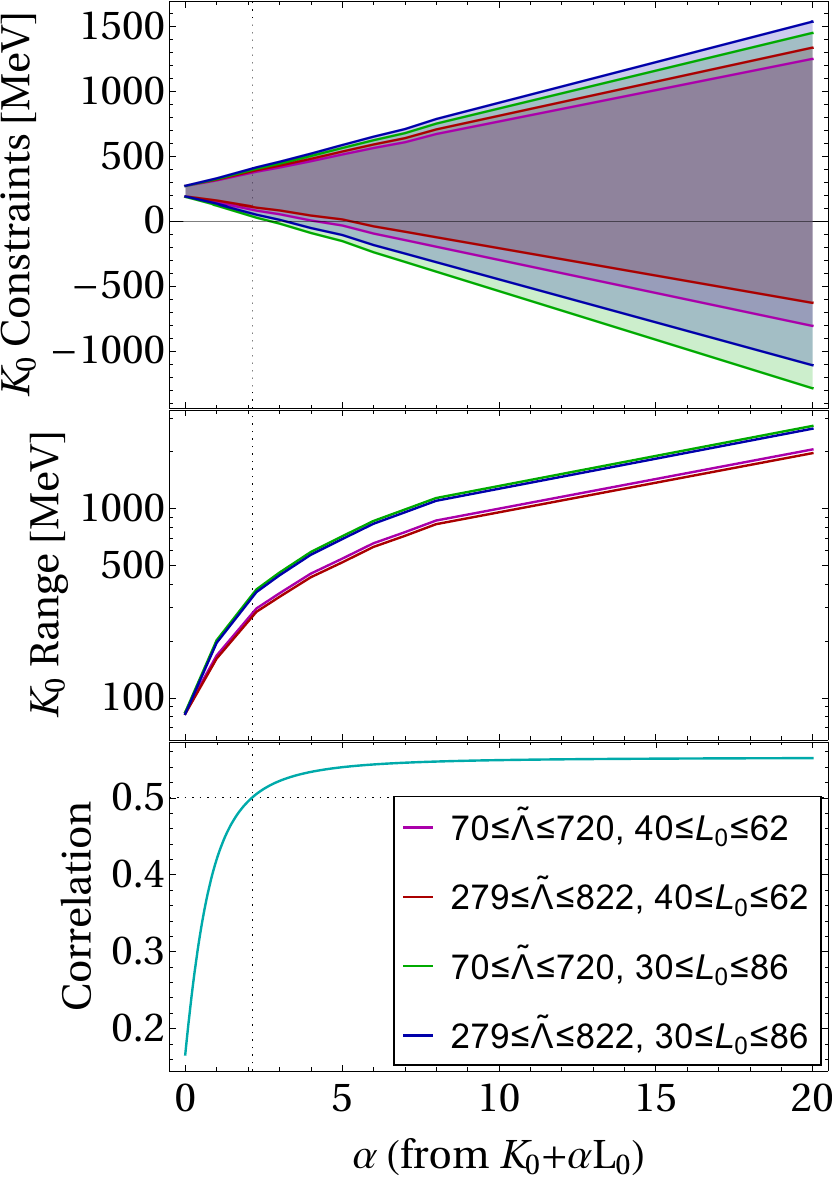}
\includegraphics[width=.32\textwidth]{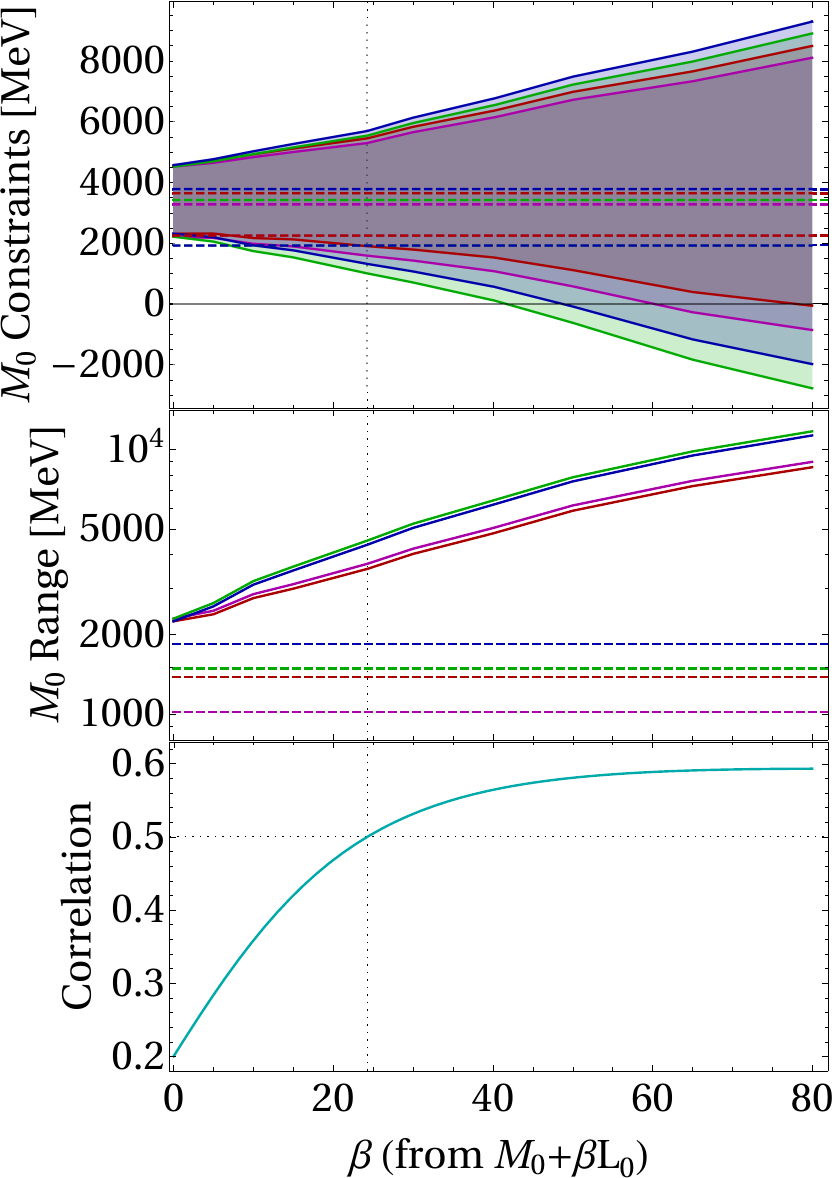}
\includegraphics[width=.32\textwidth]{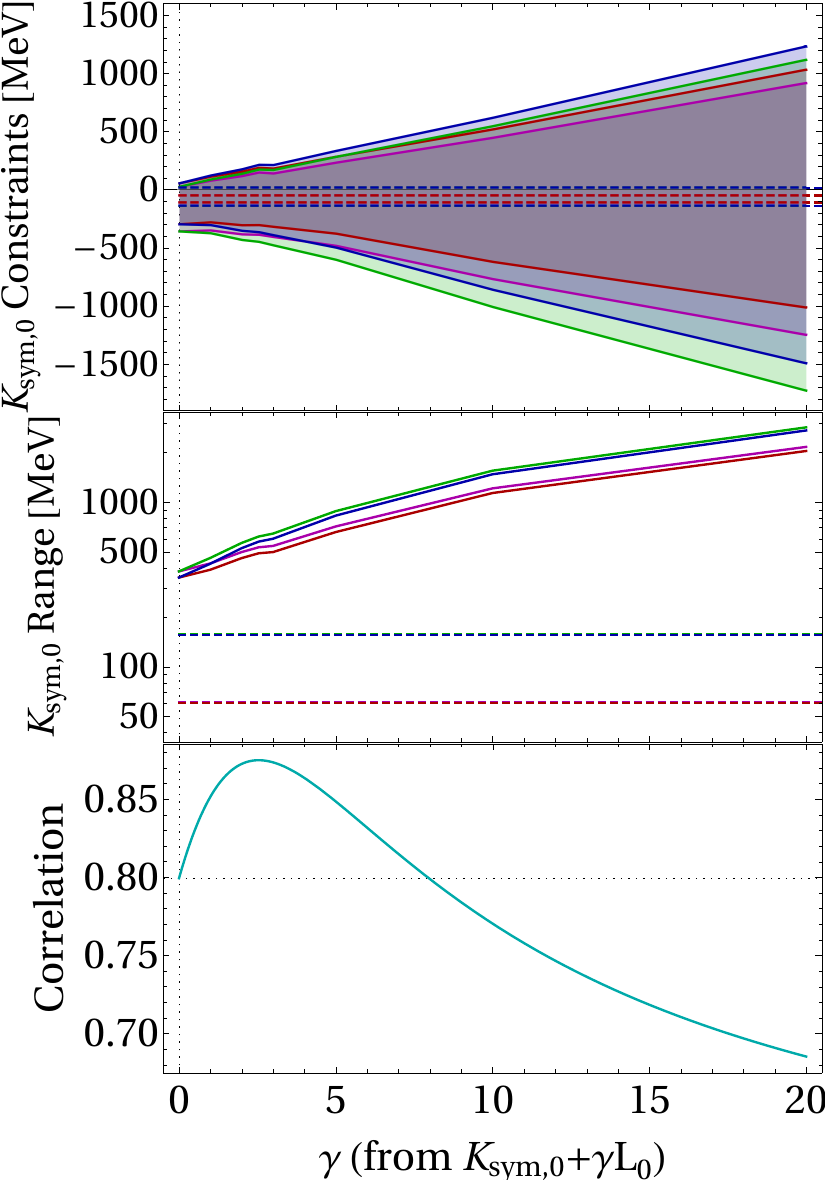}
\end{center}
\caption{
Comparisons between nuclear parameter constraints and correlations with $\tilde{\Lambda}$, evaluated at $q=0.87$, as functions of $\alpha$, $\beta$, and $\gamma$ for $K_0$ (left), $M_0$ (middle), and $K_{\text{sym},0}$ (right). 
(Top) Estimated nuclear parameter constraints for different combinations of priors: (i) $70 \leq \tilde{\Lambda} \leq 720$, $40 \leq L_0 \leq 62$ (magenta); (ii) $279 \leq \tilde{\Lambda} \leq 822$, $40 \leq L_0 \leq 62$ (red); (iii) $70 \leq \tilde{\Lambda} \leq 720$, $30 \leq L_0 \leq 86$ (green); (iv) $279 \leq \tilde{\Lambda} \leq 822$, $30 \leq L_0 \leq 86$ (blue).
Dashed horizontal lines correspond to bounds derived by~\cite{Malik2018} under similar prior assumptions.
(Middle) Constraint ranges given as the difference between upper and lower limits.
(Bottom) Correlations between $\tilde{\Lambda}$ and linear combinations of nuclear parameters. 
Dotted vertical lines represent chosen values of $\alpha$, $\beta$, and $\gamma$ for deriving final bounds on the nuclear parameters. These values for $\alpha$ and $\beta$ are chosen to give a 50\% correlation while that for $\gamma$ gives a 80\% correlation.}
\label{fig:ConstraintsOfAlpha}
\end{figure*}

Let us now use the approximate universal relations among combined nuclear parameters and $\tilde \Lambda$ to derive bounds on the former from the measurement of the latter with GW170817. 
In this section, we detail the process used to estimate nuclear parameter bounds, taking into account the EoS scattering uncertainty.
We offer two alternative methods of accomplishing this.
In Sec.~\ref{sec:linear}, we offer a crude estimation of the constraints by finding linear regressions between the nuclear parameters and $\tilde\Lambda$.
We estimate $90\%$ confidence integrals on such regressions which allows us to predict bounds on nuclear parameters.
The linear regressions provide ready-to-use type results that can easily be implemented as the measurement on $\tilde \Lambda$ from GW170817 are updated.
In Sec.~\ref{sec:posteriors}, we detail a more comprehensive analysis in which we first compute the 2-dimensional probability distribution between the nuclear parameters and $\tilde\Lambda$.
We then combine this with the probability distribution on $\tilde\Lambda$ computed by Ref.~\cite{Abbott:LTposterior} to estimate the posterior distribution on nuclear parameters $K_0$, $M_0$, and $K_{\text{sym},0}$.

\subsection{Constraint Estimation via Linear Regressions}\label{sec:linear}
In this simple error analysis, we first construct linear regressions of the form $(a\pm \delta_a^{\pm})\tilde{\Lambda} + (b \pm \delta_b^{\pm})$ on the relations evaluated at the central mass ratio of $q=0.87$ with the ``90\%'' error on the slope and $y$-intercept as follows:

\begin{equation}
\label{eq:regressions1}
\frac{K_0}{\text{MeV}} + \alpha \frac{L_0}{\text{MeV}} = 0.1086^{+0.02172}_{-0.02064}\, \tilde{\Lambda} + 299.1^{+72.97}_{-64.60}\,,
\end{equation}
\vspace{-5mm}
\begin{equation}
\label{eq:regressions2}
\frac{M_0}{\text{MeV}} + \beta \frac{L_0}{\text{MeV}} = 1.488^{+0.2456}_{-0.2038}\, \tilde{\Lambda} + 3929^{+1226}_{-990.2}\,,
\end{equation}
\vspace{-5mm}
\begin{equation}
\label{eq:regressions3}
\frac{K_{\text{sym},0}}{\text{MeV}} + \gamma \frac{L_0}{\text{MeV}} = 0.2915^{+0.007287}_{-0.004080}\, \tilde{\Lambda} - 259.1^{+67.36}_{-118.9}\,.
\end{equation}

The uncertainties on the slope and $y$-intercept, $\delta_a^{\pm}$ and $\delta_b^{\pm}$, are found by varying the upper and lower error bars throughout the parameter space, selecting only combinations of $\delta_a^{\pm}$ and $\delta_b^{\pm}$ which form ``90\% error lines" $(a \pm \delta_a^{\pm})\tilde{\Lambda} + (b \pm \delta_b^{\pm})$ containing 90\% of the data points between them.
Further, we choose the ``best fit'' 90\% error lines by minimizing the residual sum of squares, $\sum_{1=1}^{n}(y_i-f(x_i))^2$, as denoted by the dashed black lines in Figs.~\ref{fig:combinedKsymCorrelations}, and~\ref{fig:combinedLinearCorrelations}.
For reference, the covariances $\sigma_{ab}$ from Eq.~\eqref{eq:correlation} between $a$ and $b$ are found to be approximately $0.7274$, $124.5$, and $0.4235$ for Eqs.~\eqref{eq:regressions1}--\eqref{eq:regressions3}, respectively.
Using this method of uncertainty prediction, we find a 90\% confidence interval on the value of $b$ and $a$, allowing us to account for the EoS scatter in the universal relations when deriving bounds on nuclear parameters from GW170817, as we will study next.

Let us now use Eqs.~\eqref{eq:regressions1}--\eqref{eq:regressions3} to derive bounds on $K_0$, $M_0$, and $K_{\text{sym},0}$, as was done in Ref.~\cite{Malik2018}. 
We utilize prior bounds obtained from nuclear experiments and astrophysical observations as $L_0 \in \lbrack 40 , 62 \rbrack$ MeV~\cite{Lattimer2013} and $L_0 \in \lbrack 30 , 86 \rbrack$ MeV~\cite{Oertel2017,Lattimer2014,Tews2017}, as well as tidal deformability ranges of $\tilde{\Lambda} \in \lbrack 70 , 720 \rbrack$~\cite{Abbott2018}  and $\tilde{\Lambda} \in \lbrack 279 , 822 \rbrack$~\cite{Coughlin:2018fis}.
Utilizing the 90\% confidence interval's range on y-intercepts, we find constraints on $K_0$, $M_0$, and $K_{\text{sym},0}$ within priors of $L_0$ and $\tilde{\Lambda}$ such that minimal and maximal values of nuclear parameters are obtained. 
Therefore, 2 constraints on $\tilde{\Lambda}$ and 2 constraints on $L_0$ allow us to derive 4 possible constraints on each nuclear parameter $K_0$, $M_0$, and $K_{\text{sym},0}$. 
This particular method of estimating the probability distribution is conservative by nature, and also takes into account the uncertainty from scatter in our relations.

The top panels of Fig.~\ref{fig:ConstraintsOfAlpha} show comparisons between estimated nuclear parameter limits, while the central panels show constraint ranges (maximum value minus minimum value) as the linear combination coefficient ($\alpha$, $\beta$, or $\gamma$) is increased. The bounds are stronger if the ranges are smaller. For comparison, the bottom panels display the correlation between the nuclear parameter combinations and $\tilde{\Lambda}$. Observe that the bounds become weaker as one increases the coefficients, as we are introducing an additional source of uncertainty from $L_0$. Does this mean that it is always better to set the coefficients to 0 and consider universal relations with individual nuclear parameters? The answer is no because correlations are too small when $\alpha=\beta=0$, as can be seen from the bottom panels of Fig.~\ref{fig:ConstraintsOfAlpha}. If such correlations are too small, the relations can easily be affected by the addition of new EoSs and the bounds derived from these relations become unreliable. 

Therefore, we need to find the balance between having large enough correlations and yet to have reasonable bounds on the nuclear parameters.
Regarding $\alpha$ and $\beta$, notice that bounds on $K_0$ and $M_0$ increase approximately linearly with the coefficients, while correlations with $\tilde{\Lambda}$ quickly asymptote to values of $\sim 0.60$.
Thus we choose $\alpha=2.27$ and $\beta = 24.28$ such that correlations evaluated at central mass ratio $q=0.87$ are an arbitrary value of $C=0.50$, chosen to keep correlations as high as possible, while keeping $\alpha$ and $\beta$ as small as possible to avoid the propagation of uncertainty in $L_0$.
Regarding $\gamma$, because $K_{\text{sym},0}$ starts off with strong correlation at $\gamma=0$, we choose this value to remove any additional uncertainty in $\gamma$ and $L_0$ from our calculations
(Note this can not be done for the cases of $K_0$ and $M_0$ due to weak individual correlations with $\tilde{\Lambda}$).
Observe that the coefficient choices discussed in Ref.~\cite{Malik2018}, to maximize correlations to the level of $0.8$ and beyond is not necessarily applicable to every situation. 
As seen in Fig.~\ref{fig:ConstraintsOfAlpha}, high correlations are unobtainable for linear combinations involving $K_0$ and $M_0$, yielding no bounds under such a selection criteria.
Instead, reducing the threshold to $0.50$ returns constraints as shown below, albeit being less reliable.

Table~\ref{tab:Constraints} summarizes the bounds on the nuclear parameters with these fiducial choices of $\alpha$, $\beta$ and $\gamma$, using both this method of constraint estimation, and the method described in Sec.~\ref{sec:posteriors}.
The constraints on $M_0$ and $K_{\mathrm{sym},0}$ are additionally visualized in Fig.~\ref{fig:constraints}.
Notice how our conservative constraints (found by using the largest-range priors on both $L_0$ and $\tilde{\Lambda}$) on the slope of incompressibility and the curvature, \Mrange{955}{5675} and \Ksymrange{-358}{23}, are much weaker than those found in Ref.~\cite{Malik2018} (see Fig.~\ref{fig:constraints}), due to the consideration of EoS scatter uncertainty, and of additional PEs with a wider range of nuclear values.
We observe that the constraints derived here on $K_{\text{sym,0}}$ show good agreement with that of Refs.~\cite{Margueron:Ksym,Mondal:Ksym}.
Let us emphasize that the bounds on $K_0$ and $M_0$ should be considered as rough estimates, as the correlation of 0.50 is not very large; thus these bounds are more easily affected by inclusion of yet additional EoSs than the bounds on $K_\mathrm{sym,0}$.

\begin{table*}
\caption{
GW170817 constraints on the incompressibility $K_0$ (top row), its slope $M_0$ (middle row), and the symmetry energy curvature $K_{\text{sym},0}$ for 4 different sets of priors on $L_0$~\cite{Lattimer2013,Oertel2017,Lattimer2014,Tews2017}, and $\tilde{\Lambda}$~\cite{Coughlin:2018fis,Abbott2018}.
These quantities are computed using two different methods: (i) a simple linear regression estimation described in Sec.~\ref{sec:linear} (labeled ``Method 1" on the right column), and (ii) a comprehensive computation of the nuclear parameter posterior probability distributions described in Sec.~\ref{sec:posteriors} (labeled ``Method 2" on the left column).
The two methods show moderate agreement, although the first method can be seen to over-estimate the errors -- thus we recommend the use of the more accurate distributions computed in method 2, which properly take into account the covariances between the parameters, as well as utilizes the full posterior distribution on $\tilde\Lambda$ derived by the LIGO Collaboration~\cite{Abbott:LTposterior}.
The bounds on nuclear parameter $M_0$ and $K_{\text{sym},0}$ are weaker but more reliable than those found in~\cite{Malik2018} due to the inclusion of scatter uncertainty in our linear regressions. The bounds on $K_0$ and $M_0$ should be taken as a rough estimate as the correlation in universal relations that were used to derive them are not large, and thus, may be subject to change with inclusion of further EoSs.
}\label{tab:Constraints} 
\begin{tabular}{|| C{3.3cm} || C{4.6cm} || C{4.6cm} | C{4.6cm}||@{}m{0pt}@{}} 
\cline{2-3}\cline{3-4}
\multicolumn{1}{c|}{} & Method 2 & \multicolumn{2}{c||}{Method 1}\\
\hline
\diagbox[width=3.4cm]{\hspace*{2pt} \raisebox{2pt}{$L_0$ [MeV]}}{\raisebox{-5pt}{$\tilde{\Lambda}$} \hspace*{0.5cm}} & $\tilde\Lambda$ Posterior Distribution~\cite{Abbott:LTposterior} & 70--720~\cite{Abbott2018} & 279--822~\cite{Coughlin:2018fis} \\
\hline
\hline
 40--62~\cite{Lattimer2013,Lattimer2014,Tews2017} & \makecell{\\ \Krange{69}{352}  \\ \\ \Mrange{1371}{4808} \\ \\ \Ksymrange{-285}{7} \\ \\} & \makecell{\\ \Krange{100}{375}  \\ \\ \Mrange{1538}{5433} \\ \\ \Ksymrange{-358}{23} \\ \\}  & \makecell{\\ \Krange{118}{388} \\ \\ \Mrange{1849}{5609} \\ \\ \Ksymrange{-298}{54} \\ \\ } \\
 \hline
 30--86~\cite{Oertel2017} & \makecell{\\ \Krange{123}{330}  \\ \\ \Mrange{1884}{4635} \\ \\ \Ksymrange{-285}{7} \\ \\} & \makecell{\\ \Krange{45}{398}  \\ \\ \Mrange{955}{5675} \\ \\ \Ksymrange{-358}{23} \\ \\} & \makecell{\\ \Krange{63}{411} \\  \\ \Mrange{1266}{5852} \\  \\ \Ksymrange{-298}{54}\\ \\ }\\
 \hline
\end{tabular}
\end{table*}

\subsection{Constraint Estimation via LIGO Posterior Distributions}\label{sec:posteriors}
In this section, we offer a more comprehensive method of estimating nuclear matter constraints than was found in Sec.~\ref{sec:linear}.
Previously, a rough estimate on the nuclear matter constraints was computed by finding linear regressions between $\tilde\Lambda$ and nuclear parameters.
By estimating the $90\%$ errors on these lines, bounds on the nuclear parameters were manually approximated.
In this section, we improve upon this method by (i) properly taking into account the covariance between $\tilde\Lambda$ and nuclear parameters by generating a multivariate probability distribution, and (ii) taking into account the full posterior probability distribution on $\tilde\Lambda$ as derived by the LIGO Collaboration~\cite{Abbott:LTposterior}.

\begin{figure*}
\begin{center} 
\includegraphics[width=.32\linewidth]{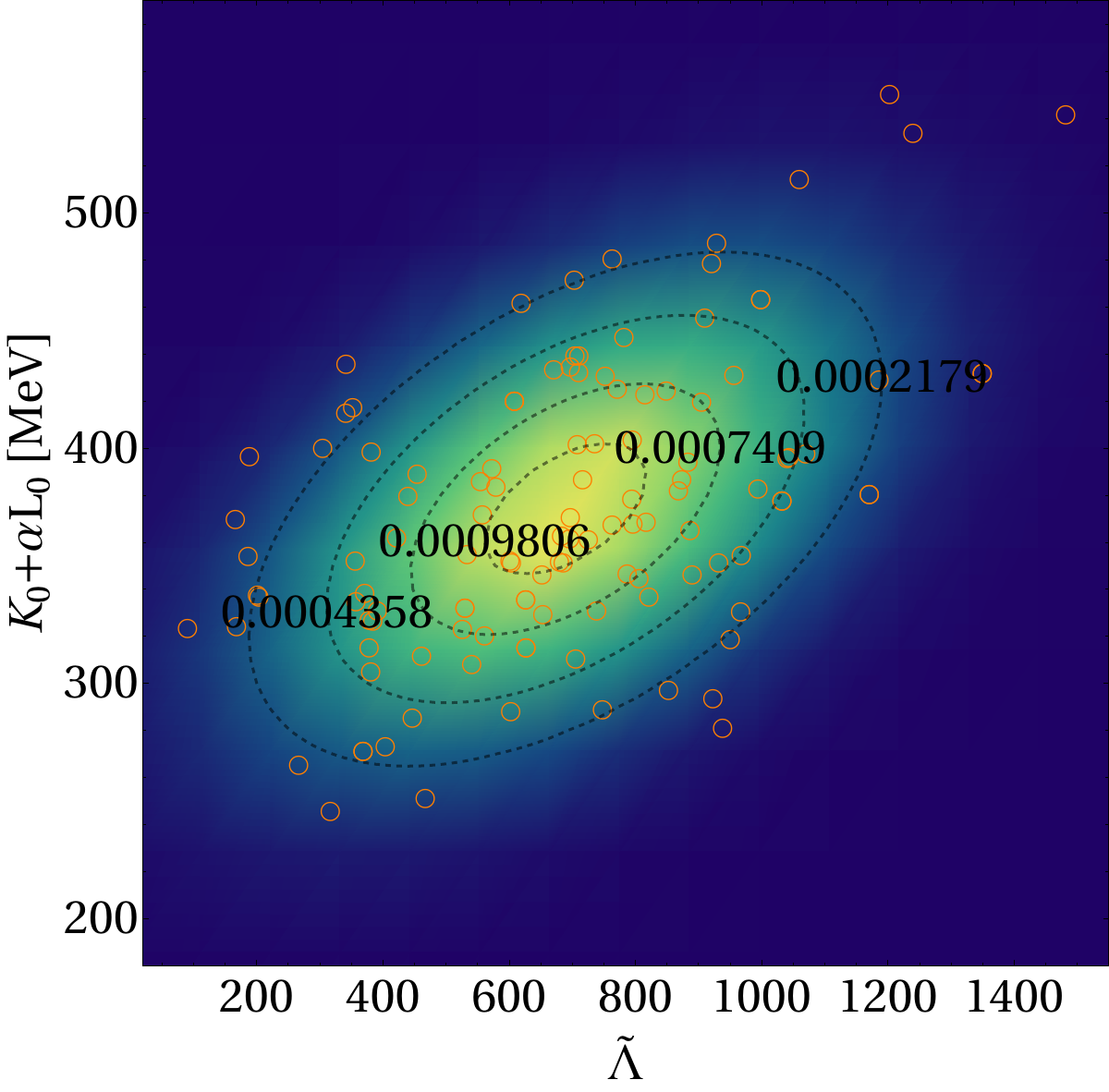}
\includegraphics[width=.32\linewidth]{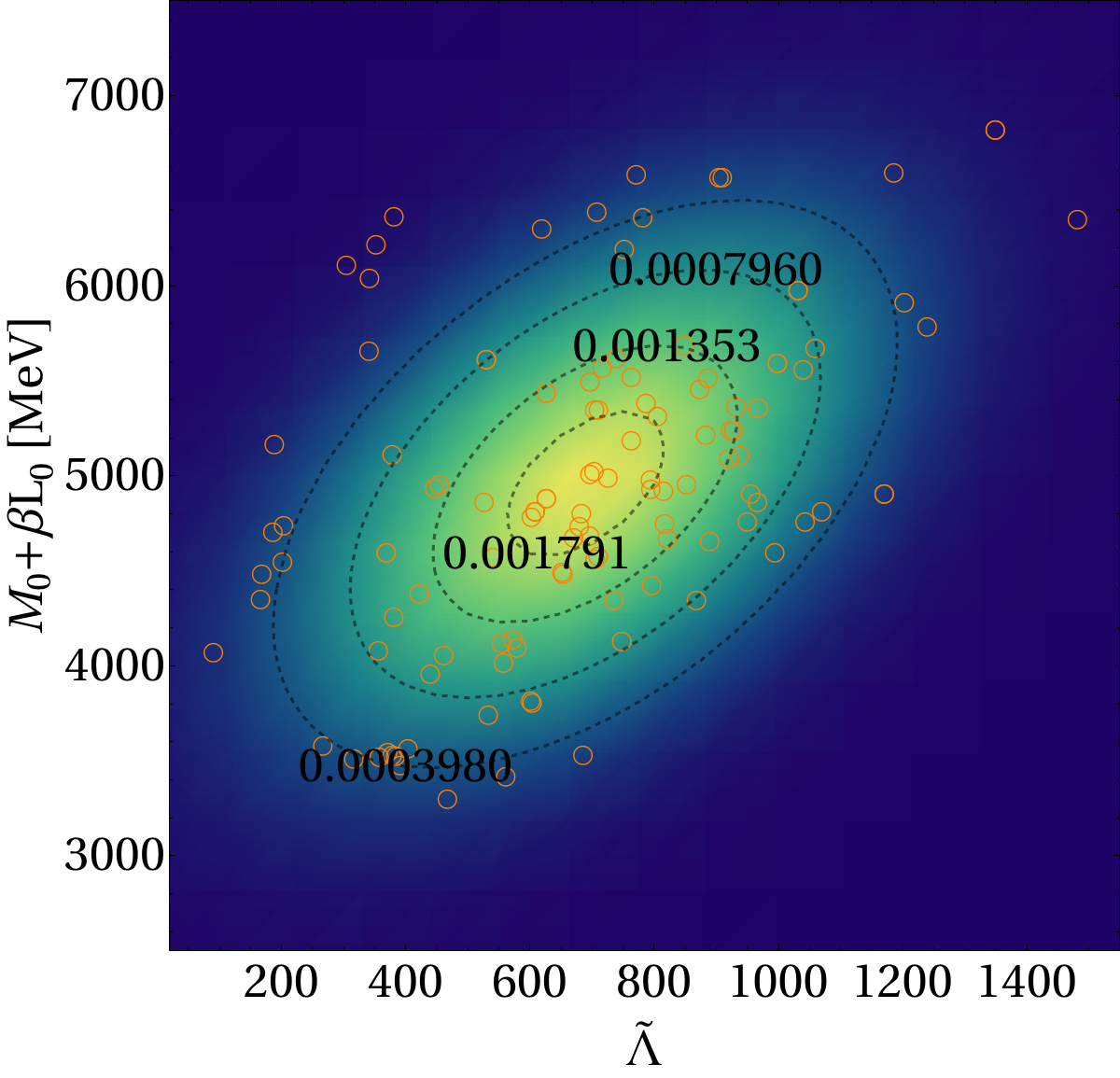}
\includegraphics[width=.32\linewidth]{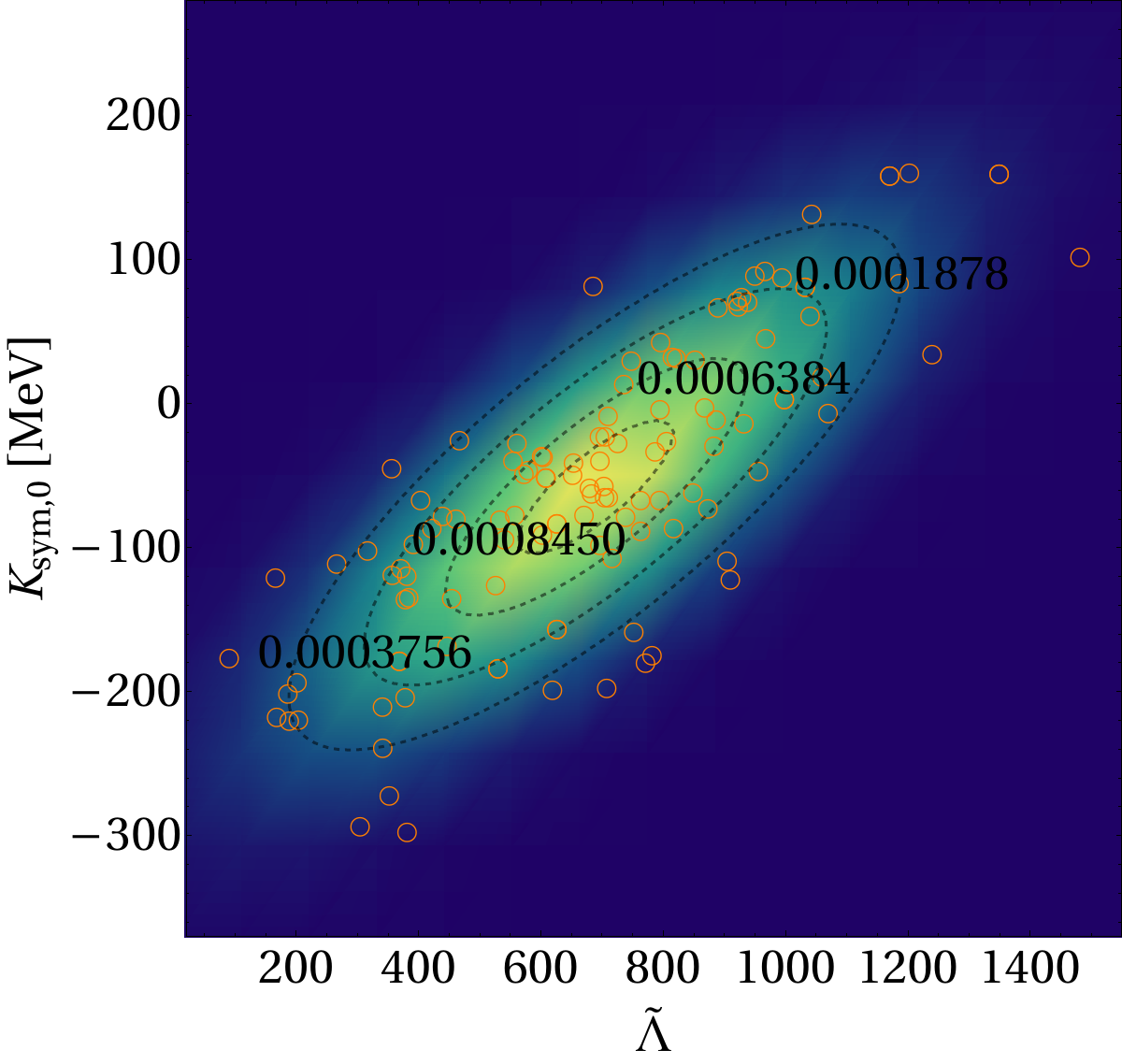}
\end{center}
\caption{2-dimensional normalized probability distributions between $\tilde\Lambda$ and nuclear parameters $K_0+\alpha L_0$ (left), $M_0+\beta L_0$ (center), and $K_{\text{sym},0}$ (right) generated via Eq.~(\ref{eq:probDist}).
Overlayed on the distributions is the set of 120 data points corresponding to each EoS used in this investigation for comparison.
Observe how the multivariate Gaussian distributions indicate high levels of covariance between the variables, indicating the importance of estimating bounds using this method.}
\label{fig:probDist}
\end{figure*}

We begin by generating the 2-dimensional probability distribution between $\tilde\Lambda$ and the nuclear parameters, taking into account the specific covariances between them.
For the example of $K_{\text{sym},0}$ the distribution is given by:
\begin{equation}\label{eq:probDist}
P(\tilde{\Lambda},K_{\text{sym},0})=\frac{1}{2\pi\sqrt{|\Sigma|}}e^{-\frac{1}{2}(x-\mu)^T\Sigma^{-1}(x-\mu)},
\end{equation}
where $x$ is the 2-dimensional vector containing $\tilde\Lambda$ and the given nuclear parameter, $\mu$ is the 2-dimensional vector containing the expected values of $x$, and $\Sigma$ is the $2\times2$ covariance matrix defined with elements given by Eq.~(\ref{eq:covariance}).
This distribution is displayed in Fig.~\ref{fig:probDist} for each nuclear parameter.
Notice here the high degree of covariance between the variables used in this analysis - indicative of the importance for using this method of constraint extraction.

Following this, we compute the conditional probability distributions on nuclear matter parameters given a tidal observation of $\tilde\Lambda_\text{obs}$.
Following Ref.~\cite{jensen_2007}, the one-dimensional conditional probability distribution on nuclear parameter $Y$ is then given by
\begin{equation}\label{eq:conditional}
P(Y|\tilde\Lambda_\text{obs})\sim \mathcal{N}\left(\mu_Y+\frac{\sigma_Y}{\sigma_{\tilde\Lambda_\text{obs}}}C(\tilde\Lambda_\text{obs}-\mu_{{\tilde\Lambda_\text{obs}}}),(1-C^2)\sigma_Y^2\right).
\end{equation}
In the above expression, $\mathcal{N}(\mu,\sigma^2)$ is the normal distribution with mean and variance $\mu$ and $\sigma^2$, and $\mu_\A$ and $\sigma_\A^2$ are the mean and variances of $Y$ and $\tilde\Lambda_\text{obs}$.

\begin{figure}
\begin{center} 
\includegraphics[width=\columnwidth]{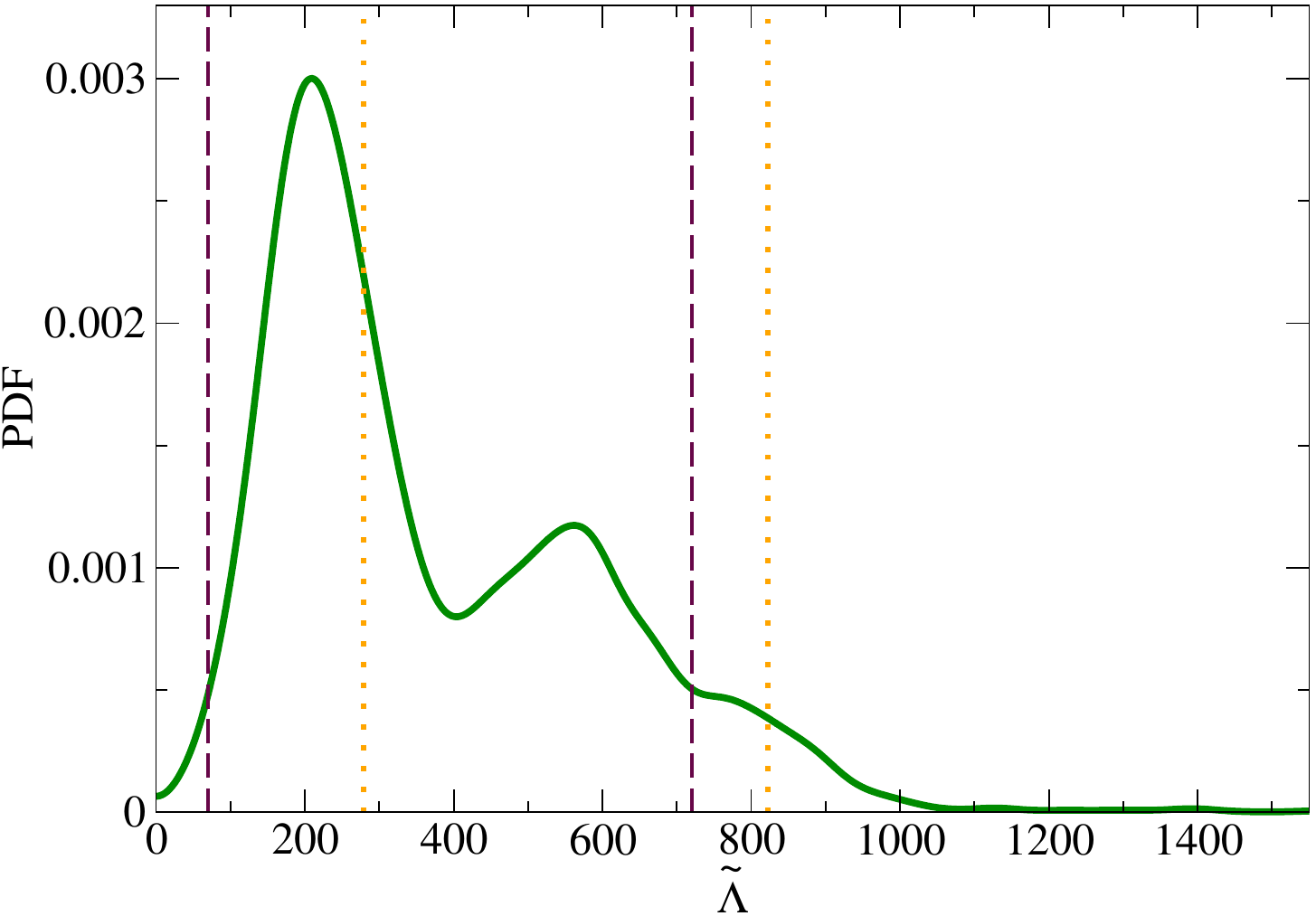}
\end{center}
\caption{Posterior probability distribution on $\tilde\Lambda$ as derived by the LIGO Collaboration in Ref.~\cite{Abbott:LTposterior}. We take this as a prior distribution when computing the posteriors on nuclear parameters. Additionally shown are the GW and EM counterpart bounds of $70 \leq \tilde{\Lambda} \leq 720$~\cite{Abbott2018} (dashed maroon) and $279 \leq \tilde{\Lambda} \leq 822$~\cite{Coughlin:2018fis} (dotted orange) for comparison.}
\label{fig:LigoPrior}
\end{figure}

Next, we extract the one-dimensional probability distributions on $K_0+\alpha L_0$, $M_0 + \beta L_0$, and $K_{\text{sym},0}$ by combining the one-dimensional conditional distributions $P(Y|\tilde\Lambda)$ found in Eq.~\eqref{eq:conditional} with the probability distribution $P_{\text{LIGO}}(\tilde\Lambda)$ on $\tilde\Lambda$  derived by the LIGO Collaboration in Ref.~\cite{Abbott:LTposterior} for GW170817, shown in Fig.~\ref{fig:LigoPrior}.
For example, the posterior probability distribution on $K_0+\alpha L_0$ is given by:
\begin{equation}
P(K_0+\alpha L_0)=\int\limits_{-\infty}^{\infty}P(K_0+\alpha L_0|\tilde{\Lambda})P_{\text{LIGO}}(\tilde{\Lambda})d\tilde{\Lambda},
\end{equation}
and similarly for $M_0+\beta L_0$ and $K_{\text{sym},0}$.
Additionally, to find the probability distributions on $K_0$ and $M_0$, we perform one last integration over the prior probability distribution of $L_0$, assumed to be Gaussian with standard deviation $\sigma=\frac{1}{2}(80+36)$ and mean $\mu=\frac{1}{2}(80-36)$~\cite{Oertel2017} (or $\sigma=\frac{1}{2}(62+40)$ and $\mu=\frac{1}{2}(62-40)$~\cite{Lattimer2013,Lattimer2014,Tews2017} for the alternative priors on $L_0$).
For example, the probability distribution on $K_0$ is given by:
\vspace{-.1cm}
\begin{equation}
P(K_0)=\int\limits_{-\infty}^{\infty}P(K_0+\alpha L_0)P(L_0)dL_0,
\end{equation}
with $\alpha=2.27$.

\begin{figure*}
\begin{center} 
\includegraphics[width=.32\linewidth]{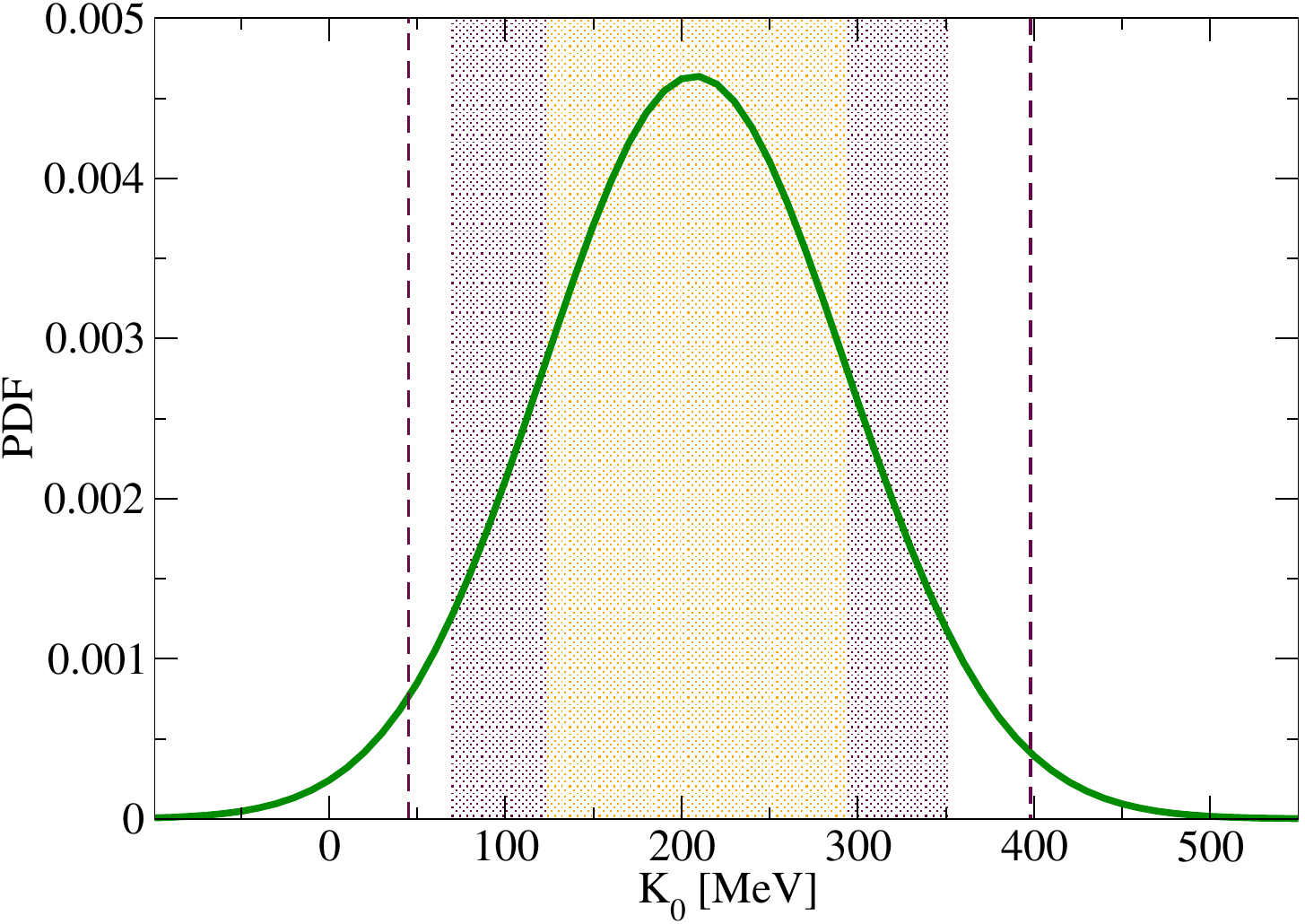}
\includegraphics[width=.32\linewidth]{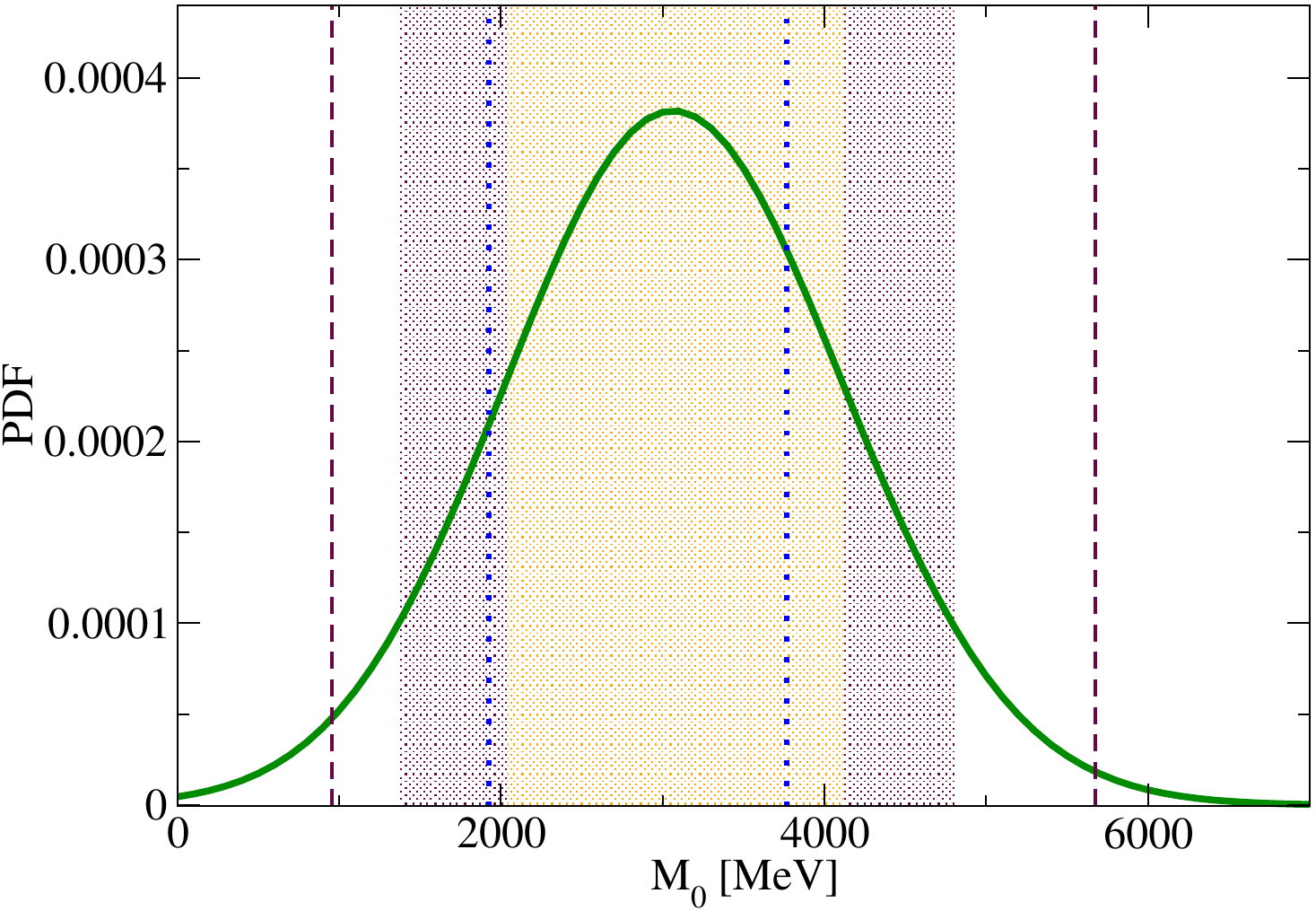}
\includegraphics[width=.32\linewidth]{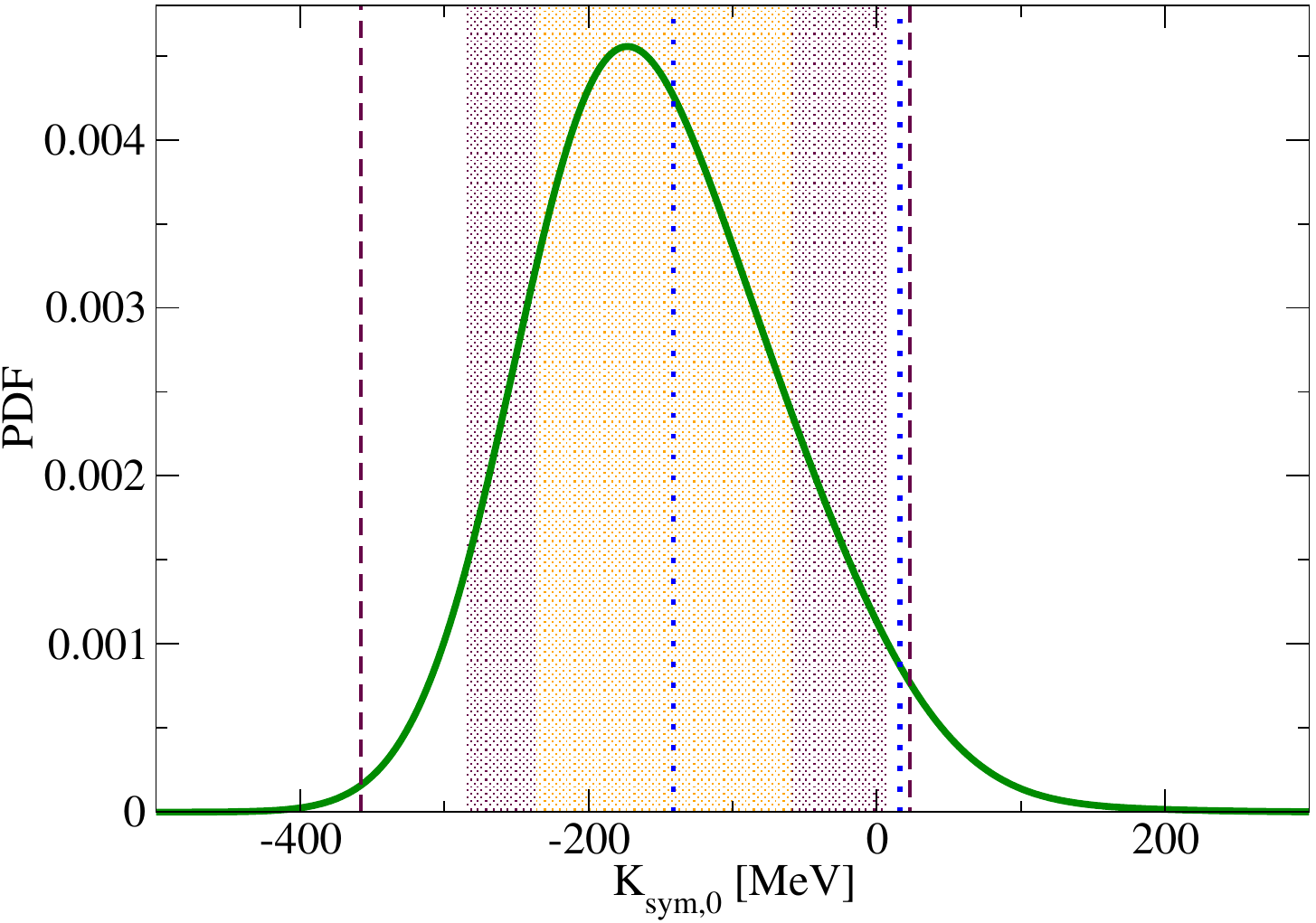}
\end{center}
\caption{Resulting posterior distributions on the nuclear incompressibility $K_0$ and its slope $M_0$, and the curvature of symmetry energy $K_{\text{sym},0}$ derived by integrating over the product of one-dimensional conditional probability distributions ($P(K_0 + \alpha L_0|\tilde{\Lambda})$, $P(M_0+\beta L_0|\tilde{\Lambda})$, and $P(K_{\text{sym},0}|\tilde{\Lambda})$) and $P_{\text{LIGO}}(\tilde{\Lambda})$ in Fig.~\ref{fig:LigoPrior}.
Further, for the linear combinations of $K_0 + \alpha L_0$ and $M_0 + \beta L_0$, one more integration over the probability distribution of $30 \text{ MeV}\leq L_0 \leq 86 \text{ MeV}$ was required to directly find the posterior distributions on $K_0$ and $M_0$. 
Overlayed are the resulting 68\% and 90\% confidence intervals in orange and maroon respectively, as well as the corresponding bounds calculated in Sec.~\ref{sec:linear} shown by dashed maroon vertical lines.
Additionally shown in dotted blue are the corresponding bounds on $M_0$ and $K_{\text{sym},0}$ computed by Ref.~\cite{Malik2018}, using priors of $\tilde{\Lambda} \in \lbrack 70,720 \rbrack$ and $L_0 \in \lbrack 30, 86 \rbrack$ MeV.
Observe how the results for the 90\% confidence intervals obtained in this section are slightly smaller than those found in Sec.~\ref{sec:linear}, indicating that previously the error was slightly overestimated (as the probability distribution on $\tilde\Lambda$ and the covariances between $\tilde\Lambda$ and the nuclear parameters were not properly taken into account).}
\label{fig:Posteriors}
\end{figure*}

The results of these computations are shown in Fig.~\ref{fig:Posteriors} for the more conservative priors on $L_0$.
We observe that $K_0$, $M_0$, and $K_{\text{sym},0}$ now obey distributions that look like skewed Gaussians centered at $K_0=208^{+86}_{-85} \text{ MeV}$, $M_0=3075^{+1045}_{-1033}\text{ MeV}$, and $K_{\text{sym},0}=-156^{+97}_{-81} \text{ MeV}$ ($68\%$ standard deviations).
This results in  $90\%$ confidence intervals of \Krange{69}{352}, \Mrange{1371}{4808}, and \Ksymrange{-285}{7}.
We tabulate these values for both priors on $L_0$ in Table~\ref{tab:Constraints} for comparison to the simple method described in Sec.~\ref{sec:linear}.
These constraints on the nuclear parameters are comparable to, yet smaller than that found in Sec.~\ref{sec:linear}, although are much more accurate because the covariances between $\tilde\Lambda$ and such nuclear parameters were properly taken into account, as well as considering the true probability distribution on $\tilde\Lambda$ from GW170817 as derived by the LIGO Collaboration.

How much does the addition of PEs affect the bounds on $K_0$, $M_0$, and $K_{\text{sym},0}$? To address this, we repeat our analysis without including these additional EoSs (see Appendix~\ref{appendix:sansHybrid} for more details). We find that the removal of such EoSs gives strong improvement in both correlations and nuclear constraints for low-order nuclear parameters $K_0$ and $M_0$, and the results are consistent with those in Ref.~\cite{Malik2018}. This further illuminates the need to study a wider variety of EoSs for use in universal relations to properly account for systematic errors.

\section{Conclusion and Discussion}\label{sec:conclusion}

The recent GW observation GW170817 coupled with the IR/UV/optical counterpart placed upper and lower bounds on the mass-weighted average tidal deformability $\tilde{\Lambda}$. 
We take advantage of this by selecting a diverse set of NS EoSs encompassing non-relativistic Skyrme-type interactions, RMF interactions and phenomenological variation of nuclear parameter models in order to constrain the nuclear matter parameters which are vital to limiting physically valid EoSs. 
We first found that approximate universal relations exist between linear combinations of nuclear parameters and $\tilde \Lambda$ for all values of mass ratio $q$ allowed from GW170817. 
We next constructed 2-dimensional probability distributions between $\tilde\Lambda$ and such nuclear parameters, converted them into one-dimensional conditional probability distributions on $K_\text{sym,0}$ given observations of $\tilde\Lambda$, and finally combined them with a posterior probability distribution on $\tilde\Lambda$ from LIGO and integrated them over $\tilde \Lambda$ in order to obtain posterior distributions on the nuclear parameters. 
From these posterior distributions, we derived 90\% confidence intervals on the incompressibility $K_0$, its slope $M_0$, and the curvature of symmetry energy $K_{\text{sym},0}$ at saturation density as \Krange{69}{352}, \Mrange{1371}{4808}, and \Ksymrange{-285}{7}.
The bounds on $M_0$ and $K_{\text{sym},0}$ are more conservative and safer to quote than those found in~\cite{Malik2018}. 
In addition, the constraints derived on $K_{\text{sym,0}}$ shows agreement with those in Refs.~\cite{Margueron:Ksym,Mondal:Ksym}.
We also note that bounds on $K_0$ and $M_0$ are less reliable than those on $K_\mathrm{sym,0}$ due to smaller correlations in the universal relations. 

The bounds derived in this paper are only valid for NSs and may not be valid for hybrid stars (HSs) with quark core and nuclear matter envelope. 
We discuss this point in more detail in Appendix~\ref{sec:hybrid}.

Future work on this subject includes investigation into combinations of nuclear parameters other than the linear ones studied here, to see if the correlations among such new combinations against $\tilde \Lambda$ improves. For example, one can consider ``multiplicative'' combinations of the form $K_0 L_0^{\eta}$ with constant $\eta$, in a similar spirit to~\cite{Sotani:2013dga,Silva:2016myw}.
Furthermore, one can study how the universal relations considered in this paper change as a function of the chirp mass, which may be useful for future binary NS merger events.
We also plan to study how the bounds derived here on nuclear parameters will improve in the future by considering upgraded ground-based GW detectors, such as aLIGO with its design sensitivity~\cite{aLIGO}, A\texttt{+}~\cite{Ap_Voyager_CE}, Voyager~\cite{Ap_Voyager_CE}, Einstein Telescope~\cite{ET} and Cosmic Explorer~\cite{Ap_Voyager_CE}, in particular by combining multiple events, and at what point systematic errors due to the EoS variation in the universal relations dominate statistical errors on $\tilde \Lambda$.
Work along these directions is currently in progress~\cite{Zack:futureNuclearConstraints}.

\section*{Acknowledgments}\label{acknowledgments}
We thank David Nichols for his illuminating advice on conditional probability distributions.
K.Y. acknowledges support from NSF Award PHY-1806776. 
K.Y. would like to also acknowledge networking support by the COST Action
GWverse CA16104. A.W.S. was supported by
NSF grant PHY 1554876 and by the U.S. DOE Office of Nuclear Physics.

\appendix
\section{$\tilde \Lambda$ versus $\Lambda_{1.4}$}\label{appendix:LTL14}

Malik \textit{et al}.~\cite{Malik2018} first studied correlations between nuclear and tidal parameters for individual NSs. Given that the tidal parameter measured from GW observations is $\tilde \Lambda$, corresponding to the mass-weighted average of two tidal parameters in a binary, the authors of Ref.~\cite{Malik2018} assumed the masses of the two NSs in GW170817 to be $m_1 = 1.4M_\odot$ and $m_2 = 1.33 M_\odot$. Next they studied correlations between $\tilde \Lambda$ in such a binary and $\Lambda_{1.4}$, representing the tidal deformability for an individual NS with a mass of $1.4M_\odot$. 

The above assumption can be dangerous because the individual mass measurements of GW170817 are not very accurate. Although the chirp mass has been measured with great accuracy as $\mathcal{M}=(m_1m_2)^{3/5} (m_1 + m_2)^{-1/5}=1.188^{+0.004}_{-0.002} \text{ M}_{\odot}$, the mass ratio varies as $q = m_2/m_1 \in \lbrack 0.73 , 1.00 \rbrack$~\cite{Abbott2017}. 

\begin{figure}
\begin{center} 
\includegraphics[width=\columnwidth]{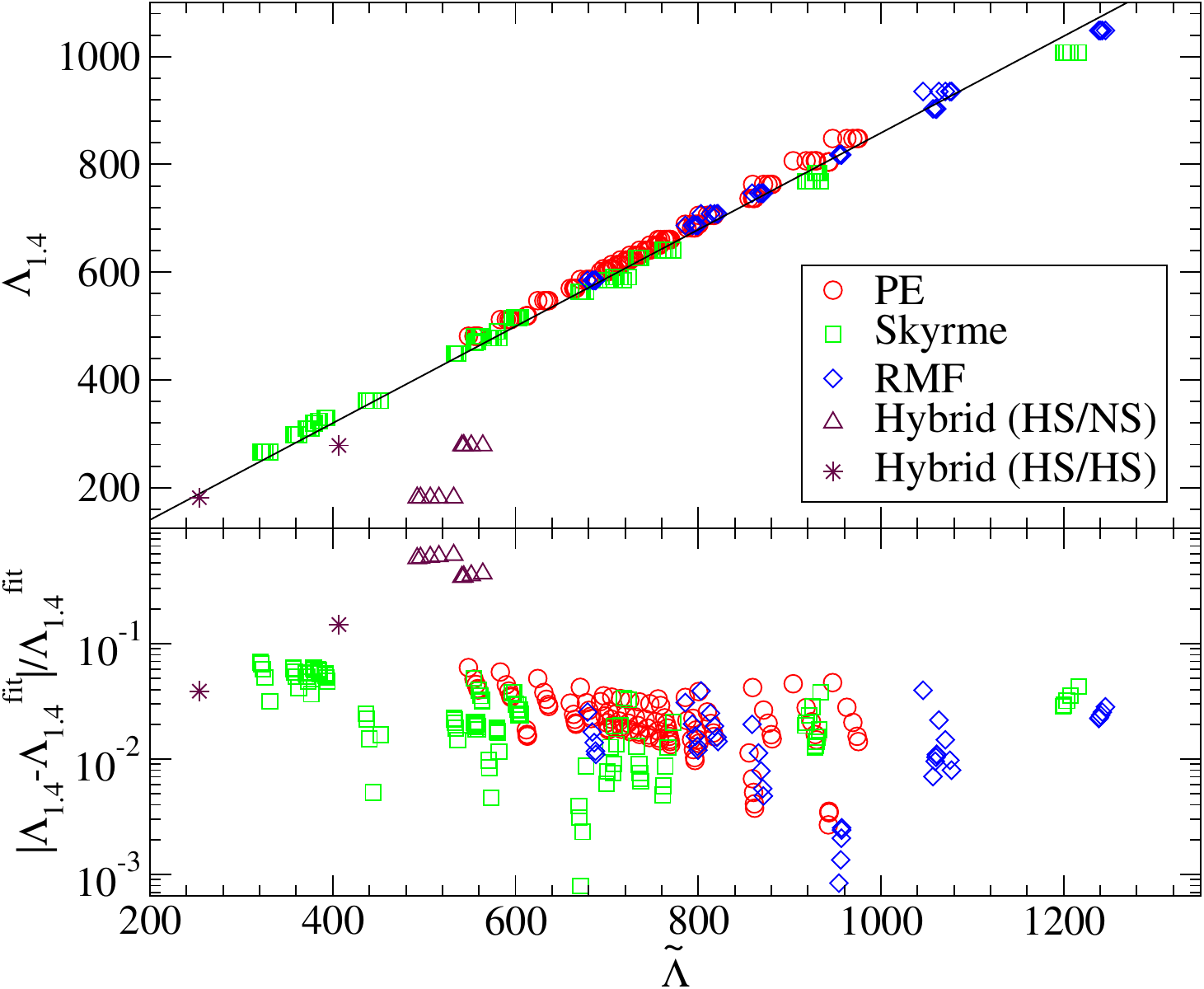}
\end{center}
\caption{(Top) Correlation between mass-weighted average tidal deformability $\tilde{\Lambda}$ and $\Lambda_{1.4}$ (individual tidal deformability at the mass of $1.4 \text{ M}_{\odot}$) for various EoSs each evaluated at mass ratios $q=0.73$, $0.80$, $0.87$, $0.93$ and $1.00$. The chirp mass is fixed to be the measured value of $\mathcal{M} = 1.188M_\odot$. (Bottom) Fractional difference from the fit for each EoS. Notice how the HS EoSs interrupt the universality between the two parameters by up to 60\% (5\% maximal percent difference in the absence of hybrid EoSs). 
}
\label{fig:LT_L14}
\end{figure}

The top panel of Fig.~\ref{fig:LT_L14} presents the $\tilde \Lambda$--$\Lambda_{1.4}$ correlation for various $q$ within the above range with the chirp mass fixed to $\mathcal{M} = 1.188M_\odot$, while the bottom panel shows the absolute fractional difference from the linear fit. Observe that a strong correlation exists between $\tilde \Lambda$ and $\Lambda_{1.4}$ for any $q$. The maximum fractional error for this case is $\sim 5$\%, with a correlation coefficient of $C=0.998$. 
On the other hand, once we include the hybrid EoSs discussed in more detail in Appendix~\ref{sec:hybrid}, one clearly sees a large deviation from the correlation with other EoSs, with the fractional difference reaching up to 60\%.
\setlength{\belowcaptionskip}{-10pt}
\setlength{\belowcaptionskip}{0pt}

\begin{figure}
\begin{center} 
\includegraphics[width=\columnwidth]{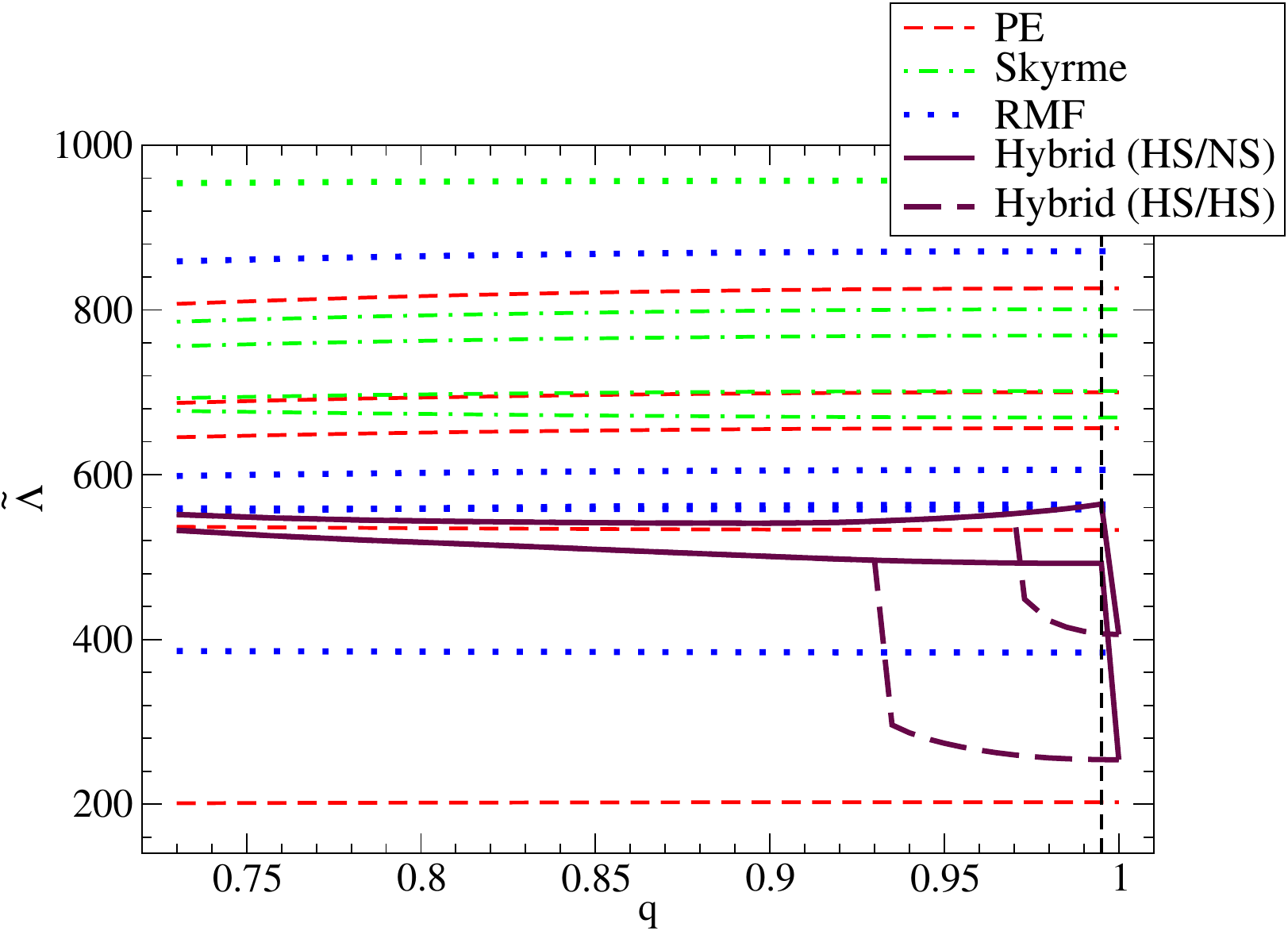}
\end{center}
\caption{$\tilde{\Lambda}$ for a representative set of EoSs as a function of mass ratio $q$ in GW170817's observed range of $q \in \lbrack 0.73 , 1.00 \rbrack$ with the chirp mass fixed to $\mathcal M = 1.188M_\odot$. Notice how $\tilde{\Lambda}$ only varies slightly in this region of interest for Skyrme, RMF, and PEs. Hybrid EoSs on the other hand admit two different configurations for GW170817, HS/NS (solid maroon) and HS/HS (dashed maroon), with the former giving a significant variation in $\tilde{\Lambda}$.
For demonstration purposes, the black vertical line corresponds to mass ratio $q=0.995$, where it can be seen that two different binary configurations emerge, discussed in more detail in Appendix~\ref{sec:hybrid}.
}
\label{fig:LT}
\end{figure}

The behavior in Fig.~\ref{fig:LT_L14} can be understood from Fig.~\ref{fig:LT}, where we show $\tilde \Lambda$ against $q$ with $\mathcal{M}$ fixed to the measured value for GW170817. If we do not consider hybrid EoSs, $\tilde \Lambda$ is insensitive to $q$~\cite{Radice2018,Burgio2018}, which is the origin of the strong correlation in the $\tilde \Lambda$--$\Lambda_{1.4}$ relation. On the other hand, for hybrid EoSs considered here, GW170817 can be either HS/HS or HS/NS when the mass ratio is close to unity\footnote{We note that hybrid EoSs considered in~\cite{Paschalidis2018} admit either NS/NS or HS/NS for GW170817.}. 
Thus, one finds a significant drop in $\tilde \Lambda$ as one increases $q$~\cite{Burgio2018}, which changes the $\tilde \Lambda$--$\Lambda_{1.4}$ relation drastically.

\section{Repeated Analysis without PEs}\label{appendix:sansHybrid}

\begin{figure}
\begin{center} 
\includegraphics[width=\columnwidth]{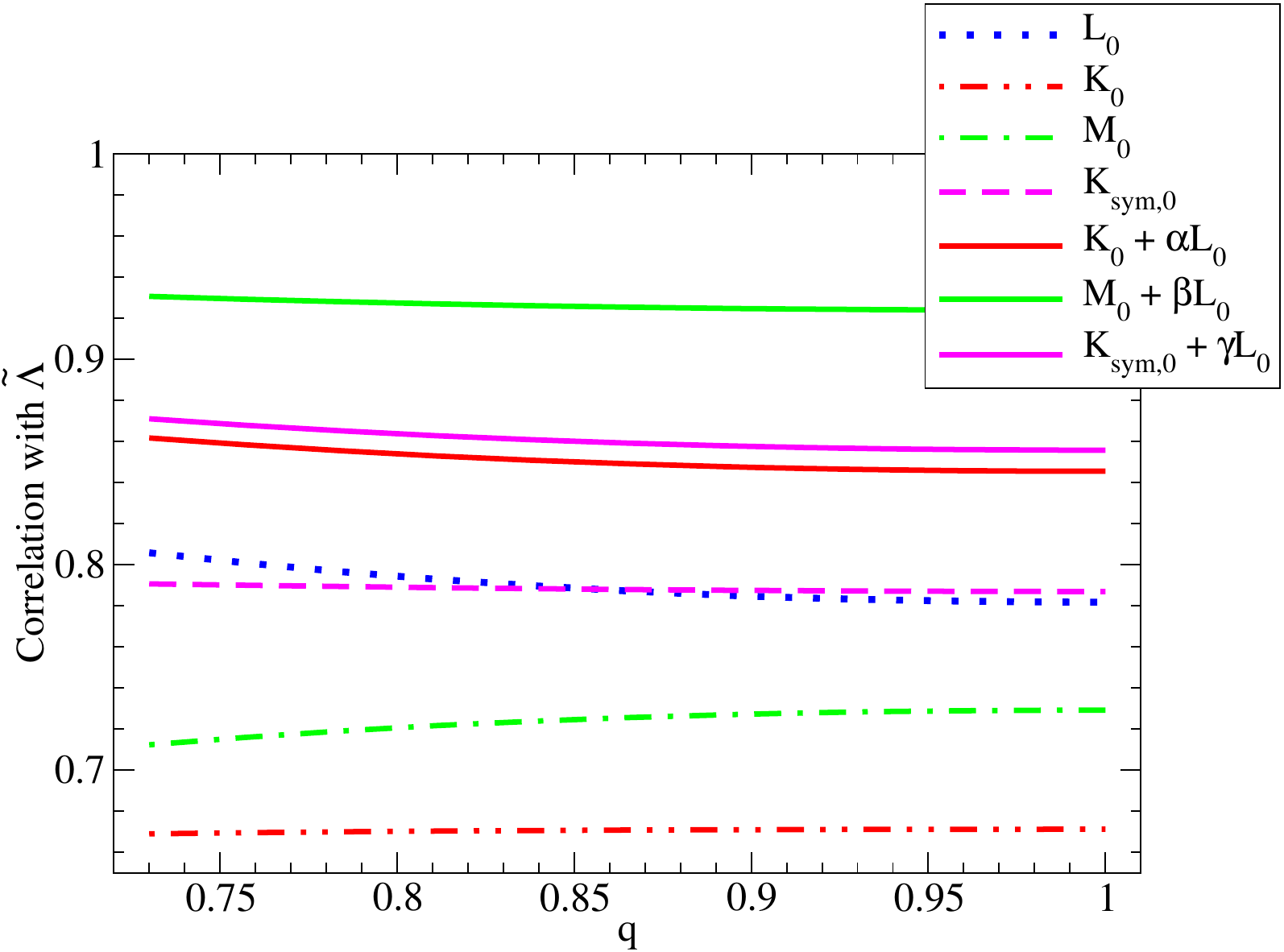}
\end{center}
\caption{
Similar to Fig.~\ref{fig:correlation} upon the removal of PEs. Observe that the correlations for linear combinations involving lower order parameters improve by up to 55\%, while linear combinations with high order parameter $K_{\text{sym},0}$ shows slightly diminished, yet comparable, correlations. Observe also that the correlations are insensitive to $q$.
}
\label{fig:alamCorrelations}
\end{figure} 

\begin{table*}
\caption{
Similar to Table~\ref{tab:Constraints}, when excluding PEs, and only considering the ``first method" of computing nuclear parameter constraints.
Observe how the bounds upon removal of PEs show drastic improvement - showing closer agreement with Ref.~\cite{Malik2018} (in addition to uncertainty from EoS variation), and also highlighting the effect of utilizing a large set of additional EoSs.
The exception is high order nuclear parameter $K_{\text{sym},0}$ - showing weakened constraints due to the inclusion of uncertainty in $L_0$.
}\label{tab:alamConstraints} 
\vspace{3mm}
\begin{tabular}{|| C{3.4cm} || C{5.6cm} | C{5.6cm}||@{}m{0pt}@{}} 
 \hline
 \diagbox[width=3.5cm]{\hspace*{2pt} \raisebox{2pt}{$L_0$ [MeV]}}{\raisebox{-5pt}{$\tilde{\Lambda}$} \hspace*{0.5cm}} & 70-720~\cite{Abbott2018} & 279--822~\cite{Coughlin:2018fis} \\
\hline
\hline
 40--62~\cite{Lattimer2013,Lattimer2014,Tews2017} & \makecell{\\ \Krange{161}{309}  \\ \\ \Mrange{1506}{3506} \\ \\ \Ksymrange{-327}{140} \\ \\}  & \makecell{\\ \Krange{182}{324} \\ \\ \Mrange{1851}{3723} \\ \\ \Ksymrange{-246}{190} \\ \\ } \\
 \hline
 30--86~\cite{Oertel2017} & \makecell{\\ \Krange{134}{320}  \\ \\ \Mrange{1131}{3662} \\ \\ \Ksymrange{-394}{168} \\ \\} & \makecell{\\ \Krange{155}{335} \\  \\ \Mrange{1476}{3880} \\  \\ \Ksymrange{-313}{218}\\ \\ }\\
 \hline
\end{tabular}
\end{table*}

In this appendix, we study the effect of PEs on nuclear parameter bounds by re-analyzing them without including such EoSs.
This way,  we can directly compare our results with those in Ref.~\cite{Malik2018} which did not include these additional EoSs.
Figure~\ref{fig:alamCorrelations} once again presents correlations between $\tilde{\Lambda}$ and 
linear combinations of nuclear parameters as a function of  mass ratio. 
Here, for comparison purposes we choose $\alpha=1.10$, $\beta=15.62$, and $\gamma=2.81$ such that correlations become maximum, as was done in Ref.~\cite{Malik2018}.
Observe that correlations with $\tilde{\Lambda}$ remain almost constant throughout the entire region of allowable mass ratios.
In addition, note how correlations for linear combination involving $K_0$ and $M_0$ are increased by up to  $55\%$ from Fig.~\ref{fig:correlation} which includes PEs, while linear combinations with higher order nuclear parameter $K_{\text{sym},0}$ interestingly shows a small decrease in correlation, yet remains comparable.
This is revealing of the flexible nature of the $K_{\text{sym},0}$ nuclear parameter.

We now derive constraints on nuclear parameters without PEs. 
Following the procedure outlined in Sec.~\ref{sec:constraints}, new bounds on $K_0$, $M_0$, and $K_{\text{sym},0}$ are calculated for a central mass ratio of $q=0.87$, and summarized in Table~\ref{tab:alamConstraints}. Comparing this with Table~\ref{tab:Constraints}, one sees that the additional PEs significantly weaken estimated constraints for low order nuclear parameters $K_0$ and $M_0$, and interestingly, improve them for high-order nuclear parameter $K_\mathrm{sym,0}$.
Here we find results somewhat agreeable to what was found in Malik et al~\cite{Malik2018}, however enlarged due to the addition of EoS variation uncertainties. 

\section{Hybrid Quark-hadron Stars}\label{sec:hybrid}
In this appendix, we investigate the use of an additional valid class of EoS: hybrid quark-hadron stars based on Ref.~\cite{Paschalidis2018}.
Here, the low-density nucleonic matter region of PEs transition into a high-density quark matter phase in a given transitional energy density region $\epsilon_1 \leq \epsilon \leq \epsilon_2$. 
For our purposes, we consider Set I quark matter EoSs, where the pressure following transition is given by~\cite{Alford:2017qgh} (see also~\cite{Montana:2018bkb, 1971SvA....15..347S, Zdunik:2012dj, Alford:2013aca}):
\begin{equation}
P(\epsilon) =    \left\{
\begin{array}{ll}
      P_{\text{tr}} & (\epsilon_1  \leq \epsilon \leq \epsilon_2) \\
      P_{\text{tr}}+c_{\text{s}}^2(\epsilon-\epsilon_2) & (\epsilon > \epsilon_2)
\end{array} 
\right. 
\end{equation}
with $c_s$ being the constant speed of sound in the quark matter, $\epsilon_1$ and $\epsilon_2$ characterizing the energy density ``jump" $\epsilon_2 - \epsilon_1 \equiv \epsilon_1 j$, and $P_{\text{tr}}$ representing the transition pressure, such that the low density hadronic matter's energy density equals $\epsilon_1$. 
In this paper, we adopt the ACS-II parameterization in~\cite{Paschalidis2018} as $P_{\text{tr}} = 1.7 \times 10^{35}$dyn/cm$^2$, $\epsilon_2 = 8.34 \times 10^{14}$g/cm$^3$ and $c_s^2=0.8$ with $j=0.8$ or 1. 
 
As we show in Fig.~\ref{fig:LT}, strong phase transitions in the star admit a secondary stable HS configuration (denoted HS/HS).
HSs  evaluate to a reduction in tidal deformability $\tilde{\Lambda}$ from their NS-branch counterparts, thus altering universal relations accordingly.
Here, we examine how this additional possibility of binary HSs and the choice of fiducial nuclear matter EoS impacts correlations between $\tilde{\Lambda}$ and nuclear parameters.

\begin{figure*}[!t]
\begin{center} 
\includegraphics[width=\textwidth]{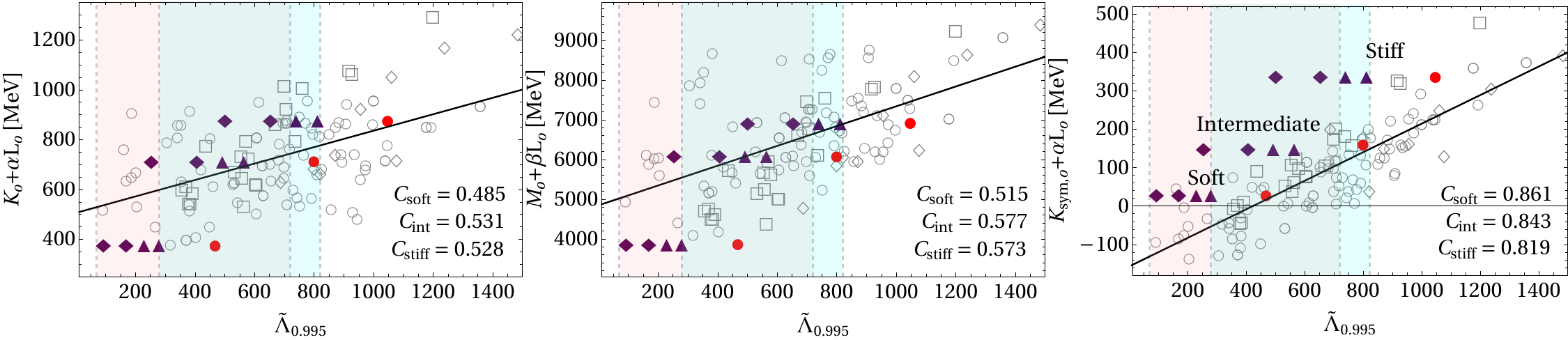}
\end{center}
\caption{
Scatter plots demonstrating variation in correlation at $q=0.995$ (with $\mathcal{M} = 1.188M_\odot$) when introducing HS EoSs based on 3 different fiducial nuclear matter EoSs with $\tilde \Lambda \approx 465$, $\tilde \Lambda \approx 800$, $\tilde \Lambda \approx 1045$, represented by filled red circles. 
These correspond to soft, intermediate and stiff fiducial EoSs respectively. 
These are followed by two different star configurations of hybrid HS/NS values (purple triangle), and HS/HS (purple diamond) with a reduction in $\tilde{\Lambda}$, as is shown in Fig.~\ref{fig:LT} for two different HS EoSs, corresponding to the $j=0.8$ and $j=1.0$ configurations.
As demonstrated in Fig.~\ref{fig:LT}, $q=0.995$ clearly admits both HS/NS and HS/HS binary configurations.
Shown in gray as reference are the PE, Skyrme, and RMF EoSs, irrelevant to this investigation.
Displayed in the bottom right corner is the correlation between nuclear parameter combinations and $\tilde{\Lambda}$ when imposing soft, intermediate, and stiff fiducial EoSs in the generation of HS structure.
Notice how the choice of fiducial EoS alters correlations between $\tilde{\Lambda}$ and combinations of nuclear parameters differently.
This indicates that potential HS EoSs could impact nuclear bounds significantly.
}
\label{fig:fiducialVariation}
\end{figure*}

Figure~\ref{fig:fiducialVariation} investigates this phenomena by choosing 3 different fiducial nuclear matter EoSs with soft ($\tilde{\Lambda}\approx 465$), intermediate ($\tilde{\Lambda} \approx 800$), and stiff ($\tilde{\Lambda} \approx 1045$) representative values of tidal deformability for $\mathcal{M} = 1.188M_\odot$ and $q=0.995$. 
Next, HS EoSs are formulated, and new universal relations are derived - including both stellar configurations at high values of $q$, as can clearly be seen by the dashed vertical line in Fig.~\ref{fig:LT}.
Observe how the choice of fiducial nuclear matter EoS impacts the universal relations differently depending on which combination of nuclear parameters is used.
For example, use of the stiff fiducial EoS compared to the intermediate one results in a small decrease in correlation for $K_0 + \alpha L_0$, a negligible decrease for $M_0 + \beta L_0$, and a large decrease for $K_{\text{sym},0} + \alpha L_0$.
Alternatively, choice of the soft fiducial EoS results in medium decreases in correlation for $K_0 + \alpha L_0$ and $M_0 + \alpha L_0$, and an \textit{increase} in correlation for $K_{\text{sym},0} + \alpha L_0$.

In conclusion, we find that the use of valid hybrid quark-hadron star EoSs in universal relations can influence universality in unexpected ways. Thus, the bounds derived in Table~\ref{tab:Constraints} are strictly valid only for NSs, and they are subject to change once one includes the possibility for HSs.
Refer also to Ref.~\cite{Montana:2018bkb} for a more detailed analysis of hybrid star EoSs in conjunction with GW170817.

\vspace*{8cm}

\bibliography{Zack}

 \newcommand{\noop}[1]{}
\begin{thebibliography}{79}%
\makeatletter
\providecommand \@ifxundefined [1]{%
 \@ifx{#1\undefined}
}%
\providecommand \@ifnum [1]{%
 \ifnum #1\expandafter \@firstoftwo
 \else \expandafter \@secondoftwo
 \fi
}%
\providecommand \@ifx [1]{%
 \ifx #1\expandafter \@firstoftwo
 \else \expandafter \@secondoftwo
 \fi
}%
\providecommand \natexlab [1]{#1}%
\providecommand \enquote  [1]{``#1''}%
\providecommand \bibnamefont  [1]{#1}%
\providecommand \bibfnamefont [1]{#1}%
\providecommand \citenamefont [1]{#1}%
\providecommand \href@noop [0]{\@secondoftwo}%
\providecommand \href [0]{\begingroup \@sanitize@url \@href}%
\providecommand \@href[1]{\@@startlink{#1}\@@href}%
\providecommand \@@href[1]{\endgroup#1\@@endlink}%
\providecommand \@sanitize@url [0]{\catcode `\\12\catcode `\$12\catcode
  `\&12\catcode `\#12\catcode `\^12\catcode `\_12\catcode `\%12\relax}%
\providecommand \@@startlink[1]{}%
\providecommand \@@endlink[0]{}%
\providecommand \url  [0]{\begingroup\@sanitize@url \@url }%
\providecommand \@url [1]{\endgroup\@href {#1}{\urlprefix }}%
\providecommand \urlprefix  [0]{URL }%
\providecommand \Eprint [0]{\href }%
\providecommand \doibase [0]{http://dx.doi.org/}%
\providecommand \selectlanguage [0]{\@gobble}%
\providecommand \bibinfo  [0]{\@secondoftwo}%
\providecommand \bibfield  [0]{\@secondoftwo}%
\providecommand \translation [1]{[#1]}%
\providecommand \BibitemOpen [0]{}%
\providecommand \bibitemStop [0]{}%
\providecommand \bibitemNoStop [0]{.\EOS\space}%
\providecommand \EOS [0]{\spacefactor3000\relax}%
\providecommand \BibitemShut  [1]{\csname bibitem#1\endcsname}%
\let\auto@bib@innerbib\@empty
\bibitem [{\citenamefont {Guver}\ and\ \citenamefont {Ozel}(2013)}]{guver}%
  \BibitemOpen
  \bibfield  {author} {\bibinfo {author} {\bibfnamefont {T.}~\bibnamefont
  {Guver}}\ and\ \bibinfo {author} {\bibfnamefont {F.}~\bibnamefont {Ozel}},\
  }\href {\doibase 10.1088/2041-8205/765/1/L1} {\bibfield  {journal} {\bibinfo
  {journal} {Astrophys. J.}\ }\textbf {\bibinfo {volume} {765}},\ \bibinfo
  {pages} {L1} (\bibinfo {year} {2013})},\ \Eprint
  {http://arxiv.org/abs/1301.0831} {arXiv:1301.0831 [astro-ph.HE]} \BibitemShut
  {NoStop}%
\bibitem [{\citenamefont {Ozel}\ \emph {et~al.}(2010)\citenamefont {Ozel},
  \citenamefont {Baym},\ and\ \citenamefont {Guver}}]{ozel-baym-guver}%
  \BibitemOpen
  \bibfield  {author} {\bibinfo {author} {\bibfnamefont {F.}~\bibnamefont
  {Ozel}}, \bibinfo {author} {\bibfnamefont {G.}~\bibnamefont {Baym}}, \ and\
  \bibinfo {author} {\bibfnamefont {T.}~\bibnamefont {Guver}},\ }\href
  {\doibase 10.1103/PhysRevD.82.101301} {\bibfield  {journal} {\bibinfo
  {journal} {Phys.Rev.}\ }\textbf {\bibinfo {volume} {D82}},\ \bibinfo {pages}
  {101301} (\bibinfo {year} {2010})},\ \Eprint {http://arxiv.org/abs/1002.3153}
  {arXiv:1002.3153 [astro-ph.HE]} \BibitemShut {NoStop}%
\bibitem [{\citenamefont {Steiner}\ \emph {et~al.}(2010)\citenamefont
  {Steiner}, \citenamefont {Lattimer},\ and\ \citenamefont
  {Brown}}]{steiner-lattimer-brown}%
  \BibitemOpen
  \bibfield  {author} {\bibinfo {author} {\bibfnamefont {A.~W.}\ \bibnamefont
  {Steiner}}, \bibinfo {author} {\bibfnamefont {J.~M.}\ \bibnamefont
  {Lattimer}}, \ and\ \bibinfo {author} {\bibfnamefont {E.~F.}\ \bibnamefont
  {Brown}},\ }\href {\doibase 10.1088/0004-637X/722/1/33} {\bibfield  {journal}
  {\bibinfo  {journal} {Astrophys.J.}\ }\textbf {\bibinfo {volume} {722}},\
  \bibinfo {pages} {33} (\bibinfo {year} {2010})}\BibitemShut {NoStop}%
\bibitem [{\citenamefont {Lattimer}\ and\ \citenamefont
  {Steiner}(2014)}]{Lattimer2014}%
  \BibitemOpen
  \bibfield  {author} {\bibinfo {author} {\bibfnamefont {J.~M.}\ \bibnamefont
  {Lattimer}}\ and\ \bibinfo {author} {\bibfnamefont {A.~W.}\ \bibnamefont
  {Steiner}},\ }\href {\doibase 10.1140/epja/i2014-14040-y} {\bibfield
  {journal} {\bibinfo  {journal} {The European Physical Journal A}\ }\textbf
  {\bibinfo {volume} {50}} (\bibinfo {year} {2014}),\
  10.1140/epja/i2014-14040-y}\BibitemShut {NoStop}%
\bibitem [{\citenamefont {Ozel}\ and\ \citenamefont
  {Freire}(2016)}]{Ozel:2016oaf}%
  \BibitemOpen
  \bibfield  {author} {\bibinfo {author} {\bibfnamefont {F.}~\bibnamefont
  {Ozel}}\ and\ \bibinfo {author} {\bibfnamefont {P.}~\bibnamefont {Freire}},\
  }\href {\doibase 10.1146/annurev-astro-081915-023322} {\bibfield  {journal}
  {\bibinfo  {journal} {Ann. Rev. Astron. Astrophys.}\ }\textbf {\bibinfo
  {volume} {54}},\ \bibinfo {pages} {401} (\bibinfo {year} {2016})},\ \Eprint
  {http://arxiv.org/abs/1603.02698} {arXiv:1603.02698 [astro-ph.HE]}
  \BibitemShut {NoStop}%
\bibitem [{\citenamefont {Abbott}\ \emph
  {et~al.}(2017{\natexlab{a}})\citenamefont {Abbott} \emph
  {et~al.}}]{TheLIGOScientific:2017qsa}%
  \BibitemOpen
  \bibfield  {author} {\bibinfo {author} {\bibfnamefont {B.~P.}\ \bibnamefont
  {Abbott}} \emph {et~al.} (\bibinfo {collaboration} {Virgo, LIGO
  Scientific}),\ }\href {\doibase 10.1103/PhysRevLett.119.161101} {\bibfield
  {journal} {\bibinfo  {journal} {Phys. Rev. Lett.}\ }\textbf {\bibinfo
  {volume} {119}},\ \bibinfo {pages} {161101} (\bibinfo {year}
  {2017}{\natexlab{a}})},\ \Eprint {http://arxiv.org/abs/1710.05832}
  {arXiv:1710.05832 [gr-qc]} \BibitemShut {NoStop}%
\bibitem [{\citenamefont {Abbott}\ \emph
  {et~al.}(2019{\natexlab{a}})\citenamefont {Abbott} \emph
  {et~al.}}]{Abbott2018}%
  \BibitemOpen
  \bibfield  {author} {\bibinfo {author} {\bibfnamefont {B.~P.}\ \bibnamefont
  {Abbott}} \emph {et~al.} (\bibinfo {collaboration} {LIGO Scientific,
  Virgo}),\ }\href {\doibase 10.1103/PhysRevX.9.011001} {\bibfield  {journal}
  {\bibinfo  {journal} {Phys. Rev.}\ }\textbf {\bibinfo {volume} {X9}},\
  \bibinfo {pages} {011001} (\bibinfo {year} {2019}{\natexlab{a}})},\ \Eprint
  {http://arxiv.org/abs/1805.11579} {arXiv:1805.11579 [gr-qc]} \BibitemShut
  {NoStop}%
\bibitem [{\citenamefont {Abbott}\ \emph {et~al.}(2018)\citenamefont {Abbott}
  \emph {et~al.}}]{Abbott:2018exr}%
  \BibitemOpen
  \bibfield  {author} {\bibinfo {author} {\bibfnamefont {B.~P.}\ \bibnamefont
  {Abbott}} \emph {et~al.} (\bibinfo {collaboration} {LIGO Scientific,
  Virgo}),\ }\href {\doibase 10.1103/PhysRevLett.121.161101} {\bibfield
  {journal} {\bibinfo  {journal} {Phys. Rev. Lett.}\ }\textbf {\bibinfo
  {volume} {121}},\ \bibinfo {pages} {161101} (\bibinfo {year} {2018})},\
  \Eprint {http://arxiv.org/abs/1805.11581} {arXiv:1805.11581 [gr-qc]}
  \BibitemShut {NoStop}%
\bibitem [{\citenamefont {Paschalidis}\ \emph {et~al.}(2018)\citenamefont
  {Paschalidis}, \citenamefont {Yagi}, \citenamefont {Alvarez-Castillo},
  \citenamefont {Blaschke},\ and\ \citenamefont {Sedrakian}}]{Paschalidis2018}%
  \BibitemOpen
  \bibfield  {author} {\bibinfo {author} {\bibfnamefont {V.}~\bibnamefont
  {Paschalidis}}, \bibinfo {author} {\bibfnamefont {K.}~\bibnamefont {Yagi}},
  \bibinfo {author} {\bibfnamefont {D.}~\bibnamefont {Alvarez-Castillo}},
  \bibinfo {author} {\bibfnamefont {D.~B.}\ \bibnamefont {Blaschke}}, \ and\
  \bibinfo {author} {\bibfnamefont {A.}~\bibnamefont {Sedrakian}},\ }\href
  {\doibase 10.1103/physrevd.97.084038} {\bibfield  {journal} {\bibinfo
  {journal} {Physical Review D}\ }\textbf {\bibinfo {volume} {97}} (\bibinfo
  {year} {2018}),\ 10.1103/physrevd.97.084038}\BibitemShut {NoStop}%
\bibitem [{\citenamefont {Burgio}\ \emph {et~al.}(2018)\citenamefont {Burgio},
  \citenamefont {Drago}, \citenamefont {Pagliara}, \citenamefont {Schulze},\
  and\ \citenamefont {Wei}}]{Burgio2018}%
  \BibitemOpen
  \bibfield  {author} {\bibinfo {author} {\bibfnamefont {G.~F.}\ \bibnamefont
  {Burgio}}, \bibinfo {author} {\bibfnamefont {A.}~\bibnamefont {Drago}},
  \bibinfo {author} {\bibfnamefont {G.}~\bibnamefont {Pagliara}}, \bibinfo
  {author} {\bibfnamefont {H.~J.}\ \bibnamefont {Schulze}}, \ and\ \bibinfo
  {author} {\bibfnamefont {J.~B.}\ \bibnamefont {Wei}},\ }\href
  {http://arxiv.org/pdf/1803.09696v1} {\bibfield  {journal} {\bibinfo
  {journal} {Arxiv}\ } (\bibinfo {year} {2018})},\ \Eprint
  {http://arxiv.org/abs/1803.09696v1} {1803.09696v1} \BibitemShut {NoStop}%
\bibitem [{\citenamefont {Malik}\ \emph {et~al.}(2018)\citenamefont {Malik},
  \citenamefont {Alam}, \citenamefont {Fortin}, \citenamefont {ProvidÃªncia},
  \citenamefont {Agrawal}, \citenamefont {Jha}, \citenamefont {Kumar},\ and\
  \citenamefont {Patra}}]{Malik2018}%
  \BibitemOpen
  \bibfield  {author} {\bibinfo {author} {\bibfnamefont {T.}~\bibnamefont
  {Malik}}, \bibinfo {author} {\bibfnamefont {N.}~\bibnamefont {Alam}},
  \bibinfo {author} {\bibfnamefont {M.}~\bibnamefont {Fortin}}, \bibinfo
  {author} {\bibfnamefont {C.}~\bibnamefont {ProvidÃªncia}}, \bibinfo
  {author} {\bibfnamefont {B.~K.}\ \bibnamefont {Agrawal}}, \bibinfo {author}
  {\bibfnamefont {T.~K.}\ \bibnamefont {Jha}}, \bibinfo {author} {\bibfnamefont
  {B.}~\bibnamefont {Kumar}}, \ and\ \bibinfo {author} {\bibfnamefont {S.~K.}\
  \bibnamefont {Patra}},\ }\href {\doibase 10.1103/PhysRevC.98.035804}
  {\bibfield  {journal} {\bibinfo  {journal} {Phys. Rev.}\ }\textbf {\bibinfo
  {volume} {C98}},\ \bibinfo {pages} {035804} (\bibinfo {year} {2018})},\
  \Eprint {http://arxiv.org/abs/1805.11963} {arXiv:1805.11963 [nucl-th]}
  \BibitemShut {NoStop}%
\bibitem [{\citenamefont {Flanagan}\ and\ \citenamefont
  {Hinderer}(2008)}]{Flanagan2008}%
  \BibitemOpen
  \bibfield  {author} {\bibinfo {author} {\bibfnamefont {{\'{E}}.~{\'{E}}.}\
  \bibnamefont {Flanagan}}\ and\ \bibinfo {author} {\bibfnamefont
  {T.}~\bibnamefont {Hinderer}},\ }\href {\doibase 10.1103/physrevd.77.021502}
  {\bibfield  {journal} {\bibinfo  {journal} {Physical Review D}\ }\textbf
  {\bibinfo {volume} {77}} (\bibinfo {year} {2008}),\
  10.1103/physrevd.77.021502}\BibitemShut {NoStop}%
\bibitem [{\citenamefont {De}\ \emph {et~al.}(2018)\citenamefont {De},
  \citenamefont {Finstad}, \citenamefont {Lattimer}, \citenamefont {Brown},
  \citenamefont {Berger},\ and\ \citenamefont {Biwer}}]{De:2018uhw}%
  \BibitemOpen
  \bibfield  {author} {\bibinfo {author} {\bibfnamefont {S.}~\bibnamefont
  {De}}, \bibinfo {author} {\bibfnamefont {D.}~\bibnamefont {Finstad}},
  \bibinfo {author} {\bibfnamefont {J.~M.}\ \bibnamefont {Lattimer}}, \bibinfo
  {author} {\bibfnamefont {D.~A.}\ \bibnamefont {Brown}}, \bibinfo {author}
  {\bibfnamefont {E.}~\bibnamefont {Berger}}, \ and\ \bibinfo {author}
  {\bibfnamefont {C.~M.}\ \bibnamefont {Biwer}},\ }\href {\doibase
  10.1103/PhysRevLett.121.259902, 10.1103/PhysRevLett.121.091102} {\bibfield
  {journal} {\bibinfo  {journal} {Phys. Rev. Lett.}\ }\textbf {\bibinfo
  {volume} {121}},\ \bibinfo {pages} {091102} (\bibinfo {year} {2018})},\
  \bibinfo {note} {[Erratum: Phys. Rev. Lett.121,no.25,259902(2018)]},\ \Eprint
  {http://arxiv.org/abs/1804.08583} {arXiv:1804.08583 [astro-ph.HE]}
  \BibitemShut {NoStop}%
\bibitem [{\citenamefont {Coughlin}\ \emph {et~al.}(2019)\citenamefont
  {Coughlin}, \citenamefont {Dietrich}, \citenamefont {Margalit},\ and\
  \citenamefont {Metzger}}]{Coughlin:2018fis}%
  \BibitemOpen
  \bibfield  {author} {\bibinfo {author} {\bibfnamefont {M.~W.}\ \bibnamefont
  {Coughlin}}, \bibinfo {author} {\bibfnamefont {T.}~\bibnamefont {Dietrich}},
  \bibinfo {author} {\bibfnamefont {B.}~\bibnamefont {Margalit}}, \ and\
  \bibinfo {author} {\bibfnamefont {B.~D.}\ \bibnamefont {Metzger}},\ }\href
  {\doibase 10.1093/mnrasl/slz133} {\bibfield  {journal} {\bibinfo  {journal}
  {Mon. Not. Roy. Astron. Soc.}\ }\textbf {\bibinfo {volume} {489}},\ \bibinfo
  {pages} {L91} (\bibinfo {year} {2019})},\ \Eprint
  {http://arxiv.org/abs/1812.04803} {arXiv:1812.04803 [astro-ph.HE]}
  \BibitemShut {NoStop}%
\bibitem [{\citenamefont {Radice}\ \emph {et~al.}(2018)\citenamefont {Radice},
  \citenamefont {Perego}, \citenamefont {Zappa},\ and\ \citenamefont
  {Bernuzzi}}]{Radice2018}%
  \BibitemOpen
  \bibfield  {author} {\bibinfo {author} {\bibfnamefont {D.}~\bibnamefont
  {Radice}}, \bibinfo {author} {\bibfnamefont {A.}~\bibnamefont {Perego}},
  \bibinfo {author} {\bibfnamefont {F.}~\bibnamefont {Zappa}}, \ and\ \bibinfo
  {author} {\bibfnamefont {S.}~\bibnamefont {Bernuzzi}},\ }\href {\doibase
  10.3847/2041-8213/aaa402} {\bibfield  {journal} {\bibinfo  {journal} {The
  Astrophysical Journal}\ }\textbf {\bibinfo {volume} {852}},\ \bibinfo {pages}
  {L29} (\bibinfo {year} {2018})}\BibitemShut {NoStop}%
\bibitem [{\citenamefont {Annala}\ \emph {et~al.}(2018)\citenamefont {Annala},
  \citenamefont {Gorda}, \citenamefont {Kurkela},\ and\ \citenamefont
  {Vuorinen}}]{Annala:2017llu}%
  \BibitemOpen
  \bibfield  {author} {\bibinfo {author} {\bibfnamefont {E.}~\bibnamefont
  {Annala}}, \bibinfo {author} {\bibfnamefont {T.}~\bibnamefont {Gorda}},
  \bibinfo {author} {\bibfnamefont {A.}~\bibnamefont {Kurkela}}, \ and\
  \bibinfo {author} {\bibfnamefont {A.}~\bibnamefont {Vuorinen}},\ }\href
  {\doibase 10.1103/PhysRevLett.120.172703} {\bibfield  {journal} {\bibinfo
  {journal} {Phys. Rev. Lett.}\ }\textbf {\bibinfo {volume} {120}},\ \bibinfo
  {pages} {172703} (\bibinfo {year} {2018})},\ \Eprint
  {http://arxiv.org/abs/1711.02644} {arXiv:1711.02644 [astro-ph.HE]}
  \BibitemShut {NoStop}%
\bibitem [{\citenamefont {Lim}\ and\ \citenamefont {Holt}(2018)}]{Lim:2018bkq}%
  \BibitemOpen
  \bibfield  {author} {\bibinfo {author} {\bibfnamefont {Y.}~\bibnamefont
  {Lim}}\ and\ \bibinfo {author} {\bibfnamefont {J.~W.}\ \bibnamefont {Holt}},\
  }\href {\doibase 10.1103/PhysRevLett.121.062701} {\bibfield  {journal}
  {\bibinfo  {journal} {Phys. Rev. Lett.}\ }\textbf {\bibinfo {volume} {121}},\
  \bibinfo {pages} {062701} (\bibinfo {year} {2018})},\ \Eprint
  {http://arxiv.org/abs/1803.02803} {arXiv:1803.02803 [nucl-th]} \BibitemShut
  {NoStop}%
\bibitem [{\citenamefont {Bauswein}\ \emph {et~al.}(2017)\citenamefont
  {Bauswein}, \citenamefont {Just}, \citenamefont {Janka},\ and\ \citenamefont
  {Stergioulas}}]{Bauswein:2017vtn}%
  \BibitemOpen
  \bibfield  {author} {\bibinfo {author} {\bibfnamefont {A.}~\bibnamefont
  {Bauswein}}, \bibinfo {author} {\bibfnamefont {O.}~\bibnamefont {Just}},
  \bibinfo {author} {\bibfnamefont {H.-T.}\ \bibnamefont {Janka}}, \ and\
  \bibinfo {author} {\bibfnamefont {N.}~\bibnamefont {Stergioulas}},\ }\href
  {\doibase 10.3847/2041-8213/aa9994} {\bibfield  {journal} {\bibinfo
  {journal} {Astrophys. J.}\ }\textbf {\bibinfo {volume} {850}},\ \bibinfo
  {pages} {L34} (\bibinfo {year} {2017})},\ \Eprint
  {http://arxiv.org/abs/1710.06843} {arXiv:1710.06843 [astro-ph.HE]}
  \BibitemShut {NoStop}%
\bibitem [{\citenamefont {Most}\ \emph {et~al.}(2018)\citenamefont {Most},
  \citenamefont {Weih}, \citenamefont {Rezzolla},\ and\ \citenamefont
  {Schaffner-Bielich}}]{Most:2018hfd}%
  \BibitemOpen
  \bibfield  {author} {\bibinfo {author} {\bibfnamefont {E.~R.}\ \bibnamefont
  {Most}}, \bibinfo {author} {\bibfnamefont {L.~R.}\ \bibnamefont {Weih}},
  \bibinfo {author} {\bibfnamefont {L.}~\bibnamefont {Rezzolla}}, \ and\
  \bibinfo {author} {\bibfnamefont {J.}~\bibnamefont {Schaffner-Bielich}},\
  }\href {\doibase 10.1103/PhysRevLett.120.261103} {\bibfield  {journal}
  {\bibinfo  {journal} {Phys. Rev. Lett.}\ }\textbf {\bibinfo {volume} {120}},\
  \bibinfo {pages} {261103} (\bibinfo {year} {2018})},\ \Eprint
  {http://arxiv.org/abs/1803.00549} {arXiv:1803.00549 [gr-qc]} \BibitemShut
  {NoStop}%
\bibitem [{\citenamefont {Read}\ \emph {et~al.}(2009)\citenamefont {Read},
  \citenamefont {Lackey}, \citenamefont {Owen},\ and\ \citenamefont
  {Friedman}}]{Read2009}%
  \BibitemOpen
  \bibfield  {author} {\bibinfo {author} {\bibfnamefont {J.~S.}\ \bibnamefont
  {Read}}, \bibinfo {author} {\bibfnamefont {B.~D.}\ \bibnamefont {Lackey}},
  \bibinfo {author} {\bibfnamefont {B.~J.}\ \bibnamefont {Owen}}, \ and\
  \bibinfo {author} {\bibfnamefont {J.~L.}\ \bibnamefont {Friedman}},\ }\href
  {\doibase 10.1103/physrevd.79.124032} {\bibfield  {journal} {\bibinfo
  {journal} {Physical Review D}\ }\textbf {\bibinfo {volume} {79}} (\bibinfo
  {year} {2009}),\ 10.1103/physrevd.79.124032}\BibitemShut {NoStop}%
\bibitem [{\citenamefont {Lackey}\ and\ \citenamefont
  {Wade}(2015)}]{Lackey:2014fwa}%
  \BibitemOpen
  \bibfield  {author} {\bibinfo {author} {\bibfnamefont {B.~D.}\ \bibnamefont
  {Lackey}}\ and\ \bibinfo {author} {\bibfnamefont {L.}~\bibnamefont {Wade}},\
  }\href {\doibase 10.1103/PhysRevD.91.043002} {\bibfield  {journal} {\bibinfo
  {journal} {Phys. Rev.}\ }\textbf {\bibinfo {volume} {D91}},\ \bibinfo {pages}
  {043002} (\bibinfo {year} {2015})},\ \Eprint {http://arxiv.org/abs/1410.8866}
  {arXiv:1410.8866 [gr-qc]} \BibitemShut {NoStop}%
\bibitem [{\citenamefont {Carney}\ \emph {et~al.}(2018)\citenamefont {Carney},
  \citenamefont {Wade},\ and\ \citenamefont {Irwin}}]{Carney:2018sdv}%
  \BibitemOpen
  \bibfield  {author} {\bibinfo {author} {\bibfnamefont {M.~F.}\ \bibnamefont
  {Carney}}, \bibinfo {author} {\bibfnamefont {L.~E.}\ \bibnamefont {Wade}}, \
  and\ \bibinfo {author} {\bibfnamefont {B.~S.}\ \bibnamefont {Irwin}},\ }\href
  {\doibase 10.1103/PhysRevD.98.063004} {\bibfield  {journal} {\bibinfo
  {journal} {Phys. Rev.}\ }\textbf {\bibinfo {volume} {D98}},\ \bibinfo {pages}
  {063004} (\bibinfo {year} {2018})},\ \Eprint
  {http://arxiv.org/abs/1805.11217} {arXiv:1805.11217 [gr-qc]} \BibitemShut
  {NoStop}%
\bibitem [{\citenamefont {Lindblom}(2010)}]{Lindblom:2010bb}%
  \BibitemOpen
  \bibfield  {author} {\bibinfo {author} {\bibfnamefont {L.}~\bibnamefont
  {Lindblom}},\ }\href {\doibase 10.1103/PhysRevD.82.103011} {\bibfield
  {journal} {\bibinfo  {journal} {Phys. Rev.}\ }\textbf {\bibinfo {volume}
  {D82}},\ \bibinfo {pages} {103011} (\bibinfo {year} {2010})},\ \Eprint
  {http://arxiv.org/abs/1009.0738} {arXiv:1009.0738 [astro-ph.HE]} \BibitemShut
  {NoStop}%
\bibitem [{\citenamefont {Lindblom}\ and\ \citenamefont
  {Indik}(2012)}]{Lindblom:2012zi}%
  \BibitemOpen
  \bibfield  {author} {\bibinfo {author} {\bibfnamefont {L.}~\bibnamefont
  {Lindblom}}\ and\ \bibinfo {author} {\bibfnamefont {N.~M.}\ \bibnamefont
  {Indik}},\ }\href {\doibase 10.1103/PhysRevD.86.084003} {\bibfield  {journal}
  {\bibinfo  {journal} {Phys. Rev.}\ }\textbf {\bibinfo {volume} {D86}},\
  \bibinfo {pages} {084003} (\bibinfo {year} {2012})},\ \Eprint
  {http://arxiv.org/abs/1207.3744} {arXiv:1207.3744 [astro-ph.HE]} \BibitemShut
  {NoStop}%
\bibitem [{\citenamefont {Lindblom}\ and\ \citenamefont
  {Indik}(2014)}]{Lindblom:2013kra}%
  \BibitemOpen
  \bibfield  {author} {\bibinfo {author} {\bibfnamefont {L.}~\bibnamefont
  {Lindblom}}\ and\ \bibinfo {author} {\bibfnamefont {N.~M.}\ \bibnamefont
  {Indik}},\ }\href {\doibase 10.1103/PhysRevD.89.064003,
  10.1103/PhysRevD.93.129903} {\bibfield  {journal} {\bibinfo  {journal} {Phys.
  Rev.}\ }\textbf {\bibinfo {volume} {D89}},\ \bibinfo {pages} {064003}
  (\bibinfo {year} {2014})},\ \bibinfo {note} {[Erratum: Phys.
  Rev.D93,no.12,129903(2016)]},\ \Eprint {http://arxiv.org/abs/1310.0803}
  {arXiv:1310.0803 [astro-ph.HE]} \BibitemShut {NoStop}%
\bibitem [{\citenamefont {Lindblom}(2018)}]{Lindblom:2018rfr}%
  \BibitemOpen
  \bibfield  {author} {\bibinfo {author} {\bibfnamefont {L.}~\bibnamefont
  {Lindblom}},\ }\href {\doibase 10.1103/PhysRevD.97.123019} {\bibfield
  {journal} {\bibinfo  {journal} {Phys. Rev.}\ }\textbf {\bibinfo {volume}
  {D97}},\ \bibinfo {pages} {123019} (\bibinfo {year} {2018})},\ \Eprint
  {http://arxiv.org/abs/1804.04072} {arXiv:1804.04072 [astro-ph.HE]}
  \BibitemShut {NoStop}%
\bibitem [{\citenamefont {Landry}\ and\ \citenamefont
  {Essick}(2019)}]{Landry:2018prl}%
  \BibitemOpen
  \bibfield  {author} {\bibinfo {author} {\bibfnamefont {P.}~\bibnamefont
  {Landry}}\ and\ \bibinfo {author} {\bibfnamefont {R.}~\bibnamefont
  {Essick}},\ }\href {\doibase 10.1103/PhysRevD.99.084049} {\bibfield
  {journal} {\bibinfo  {journal} {Phys. Rev.}\ }\textbf {\bibinfo {volume}
  {D99}},\ \bibinfo {pages} {084049} (\bibinfo {year} {2019})},\ \Eprint
  {http://arxiv.org/abs/1811.12529} {arXiv:1811.12529 [gr-qc]} \BibitemShut
  {NoStop}%
\bibitem [{\citenamefont {Alam}\ \emph {et~al.}(2016)\citenamefont {Alam},
  \citenamefont {Agrawal}, \citenamefont {Fortin}, \citenamefont {Pais},
  \citenamefont {Provid{\^{e}}ncia}, \citenamefont {Raduta},\ and\
  \citenamefont {Sulaksono}}]{Alam2016}%
  \BibitemOpen
  \bibfield  {author} {\bibinfo {author} {\bibfnamefont {N.}~\bibnamefont
  {Alam}}, \bibinfo {author} {\bibfnamefont {B.~K.}\ \bibnamefont {Agrawal}},
  \bibinfo {author} {\bibfnamefont {M.}~\bibnamefont {Fortin}}, \bibinfo
  {author} {\bibfnamefont {H.}~\bibnamefont {Pais}}, \bibinfo {author}
  {\bibfnamefont {C.}~\bibnamefont {Provid{\^{e}}ncia}}, \bibinfo {author}
  {\bibfnamefont {A.~R.}\ \bibnamefont {Raduta}}, \ and\ \bibinfo {author}
  {\bibfnamefont {A.}~\bibnamefont {Sulaksono}},\ }\href {\doibase
  10.1103/physrevc.94.052801} {\bibfield  {journal} {\bibinfo  {journal}
  {Physical Review C}\ }\textbf {\bibinfo {volume} {94}} (\bibinfo {year}
  {2016}),\ 10.1103/physrevc.94.052801}\BibitemShut {NoStop}%
\bibitem [{\citenamefont {Sotani}\ \emph {et~al.}(2014)\citenamefont {Sotani},
  \citenamefont {Iida}, \citenamefont {Oyamatsu},\ and\ \citenamefont
  {Ohnishi}}]{Sotani:2013dga}%
  \BibitemOpen
  \bibfield  {author} {\bibinfo {author} {\bibfnamefont {H.}~\bibnamefont
  {Sotani}}, \bibinfo {author} {\bibfnamefont {K.}~\bibnamefont {Iida}},
  \bibinfo {author} {\bibfnamefont {K.}~\bibnamefont {Oyamatsu}}, \ and\
  \bibinfo {author} {\bibfnamefont {A.}~\bibnamefont {Ohnishi}},\ }\href
  {\doibase 10.1093/ptep/ptu052} {\bibfield  {journal} {\bibinfo  {journal}
  {PTEP}\ }\textbf {\bibinfo {volume} {2014}},\ \bibinfo {pages} {051E01}
  (\bibinfo {year} {2014})},\ \Eprint {http://arxiv.org/abs/1401.0161}
  {arXiv:1401.0161 [astro-ph.HE]} \BibitemShut {NoStop}%
\bibitem [{\citenamefont {Silva}\ \emph {et~al.}(2016)\citenamefont {Silva},
  \citenamefont {Sotani},\ and\ \citenamefont {Berti}}]{Silva:2016myw}%
  \BibitemOpen
  \bibfield  {author} {\bibinfo {author} {\bibfnamefont {H.~O.}\ \bibnamefont
  {Silva}}, \bibinfo {author} {\bibfnamefont {H.}~\bibnamefont {Sotani}}, \
  and\ \bibinfo {author} {\bibfnamefont {E.}~\bibnamefont {Berti}},\ }\href
  {\doibase 10.1093/mnras/stw969} {\bibfield  {journal} {\bibinfo  {journal}
  {Mon. Not. Roy. Astron. Soc.}\ }\textbf {\bibinfo {volume} {459}},\ \bibinfo
  {pages} {4378} (\bibinfo {year} {2016})},\ \Eprint
  {http://arxiv.org/abs/1601.03407} {arXiv:1601.03407 [astro-ph.HE]}
  \BibitemShut {NoStop}%
\bibitem [{\citenamefont {Abbott}\ \emph
  {et~al.}(2017{\natexlab{b}})\citenamefont {Abbott} \emph
  {et~al.}}]{Abbott2017}%
  \BibitemOpen
  \bibfield  {author} {\bibinfo {author} {\bibfnamefont {B.}~\bibnamefont
  {Abbott}} \emph {et~al.},\ }\href {\doibase 10.1103/physrevlett.119.161101}
  {\bibfield  {journal} {\bibinfo  {journal} {Physical Review Letters}\
  }\textbf {\bibinfo {volume} {119}} (\bibinfo {year} {2017}{\natexlab{b}}),\
  10.1103/physrevlett.119.161101}\BibitemShut {NoStop}%
\bibitem [{\citenamefont {Oertel}\ \emph {et~al.}(2017)\citenamefont {Oertel},
  \citenamefont {Hempel}, \citenamefont {Kl\"ahn},\ and\ \citenamefont
  {Typel}}]{Oertel2017}%
  \BibitemOpen
  \bibfield  {author} {\bibinfo {author} {\bibfnamefont {M.}~\bibnamefont
  {Oertel}}, \bibinfo {author} {\bibfnamefont {M.}~\bibnamefont {Hempel}},
  \bibinfo {author} {\bibfnamefont {T.}~\bibnamefont {Kl\"ahn}}, \ and\
  \bibinfo {author} {\bibfnamefont {S.}~\bibnamefont {Typel}},\ }\href
  {\doibase 10.1103/RevModPhys.89.015007} {\bibfield  {journal} {\bibinfo
  {journal} {Rev. Mod. Phys.}\ }\textbf {\bibinfo {volume} {89}},\ \bibinfo
  {pages} {015007} (\bibinfo {year} {2017})}\BibitemShut {NoStop}%
\bibitem [{\citenamefont {Abbott}\ \emph
  {et~al.}(2019{\natexlab{b}})\citenamefont {Abbott} \emph
  {et~al.}}]{Abbott:LTposterior}%
  \BibitemOpen
  \bibfield  {author} {\bibinfo {author} {\bibfnamefont {B.~P.}\ \bibnamefont
  {Abbott}} \emph {et~al.} (\bibinfo {collaboration} {LIGO Scientific,
  Virgo}),\ }\href {\doibase 10.1103/PhysRevX.9.011001} {\bibfield  {journal}
  {\bibinfo  {journal} {Phys. Rev.}\ }\textbf {\bibinfo {volume} {X9}},\
  \bibinfo {pages} {011001} (\bibinfo {year} {2019}{\natexlab{b}})},\ \Eprint
  {http://arxiv.org/abs/1805.11579} {arXiv:1805.11579 [gr-qc]} \BibitemShut
  {NoStop}%
\bibitem [{\citenamefont {Margueron}\ and\ \citenamefont
  {Gulminelli}(2018)}]{Margueron:Ksym}%
  \BibitemOpen
  \bibfield  {author} {\bibinfo {author} {\bibfnamefont {J.}~\bibnamefont
  {Margueron}}\ and\ \bibinfo {author} {\bibfnamefont {F.}~\bibnamefont
  {Gulminelli}},\ }\href@noop {} {\  (\bibinfo {year} {2018})},\ \Eprint
  {http://arxiv.org/abs/1807.01729} {arXiv:1807.01729 [nucl-th]} \BibitemShut
  {NoStop}%
\bibitem [{\citenamefont {Mondal}\ \emph {et~al.}(2018)\citenamefont {Mondal},
  \citenamefont {Agrawal}, \citenamefont {De},\ and\ \citenamefont
  {Samaddar}}]{Mondal:Ksym}%
  \BibitemOpen
  \bibfield  {author} {\bibinfo {author} {\bibfnamefont {C.}~\bibnamefont
  {Mondal}}, \bibinfo {author} {\bibfnamefont {B.~K.}\ \bibnamefont {Agrawal}},
  \bibinfo {author} {\bibfnamefont {J.~N.}\ \bibnamefont {De}}, \ and\ \bibinfo
  {author} {\bibfnamefont {S.~K.}\ \bibnamefont {Samaddar}},\ }\href {\doibase
  10.1142/S0218301318500787} {\bibfield  {journal} {\bibinfo  {journal} {Int.
  J. Mod. Phys.}\ }\textbf {\bibinfo {volume} {E27}},\ \bibinfo {pages}
  {1850078} (\bibinfo {year} {2018})},\ \Eprint
  {http://arxiv.org/abs/1809.05354} {arXiv:1809.05354 [nucl-th]} \BibitemShut
  {NoStop}%
\bibitem [{\citenamefont {Hinderer}(2008)}]{hinderer-love}%
  \BibitemOpen
  \bibfield  {author} {\bibinfo {author} {\bibfnamefont {T.}~\bibnamefont
  {Hinderer}},\ }\href {http://stacks.iop.org/0004-637X/677/i=2/a=1216}
  {\bibfield  {journal} {\bibinfo  {journal} {The Astrophysical Journal}\
  }\textbf {\bibinfo {volume} {677}},\ \bibinfo {pages} {1216} (\bibinfo {year}
  {2008})}\BibitemShut {NoStop}%
\bibitem [{\citenamefont {Yagi}\ and\ \citenamefont {Yunes}(2013)}]{Yagi2013}%
  \BibitemOpen
  \bibfield  {author} {\bibinfo {author} {\bibfnamefont {K.}~\bibnamefont
  {Yagi}}\ and\ \bibinfo {author} {\bibfnamefont {N.}~\bibnamefont {Yunes}},\
  }\href {\doibase 10.1103/physrevd.88.023009} {\bibfield  {journal} {\bibinfo
  {journal} {Physical Review D}\ }\textbf {\bibinfo {volume} {88}} (\bibinfo
  {year} {2013}),\ 10.1103/physrevd.88.023009}\BibitemShut {NoStop}%
\bibitem [{\citenamefont {Damour}\ and\ \citenamefont
  {Nagar}(2009)}]{damour-nagar}%
  \BibitemOpen
  \bibfield  {author} {\bibinfo {author} {\bibfnamefont {T.}~\bibnamefont
  {Damour}}\ and\ \bibinfo {author} {\bibfnamefont {A.}~\bibnamefont {Nagar}},\
  }\href {\doibase 10.1103/PhysRevD.80.084035} {\bibfield  {journal} {\bibinfo
  {journal} {Phys.Rev.}\ }\textbf {\bibinfo {volume} {D80}},\ \bibinfo {pages}
  {084035} (\bibinfo {year} {2009})},\ \Eprint {http://arxiv.org/abs/0906.0096}
  {arXiv:0906.0096 [gr-qc]} \BibitemShut {NoStop}%
\bibitem [{\citenamefont {Yagi}\ and\ \citenamefont
  {Yunes}(2016)}]{Yagi:2015pkc}%
  \BibitemOpen
  \bibfield  {author} {\bibinfo {author} {\bibfnamefont {K.}~\bibnamefont
  {Yagi}}\ and\ \bibinfo {author} {\bibfnamefont {N.}~\bibnamefont {Yunes}},\
  }\href {\doibase 10.1088/0264-9381/33/13/13LT01} {\bibfield  {journal}
  {\bibinfo  {journal} {Class. Quant. Grav.}\ }\textbf {\bibinfo {volume}
  {33}},\ \bibinfo {pages} {13LT01} (\bibinfo {year} {2016})},\ \Eprint
  {http://arxiv.org/abs/1512.02639} {arXiv:1512.02639 [gr-qc]} \BibitemShut
  {NoStop}%
\bibitem [{\citenamefont {Yagi}\ and\ \citenamefont
  {Yunes}(2017)}]{Yagi:2016qmr}%
  \BibitemOpen
  \bibfield  {author} {\bibinfo {author} {\bibfnamefont {K.}~\bibnamefont
  {Yagi}}\ and\ \bibinfo {author} {\bibfnamefont {N.}~\bibnamefont {Yunes}},\
  }\href {\doibase 10.1088/1361-6382/34/1/015006} {\bibfield  {journal}
  {\bibinfo  {journal} {Class. Quant. Grav.}\ }\textbf {\bibinfo {volume}
  {34}},\ \bibinfo {pages} {015006} (\bibinfo {year} {2017})},\ \Eprint
  {http://arxiv.org/abs/1608.06187} {arXiv:1608.06187 [gr-qc]} \BibitemShut
  {NoStop}%
\bibitem [{\citenamefont {Zhao}\ and\ \citenamefont
  {Lattimer}(2018)}]{Zhao:2018nyf}%
  \BibitemOpen
  \bibfield  {author} {\bibinfo {author} {\bibfnamefont {T.}~\bibnamefont
  {Zhao}}\ and\ \bibinfo {author} {\bibfnamefont {J.~M.}\ \bibnamefont
  {Lattimer}},\ }\href {\doibase 10.1103/PhysRevD.98.063020} {\bibfield
  {journal} {\bibinfo  {journal} {Phys. Rev.}\ }\textbf {\bibinfo {volume}
  {D98}},\ \bibinfo {pages} {063020} (\bibinfo {year} {2018})},\ \Eprint
  {http://arxiv.org/abs/1808.02858} {arXiv:1808.02858 [astro-ph.HE]}
  \BibitemShut {NoStop}%
\bibitem [{\citenamefont {Vida{\~{n}}a}\ \emph {et~al.}(2009)\citenamefont
  {Vida{\~{n}}a}, \citenamefont {Provid{\^{e}}ncia}, \citenamefont {Polls},\
  and\ \citenamefont {Rios}}]{Vidana2009}%
  \BibitemOpen
  \bibfield  {author} {\bibinfo {author} {\bibfnamefont {I.}~\bibnamefont
  {Vida{\~{n}}a}}, \bibinfo {author} {\bibfnamefont {C.}~\bibnamefont
  {Provid{\^{e}}ncia}}, \bibinfo {author} {\bibfnamefont {A.}~\bibnamefont
  {Polls}}, \ and\ \bibinfo {author} {\bibfnamefont {A.}~\bibnamefont {Rios}},\
  }\href {\doibase 10.1103/physrevc.80.045806} {\bibfield  {journal} {\bibinfo
  {journal} {Physical Review C}\ }\textbf {\bibinfo {volume} {80}} (\bibinfo
  {year} {2009}),\ 10.1103/physrevc.80.045806}\BibitemShut {NoStop}%
\bibitem [{\citenamefont {Alam}\ \emph {et~al.}(2014)\citenamefont {Alam},
  \citenamefont {Agrawal}, \citenamefont {De}, \citenamefont {Samaddar},\ and\
  \citenamefont {Col{\`{o}}}}]{Alam2014}%
  \BibitemOpen
  \bibfield  {author} {\bibinfo {author} {\bibfnamefont {N.}~\bibnamefont
  {Alam}}, \bibinfo {author} {\bibfnamefont {B.~K.}\ \bibnamefont {Agrawal}},
  \bibinfo {author} {\bibfnamefont {J.~N.}\ \bibnamefont {De}}, \bibinfo
  {author} {\bibfnamefont {S.~K.}\ \bibnamefont {Samaddar}}, \ and\ \bibinfo
  {author} {\bibfnamefont {G.}~\bibnamefont {Col{\`{o}}}},\ }\href {\doibase
  10.1103/physrevc.90.054317} {\bibfield  {journal} {\bibinfo  {journal}
  {Physical Review C}\ }\textbf {\bibinfo {volume} {90}} (\bibinfo {year}
  {2014}),\ 10.1103/physrevc.90.054317}\BibitemShut {NoStop}%
\bibitem [{\citenamefont {Lattimer}\ and\ \citenamefont
  {Lim}(2013)}]{Lattimer2013}%
  \BibitemOpen
  \bibfield  {author} {\bibinfo {author} {\bibfnamefont {J.~M.}\ \bibnamefont
  {Lattimer}}\ and\ \bibinfo {author} {\bibfnamefont {Y.}~\bibnamefont {Lim}},\
  }\href {\doibase 10.1088/0004-637x/771/1/51} {\bibfield  {journal} {\bibinfo
  {journal} {The Astrophysical Journal}\ }\textbf {\bibinfo {volume} {771}},\
  \bibinfo {pages} {51} (\bibinfo {year} {2013})}\BibitemShut {NoStop}%
\bibitem [{\citenamefont {Tews}\ \emph {et~al.}(2017)\citenamefont {Tews},
  \citenamefont {Lattimer}, \citenamefont {Ohnishi},\ and\ \citenamefont
  {Kolomeitsev}}]{Tews2017}%
  \BibitemOpen
  \bibfield  {author} {\bibinfo {author} {\bibfnamefont {I.}~\bibnamefont
  {Tews}}, \bibinfo {author} {\bibfnamefont {J.~M.}\ \bibnamefont {Lattimer}},
  \bibinfo {author} {\bibfnamefont {A.}~\bibnamefont {Ohnishi}}, \ and\
  \bibinfo {author} {\bibfnamefont {E.~E.}\ \bibnamefont {Kolomeitsev}},\
  }\href {\doibase 10.3847/1538-4357/aa8db9} {\bibfield  {journal} {\bibinfo
  {journal} {The Astrophysical Journal}\ }\textbf {\bibinfo {volume} {848}},\
  \bibinfo {pages} {105} (\bibinfo {year} {2017})}\BibitemShut {NoStop}%
\bibitem [{\citenamefont {Douchin}\ and\ \citenamefont
  {Haensel}(2001)}]{Douchin:2001sv}%
  \BibitemOpen
  \bibfield  {author} {\bibinfo {author} {\bibfnamefont {F.}~\bibnamefont
  {Douchin}}\ and\ \bibinfo {author} {\bibfnamefont {P.}~\bibnamefont
  {Haensel}},\ }\href {\doibase 10.1051/0004-6361:20011402} {\bibfield
  {journal} {\bibinfo  {journal} {Astron. Astrophys.}\ }\textbf {\bibinfo
  {volume} {380}},\ \bibinfo {pages} {151} (\bibinfo {year} {2001})},\ \Eprint
  {http://arxiv.org/abs/astro-ph/0111092} {arXiv:astro-ph/0111092 [astro-ph]}
  \BibitemShut {NoStop}%
\bibitem [{\citenamefont {KÃ¶hler}(1976)}]{Koehler1976}%
  \BibitemOpen
  \bibfield  {author} {\bibinfo {author} {\bibfnamefont {H.}~\bibnamefont
  {KÃ¶hler}},\ }\href {\doibase 10.1016/0375-9474(76)90008-7} {\bibfield
  {journal} {\bibinfo  {journal} {Nuclear Physics A}\ }\textbf {\bibinfo
  {volume} {258}},\ \bibinfo {pages} {301} (\bibinfo {year}
  {1976})}\BibitemShut {NoStop}%
\bibitem [{\citenamefont {Reinhard}\ and\ \citenamefont
  {Flocard}(1995)}]{Reinhard1995}%
  \BibitemOpen
  \bibfield  {author} {\bibinfo {author} {\bibfnamefont {P.-G.}\ \bibnamefont
  {Reinhard}}\ and\ \bibinfo {author} {\bibfnamefont {H.}~\bibnamefont
  {Flocard}},\ }\href {\doibase 10.1016/0375-9474(94)00770-n} {\bibfield
  {journal} {\bibinfo  {journal} {Nuclear Physics A}\ }\textbf {\bibinfo
  {volume} {584}},\ \bibinfo {pages} {467} (\bibinfo {year}
  {1995})}\BibitemShut {NoStop}%
\bibitem [{\citenamefont {Nazarewicz}\ \emph {et~al.}(1996)\citenamefont
  {Nazarewicz}, \citenamefont {Dobaczewski}, \citenamefont {Werner},
  \citenamefont {Maruhn}, \citenamefont {Reinhard}, \citenamefont {Rutz},
  \citenamefont {Chinn}, \citenamefont {Umar},\ and\ \citenamefont
  {Strayer}}]{Nazarewicz1996}%
  \BibitemOpen
  \bibfield  {author} {\bibinfo {author} {\bibfnamefont {W.}~\bibnamefont
  {Nazarewicz}}, \bibinfo {author} {\bibfnamefont {J.}~\bibnamefont
  {Dobaczewski}}, \bibinfo {author} {\bibfnamefont {T.~R.}\ \bibnamefont
  {Werner}}, \bibinfo {author} {\bibfnamefont {J.~A.}\ \bibnamefont {Maruhn}},
  \bibinfo {author} {\bibfnamefont {P.-G.}\ \bibnamefont {Reinhard}}, \bibinfo
  {author} {\bibfnamefont {K.}~\bibnamefont {Rutz}}, \bibinfo {author}
  {\bibfnamefont {C.~R.}\ \bibnamefont {Chinn}}, \bibinfo {author}
  {\bibfnamefont {A.~S.}\ \bibnamefont {Umar}}, \ and\ \bibinfo {author}
  {\bibfnamefont {M.~R.}\ \bibnamefont {Strayer}},\ }\href {\doibase
  10.1103/physrevc.53.740} {\bibfield  {journal} {\bibinfo  {journal} {Physical
  Review C}\ }\textbf {\bibinfo {volume} {53}},\ \bibinfo {pages} {740}
  (\bibinfo {year} {1996})}\BibitemShut {NoStop}%
\bibitem [{\citenamefont {Chabanat}\ \emph {et~al.}(1997)\citenamefont
  {Chabanat}, \citenamefont {Bonche}, \citenamefont {Haensel}, \citenamefont
  {Meyer},\ and\ \citenamefont {Schaeffer}}]{Chabanat1997}%
  \BibitemOpen
  \bibfield  {author} {\bibinfo {author} {\bibfnamefont {E.}~\bibnamefont
  {Chabanat}}, \bibinfo {author} {\bibfnamefont {P.}~\bibnamefont {Bonche}},
  \bibinfo {author} {\bibfnamefont {P.}~\bibnamefont {Haensel}}, \bibinfo
  {author} {\bibfnamefont {J.}~\bibnamefont {Meyer}}, \ and\ \bibinfo {author}
  {\bibfnamefont {R.}~\bibnamefont {Schaeffer}},\ }\href {\doibase
  10.1016/s0375-9474(97)00596-4} {\bibfield  {journal} {\bibinfo  {journal}
  {Nuclear Physics A}\ }\textbf {\bibinfo {volume} {627}},\ \bibinfo {pages}
  {710} (\bibinfo {year} {1997})}\BibitemShut {NoStop}%
\bibitem [{\citenamefont {Chabanat}(1995)}]{Chabanat1995}%
  \BibitemOpen
  \bibfield  {author} {\bibinfo {author} {\bibfnamefont {E.}~\bibnamefont
  {Chabanat}},\ }\emph {\bibinfo {title} {Interactions effectives pour des
  conditions extremes d'isospin}},\ \href@noop {} {Ph.D. thesis},\ \bibinfo
  {school} {University Claude Bernard Lyon-I} (\bibinfo {year}
  {1995})\BibitemShut {NoStop}%
\bibitem [{\citenamefont {Chabanat}\ \emph {et~al.}(1998)\citenamefont
  {Chabanat}, \citenamefont {Bonche}, \citenamefont {Haensel}, \citenamefont
  {Meyer},\ and\ \citenamefont {Schaeffer}}]{Chabanat1998}%
  \BibitemOpen
  \bibfield  {author} {\bibinfo {author} {\bibfnamefont {E.}~\bibnamefont
  {Chabanat}}, \bibinfo {author} {\bibfnamefont {P.}~\bibnamefont {Bonche}},
  \bibinfo {author} {\bibfnamefont {P.}~\bibnamefont {Haensel}}, \bibinfo
  {author} {\bibfnamefont {J.}~\bibnamefont {Meyer}}, \ and\ \bibinfo {author}
  {\bibfnamefont {R.}~\bibnamefont {Schaeffer}},\ }\href {\doibase
  10.1016/s0375-9474(98)00180-8} {\bibfield  {journal} {\bibinfo  {journal}
  {Nuclear Physics A}\ }\textbf {\bibinfo {volume} {635}},\ \bibinfo {pages}
  {231} (\bibinfo {year} {1998})}\BibitemShut {NoStop}%
\bibitem [{\citenamefont {Bennour}\ \emph {et~al.}(1989)\citenamefont
  {Bennour}, \citenamefont {Heenen}, \citenamefont {Bonche}, \citenamefont
  {Dobaczewski},\ and\ \citenamefont {Flocard}}]{Bennour1989}%
  \BibitemOpen
  \bibfield  {author} {\bibinfo {author} {\bibfnamefont {L.}~\bibnamefont
  {Bennour}}, \bibinfo {author} {\bibfnamefont {P.-H.}\ \bibnamefont {Heenen}},
  \bibinfo {author} {\bibfnamefont {P.}~\bibnamefont {Bonche}}, \bibinfo
  {author} {\bibfnamefont {J.}~\bibnamefont {Dobaczewski}}, \ and\ \bibinfo
  {author} {\bibfnamefont {H.}~\bibnamefont {Flocard}},\ }\href {\doibase
  10.1103/physrevc.40.2834} {\bibfield  {journal} {\bibinfo  {journal}
  {Physical Review C}\ }\textbf {\bibinfo {volume} {40}},\ \bibinfo {pages}
  {2834} (\bibinfo {year} {1989})}\BibitemShut {NoStop}%
\bibitem [{\citenamefont {Reinhard}(1999)}]{Reinhard1999}%
  \BibitemOpen
  \bibfield  {author} {\bibinfo {author} {\bibfnamefont {P.-G.}\ \bibnamefont
  {Reinhard}},\ }\href {\doibase 10.1016/s0375-9474(99)00076-7} {\bibfield
  {journal} {\bibinfo  {journal} {Nuclear Physics A}\ }\textbf {\bibinfo
  {volume} {649}},\ \bibinfo {pages} {305} (\bibinfo {year}
  {1999})}\BibitemShut {NoStop}%
\bibitem [{\citenamefont {Agrawal}\ \emph {et~al.}(2005)\citenamefont
  {Agrawal}, \citenamefont {Shlomo},\ and\ \citenamefont {Au}}]{Agrawal2005}%
  \BibitemOpen
  \bibfield  {author} {\bibinfo {author} {\bibfnamefont {B.~K.}\ \bibnamefont
  {Agrawal}}, \bibinfo {author} {\bibfnamefont {S.}~\bibnamefont {Shlomo}}, \
  and\ \bibinfo {author} {\bibfnamefont {V.~K.}\ \bibnamefont {Au}},\ }\href
  {\doibase 10.1103/physrevc.72.014310} {\bibfield  {journal} {\bibinfo
  {journal} {Physical Review C}\ }\textbf {\bibinfo {volume} {72}} (\bibinfo
  {year} {2005}),\ 10.1103/physrevc.72.014310}\BibitemShut {NoStop}%
\bibitem [{\citenamefont {Agrawal}\ \emph {et~al.}(2003)\citenamefont
  {Agrawal}, \citenamefont {Shlomo},\ and\ \citenamefont {Au}}]{Agrawal2003}%
  \BibitemOpen
  \bibfield  {author} {\bibinfo {author} {\bibfnamefont {B.~K.}\ \bibnamefont
  {Agrawal}}, \bibinfo {author} {\bibfnamefont {S.}~\bibnamefont {Shlomo}}, \
  and\ \bibinfo {author} {\bibfnamefont {V.~K.}\ \bibnamefont {Au}},\ }\href
  {\doibase 10.1103/physrevc.68.031304} {\bibfield  {journal} {\bibinfo
  {journal} {Physical Review C}\ }\textbf {\bibinfo {volume} {68}} (\bibinfo
  {year} {2003}),\ 10.1103/physrevc.68.031304}\BibitemShut {NoStop}%
\bibitem [{\citenamefont {Friedrich}\ and\ \citenamefont
  {Reinhard}(1986)}]{Friedrich1986}%
  \BibitemOpen
  \bibfield  {author} {\bibinfo {author} {\bibfnamefont {J.}~\bibnamefont
  {Friedrich}}\ and\ \bibinfo {author} {\bibfnamefont {P.-G.}\ \bibnamefont
  {Reinhard}},\ }\href {\doibase 10.1103/physrevc.33.335} {\bibfield  {journal}
  {\bibinfo  {journal} {Physical Review C}\ }\textbf {\bibinfo {volume} {33}},\
  \bibinfo {pages} {335} (\bibinfo {year} {1986})}\BibitemShut {NoStop}%
\bibitem [{\citenamefont {Goriely}\ \emph {et~al.}(2010)\citenamefont
  {Goriely}, \citenamefont {Chamel},\ and\ \citenamefont
  {Pearson}}]{Goriely2010}%
  \BibitemOpen
  \bibfield  {author} {\bibinfo {author} {\bibfnamefont {S.}~\bibnamefont
  {Goriely}}, \bibinfo {author} {\bibfnamefont {N.}~\bibnamefont {Chamel}}, \
  and\ \bibinfo {author} {\bibfnamefont {J.~M.}\ \bibnamefont {Pearson}},\
  }\href {\doibase 10.1103/physrevc.82.035804} {\bibfield  {journal} {\bibinfo
  {journal} {Physical Review C}\ }\textbf {\bibinfo {volume} {82}} (\bibinfo
  {year} {2010}),\ 10.1103/physrevc.82.035804}\BibitemShut {NoStop}%
\bibitem [{\citenamefont {Goriely}\ \emph {et~al.}(2013)\citenamefont
  {Goriely}, \citenamefont {Chamel},\ and\ \citenamefont
  {Pearson}}]{Goriely2013}%
  \BibitemOpen
  \bibfield  {author} {\bibinfo {author} {\bibfnamefont {S.}~\bibnamefont
  {Goriely}}, \bibinfo {author} {\bibfnamefont {N.}~\bibnamefont {Chamel}}, \
  and\ \bibinfo {author} {\bibfnamefont {J.~M.}\ \bibnamefont {Pearson}},\
  }\href {\doibase 10.1103/physrevc.88.024308} {\bibfield  {journal} {\bibinfo
  {journal} {Physical Review C}\ }\textbf {\bibinfo {volume} {88}} (\bibinfo
  {year} {2013}),\ 10.1103/physrevc.88.024308}\BibitemShut {NoStop}%
\bibitem [{\citenamefont {Dhiman}\ \emph {et~al.}(2007)\citenamefont {Dhiman},
  \citenamefont {Kumar},\ and\ \citenamefont {Agrawal}}]{Dhiman2007}%
  \BibitemOpen
  \bibfield  {author} {\bibinfo {author} {\bibfnamefont {S.~K.}\ \bibnamefont
  {Dhiman}}, \bibinfo {author} {\bibfnamefont {R.}~\bibnamefont {Kumar}}, \
  and\ \bibinfo {author} {\bibfnamefont {B.~K.}\ \bibnamefont {Agrawal}},\
  }\href {\doibase 10.1103/physrevc.76.045801} {\bibfield  {journal} {\bibinfo
  {journal} {Physical Review C}\ }\textbf {\bibinfo {volume} {76}} (\bibinfo
  {year} {2007}),\ 10.1103/physrevc.76.045801}\BibitemShut {NoStop}%
\bibitem [{\citenamefont {Agrawal}(2010)}]{Agrawal2010}%
  \BibitemOpen
  \bibfield  {author} {\bibinfo {author} {\bibfnamefont {B.~K.}\ \bibnamefont
  {Agrawal}},\ }\href {\doibase 10.1103/physrevc.81.034323} {\bibfield
  {journal} {\bibinfo  {journal} {Physical Review C}\ }\textbf {\bibinfo
  {volume} {81}} (\bibinfo {year} {2010}),\
  10.1103/physrevc.81.034323}\BibitemShut {NoStop}%
\bibitem [{\citenamefont {Glendenning}(1991)}]{Glendenning1991}%
  \BibitemOpen
  \bibfield  {author} {\bibinfo {author} {\bibfnamefont {N.}~\bibnamefont
  {Glendenning}},\ }\href {\doibase 10.1103/PhysRevLett.67.2414} {\bibfield
  {journal} {\bibinfo  {journal} {Physical Review Letters}\ } (\bibinfo {year}
  {1991}),\ 10.1103/PhysRevLett.67.2414}\BibitemShut {NoStop}%
\bibitem [{\citenamefont {Lalazissis}\ \emph {et~al.}(1997)\citenamefont
  {Lalazissis}, \citenamefont {KÃ¶nig},\ and\ \citenamefont
  {Ring}}]{Lalazissis1997}%
  \BibitemOpen
  \bibfield  {author} {\bibinfo {author} {\bibfnamefont {G.~A.}\ \bibnamefont
  {Lalazissis}}, \bibinfo {author} {\bibfnamefont {J.}~\bibnamefont
  {KÃ¶nig}}, \ and\ \bibinfo {author} {\bibfnamefont {P.}~\bibnamefont
  {Ring}},\ }\href {\doibase 10.1103/physrevc.55.540} {\bibfield  {journal}
  {\bibinfo  {journal} {Physical Review C}\ }\textbf {\bibinfo {volume} {55}},\
  \bibinfo {pages} {540} (\bibinfo {year} {1997})}\BibitemShut {NoStop}%
\bibitem [{\citenamefont {Carriere}\ \emph {et~al.}(2003)\citenamefont
  {Carriere}, \citenamefont {Horowitz},\ and\ \citenamefont
  {Piekarewicz}}]{Carriere2003}%
  \BibitemOpen
  \bibfield  {author} {\bibinfo {author} {\bibfnamefont {J.}~\bibnamefont
  {Carriere}}, \bibinfo {author} {\bibfnamefont {C.~J.}\ \bibnamefont
  {Horowitz}}, \ and\ \bibinfo {author} {\bibfnamefont {J.}~\bibnamefont
  {Piekarewicz}},\ }\href {\doibase 10.1086/376515} {\bibfield  {journal}
  {\bibinfo  {journal} {The Astrophysical Journal}\ }\textbf {\bibinfo {volume}
  {593}},\ \bibinfo {pages} {463} (\bibinfo {year} {2003})}\BibitemShut
  {NoStop}%
\bibitem [{\citenamefont {Sugahara}\ and\ \citenamefont
  {Toki}(1994)}]{Sugahara1994}%
  \BibitemOpen
  \bibfield  {author} {\bibinfo {author} {\bibfnamefont {Y.}~\bibnamefont
  {Sugahara}}\ and\ \bibinfo {author} {\bibfnamefont {H.}~\bibnamefont
  {Toki}},\ }\href {\doibase 10.1016/0375-9474(94)90923-7} {\bibfield
  {journal} {\bibinfo  {journal} {Nuclear Physics A}\ }\textbf {\bibinfo
  {volume} {579}},\ \bibinfo {pages} {557} (\bibinfo {year}
  {1994})}\BibitemShut {NoStop}%
\bibitem [{\citenamefont {Typel}\ \emph {et~al.}(2010)\citenamefont {Typel},
  \citenamefont {RÃ¶pke}, \citenamefont {KlÃ€hn}, \citenamefont
  {Blaschke},\ and\ \citenamefont {Wolter}}]{Typel2010}%
  \BibitemOpen
  \bibfield  {author} {\bibinfo {author} {\bibfnamefont {S.}~\bibnamefont
  {Typel}}, \bibinfo {author} {\bibfnamefont {G.}~\bibnamefont {RÃ¶pke}},
  \bibinfo {author} {\bibfnamefont {T.}~\bibnamefont {KlÃ€hn}}, \bibinfo
  {author} {\bibfnamefont {D.}~\bibnamefont {Blaschke}}, \ and\ \bibinfo
  {author} {\bibfnamefont {H.~H.}\ \bibnamefont {Wolter}},\ }\href {\doibase
  10.1103/physrevc.81.015803} {\bibfield  {journal} {\bibinfo  {journal}
  {Physical Review C}\ }\textbf {\bibinfo {volume} {81}} (\bibinfo {year}
  {2010}),\ 10.1103/physrevc.81.015803}\BibitemShut {NoStop}%
\bibitem [{\citenamefont {Gaitanos}\ \emph {et~al.}(2004)\citenamefont
  {Gaitanos}, \citenamefont {Toro}, \citenamefont {Typel}, \citenamefont
  {Baran}, \citenamefont {Fuchs}, \citenamefont {Greco},\ and\ \citenamefont
  {Wolter}}]{Gaitanos2004}%
  \BibitemOpen
  \bibfield  {author} {\bibinfo {author} {\bibfnamefont {T.}~\bibnamefont
  {Gaitanos}}, \bibinfo {author} {\bibfnamefont {M.~D.}\ \bibnamefont {Toro}},
  \bibinfo {author} {\bibfnamefont {S.}~\bibnamefont {Typel}}, \bibinfo
  {author} {\bibfnamefont {V.}~\bibnamefont {Baran}}, \bibinfo {author}
  {\bibfnamefont {C.}~\bibnamefont {Fuchs}}, \bibinfo {author} {\bibfnamefont
  {V.}~\bibnamefont {Greco}}, \ and\ \bibinfo {author} {\bibfnamefont
  {H.}~\bibnamefont {Wolter}},\ }\href {\doibase
  10.1016/j.nuclphysa.2003.12.001} {\bibfield  {journal} {\bibinfo  {journal}
  {Nuclear Physics A}\ }\textbf {\bibinfo {volume} {732}},\ \bibinfo {pages}
  {24} (\bibinfo {year} {2004})}\BibitemShut {NoStop}%
\bibitem [{\citenamefont {Typel}\ and\ \citenamefont
  {Wolter}(1999)}]{Typel1999}%
  \BibitemOpen
  \bibfield  {author} {\bibinfo {author} {\bibfnamefont {S.}~\bibnamefont
  {Typel}}\ and\ \bibinfo {author} {\bibfnamefont {H.}~\bibnamefont {Wolter}},\
  }\href {\doibase 10.1016/s0375-9474(99)00310-3} {\bibfield  {journal}
  {\bibinfo  {journal} {Nuclear Physics A}\ }\textbf {\bibinfo {volume}
  {656}},\ \bibinfo {pages} {331} (\bibinfo {year} {1999})}\BibitemShut
  {NoStop}%
\bibitem [{\citenamefont {Margueron}\ \emph {et~al.}(2018)\citenamefont
  {Margueron}, \citenamefont {Hoffmann~Casali},\ and\ \citenamefont
  {Gulminelli}}]{Margueron2018}%
  \BibitemOpen
  \bibfield  {author} {\bibinfo {author} {\bibfnamefont {J.}~\bibnamefont
  {Margueron}}, \bibinfo {author} {\bibfnamefont {R.}~\bibnamefont
  {Hoffmann~Casali}}, \ and\ \bibinfo {author} {\bibfnamefont {F.}~\bibnamefont
  {Gulminelli}},\ }\href {\doibase 10.1103/PhysRevC.97.025806} {\bibfield
  {journal} {\bibinfo  {journal} {Phys. Rev. C}\ }\textbf {\bibinfo {volume}
  {97}},\ \bibinfo {pages} {025806} (\bibinfo {year} {2018})}\BibitemShut
  {NoStop}%
\bibitem [{\citenamefont {Jensen}(2007)}]{jensen_2007}%
  \BibitemOpen
  \bibfield  {author} {\bibinfo {author} {\bibfnamefont {J.~L.}\ \bibnamefont
  {Jensen}},\ }\href@noop {} {\emph {\bibinfo {title} {Statistics for petroleum
  engineers and geoscientists}}}\ (\bibinfo  {publisher} {Elsevier},\ \bibinfo
  {year} {2007})\BibitemShut {NoStop}%
\bibitem [{aLI()}]{aLIGO}%
  \BibitemOpen
  \href {https://www.advancedligo.mit,.edu/} {\enquote {\bibinfo {title}
  {Advanced {LIGO}},}\ }\bibinfo {howpublished}
  {\url{https://www.advancedligo.mit.edu/}}\BibitemShut {NoStop}%
\bibitem [{Ap_()}]{Ap_Voyager_CE}%
  \BibitemOpen
  \href {https://dcc.ligo.org/ligo-T1400316/public} {\enquote {\bibinfo {title}
  {Ligo-t1400316-v4: Instrument science white paper},}\ }\bibinfo
  {howpublished} {\url{https://dcc.ligo.org/ligo-T1400316/public}}\BibitemShut
  {NoStop}%
\bibitem [{ET()}]{ET}%
  \BibitemOpen
  \href {http://www.et-gw.eu/} {\enquote {\bibinfo {title} {The {ET} project
  website},}\ }\bibinfo {howpublished} {\url{http://www.et-gw.eu/}}\BibitemShut
  {NoStop}%
\bibitem [{\citenamefont {Carson}\ \emph {et~al.}(2019)\citenamefont {Carson},
  \citenamefont {Steiner},\ and\ \citenamefont
  {Yagi}}]{Zack:futureNuclearConstraints}%
  \BibitemOpen
  \bibfield  {author} {\bibinfo {author} {\bibfnamefont {Z.}~\bibnamefont
  {Carson}}, \bibinfo {author} {\bibfnamefont {A.~W.}\ \bibnamefont {Steiner}},
  \ and\ \bibinfo {author} {\bibfnamefont {K.}~\bibnamefont {Yagi}},\ }\href
  {\doibase 10.1103/PhysRevD.100.023012} {\bibfield  {journal} {\bibinfo
  {journal} {Phys. Rev.}\ }\textbf {\bibinfo {volume} {D100}},\ \bibinfo
  {pages} {023012} (\bibinfo {year} {2019})},\ \Eprint
  {http://arxiv.org/abs/1906.05978} {arXiv:1906.05978 [gr-qc]} \BibitemShut
  {NoStop}%
\bibitem [{\citenamefont {Alford}\ and\ \citenamefont
  {Sedrakian}(2017)}]{Alford:2017qgh}%
  \BibitemOpen
  \bibfield  {author} {\bibinfo {author} {\bibfnamefont {M.~G.}\ \bibnamefont
  {Alford}}\ and\ \bibinfo {author} {\bibfnamefont {A.}~\bibnamefont
  {Sedrakian}},\ }\href {\doibase 10.1103/PhysRevLett.119.161104} {\bibfield
  {journal} {\bibinfo  {journal} {Phys. Rev. Lett.}\ }\textbf {\bibinfo
  {volume} {119}},\ \bibinfo {pages} {161104} (\bibinfo {year} {2017})},\
  \Eprint {http://arxiv.org/abs/1706.01592} {arXiv:1706.01592 [astro-ph.HE]}
  \BibitemShut {NoStop}%
\bibitem [{\citenamefont {Montana}\ \emph {et~al.}(2019)\citenamefont
  {Montana}, \citenamefont {Tolos}, \citenamefont {Hanauske},\ and\
  \citenamefont {Rezzolla}}]{Montana:2018bkb}%
  \BibitemOpen
  \bibfield  {author} {\bibinfo {author} {\bibfnamefont {G.}~\bibnamefont
  {Montana}}, \bibinfo {author} {\bibfnamefont {L.}~\bibnamefont {Tolos}},
  \bibinfo {author} {\bibfnamefont {M.}~\bibnamefont {Hanauske}}, \ and\
  \bibinfo {author} {\bibfnamefont {L.}~\bibnamefont {Rezzolla}},\ }\href
  {\doibase 10.1103/PhysRevD.99.103009} {\bibfield  {journal} {\bibinfo
  {journal} {Phys. Rev.}\ }\textbf {\bibinfo {volume} {D99}},\ \bibinfo {pages}
  {103009} (\bibinfo {year} {2019})},\ \Eprint
  {http://arxiv.org/abs/1811.10929} {arXiv:1811.10929 [astro-ph.HE]}
  \BibitemShut {NoStop}%
\bibitem [{\citenamefont {{Seidov}}(1971)}]{1971SvA....15..347S}%
  \BibitemOpen
  \bibfield  {author} {\bibinfo {author} {\bibfnamefont {Z.~F.}\ \bibnamefont
  {{Seidov}}},\ }\href@noop {} {\bibfield  {journal} {\bibinfo  {journal} {Sov.
  Ast.}\ }\textbf {\bibinfo {volume} {15}},\ \bibinfo {pages} {347} (\bibinfo
  {year} {1971})}\BibitemShut {NoStop}%
\bibitem [{\citenamefont {Zdunik}\ and\ \citenamefont
  {Haensel}(2013)}]{Zdunik:2012dj}%
  \BibitemOpen
  \bibfield  {author} {\bibinfo {author} {\bibfnamefont {J.~L.}\ \bibnamefont
  {Zdunik}}\ and\ \bibinfo {author} {\bibfnamefont {P.}~\bibnamefont
  {Haensel}},\ }\href {\doibase 10.1051/0004-6361/201220697} {\bibfield
  {journal} {\bibinfo  {journal} {Astron. Astrophys.}\ }\textbf {\bibinfo
  {volume} {551}},\ \bibinfo {pages} {A61} (\bibinfo {year} {2013})},\ \Eprint
  {http://arxiv.org/abs/1211.1231} {arXiv:1211.1231 [astro-ph.SR]} \BibitemShut
  {NoStop}%
\bibitem [{\citenamefont {Alford}\ \emph {et~al.}(2013)\citenamefont {Alford},
  \citenamefont {Han},\ and\ \citenamefont {Prakash}}]{Alford:2013aca}%
  \BibitemOpen
  \bibfield  {author} {\bibinfo {author} {\bibfnamefont {M.~G.}\ \bibnamefont
  {Alford}}, \bibinfo {author} {\bibfnamefont {S.}~\bibnamefont {Han}}, \ and\
  \bibinfo {author} {\bibfnamefont {M.}~\bibnamefont {Prakash}},\ }\href
  {\doibase 10.1103/PhysRevD.88.083013} {\bibfield  {journal} {\bibinfo
  {journal} {Phys. Rev.}\ }\textbf {\bibinfo {volume} {D88}},\ \bibinfo {pages}
  {083013} (\bibinfo {year} {2013})},\ \Eprint {http://arxiv.org/abs/1302.4732}
  {arXiv:1302.4732 [astro-ph.SR]} \BibitemShut {NoStop}%
\end{thebibliography}%
\end{document}